\documentclass[a4paper,11pt]{article}
\pdfoutput=1 

\usepackage{jcappub} 

\usepackage{xcolor}

\usepackage[T1]{fontenc} 
\usepackage{framed}

\usepackage{subcaption}

\title{\boldmath Primordial features as probes of baryogenesis from supersymmetric flat directions}

\author[a]{Yi-Peng Wu,}
\author[b]{Xingang Chen,}
\author[c]{Nino Ephremidze,}
\author[d,e,f]{Lingfeng Li}
\affiliation[a]{Institute of Physics, Academia Sinica, Taipei 115201, Taiwan}
\affiliation[b]{Institute for Theory and Computation, Harvard-Smithsonian Center for Astrophysics, 60 Garden Street, Cambridge, MA 02138, USA}
\affiliation[c]{Department of Physics, Harvard University, Cambridge, MA 02138, USA}
\affiliation[d]{International Center of Theoretical Physics-Asia Pacific, Beijing 100190, China}
\affiliation[e]{Institute of High Energy Physics, Beijing 100049, China}
\affiliation[f]{Department of Physics, Brown University, Providence, RI 02906, USA}

\emailAdd{ypwu@as.edu.tw, xingang.chen@cfa.harvard.edu, nino\_ephremidze@g.harvard.edu, l.f.li165@gmail.com}

\abstract{The Affleck–Dine mechanism is a leading baryogenesis scenario in which scalar condensates form coherently during inflation along supersymmetric flat directions that are lifted by supersymmetry-breaking effects. We update the viable parameter space for baryogenesis using recent Cosmic Microwave Background constraints on baryon-density isocurvature perturbations, taking the quantum fluctuations of the scalar condensate generated during inflation as initial conditions.
We then show that primordial features arising from the inflaton sector can serve as a unique probe of baryogenesis models, whose mechanisms are otherwise difficult to access directly due to their high energy scales. These primordial features leave correlated imprints, such as sharp feature signals and clock signals, on both the curvature and baryon-density isocurvature perturbations, providing direct evidence for the existence of both light and heavy modes involved in the Affleck–Dine mechanism.
}

\definecolor{linkcolor}{RGB}{41, 127, 255}
\hypersetup{
	pdfstartview={XYZ},
	colorlinks,
	allcolors=linkcolor,
	bookmarksopen
}

\begin{document}
\maketitle
\flushbottom

\section{Introduction}\label{Sec. introduction}
Affleck-Dine (AD) mechanism \cite{Affleck:1984fy} efficiently realizes the production of the observed baryon asymmetry in our universe today through the relaxation of a coherent scalar condensate from large initial vacuum expectation values (VEVs) developed in primordial eras. The scalar condensate, referred to as the AD field in this work, must carry some non-trivial combinations of the Standard Model baryon and lepton numbers. Baryon-number-violating terms must also arise during the relaxation process, creating a rotation of the scalar VEVs along the phase (or angular) direction of the AD field configuration. All necessary conditions for a successful AD baryogenesis have been found promising in the supersymmetric extension of the Standard Model \cite{Dine:1995uk, Dine:1995kz} (see also \cite{Dine:2003ax,Enqvist:2003gh} for reviews and references therein). 

The general existence of ``flat directions'' in supersymmetric field theories provides ideal initial conditions for the coherent production of a baryon number via the AD mechanism. First, a flat direction always includes a scalar component that contains a global $U(1)$ quantum number. During inflation, the flat direction can, therefore, foster a scalar condensate carrying some net particle numbers, as necessary for the AD baryogenesis. Second, supersymmetry must be broken in realistic scenarios for the primordial universe. During inflation, the positive but finite background energy density implies supersymmetry breaking. The finite-energy supersymmetry breaking can convert interactions between the AD field and the inflationary background field (the inflaton) to a tachyonic mass term of the order of the Hubble expansion rate. Non-renormalizable self-interactions of the AD field can lift the flat direction at large field values \cite{Ellis:1987rw,Ng:1988yn}. These are appropriate for the development of a stable scalar condensate during inflation. Third, $U(1)$-violating terms generally come together with the conversion of field interactions to classical potentials of the flat direction. These $U(1)$-violating terms can be important for defining the initial phase (the angular VEV) of the AD field, which determines the total amount of baryon number to be created at later times.

The AD baryogenesis from a flat-direction model is appealing for more reasons. The AD mechanism can provide an explanation for the origin of both ordinary and dark matter, shedding light on the so-called cosmic coincidence mystery for the similarity between the baryon and dark matter densities at present \cite{Dine:2003ax,Bell:2011tn,vonHarling:2012yn,Petraki:2013wwa,Wu:2021gtd,Wu:2021mwy,Balaji:2022rsy}. Even if supersymmetric models suggest significant dilution of the final baryon number density at lower energy scales, the flat-direction scenario can be very efficient for generating arbitrarily large baryon number density carried by the AD field. The late-time dilution then yields the correct baryon density observed today with the decay of the AD field into Standard Model particles \cite{Dine:2003ax}. Flat directions can also create initial rotations of axion fields, indicating another interesting explanation for the origin of (in)visible matter in our universe \cite{Co:2019jts,Co:2019wyp,Co:2020jtv}. 
First order phase transitions proceed through vacuum bubble nucleation in a flat-direction potential may generate a stochastic gravitational wave background \cite{Craig:2020jfv}.

However, experimental tests for the coherent production of baryons via scalar condensate developed during inflation remain challenging, as the mechanism may operate at very high energy scales. Currently, no sign of the existence of the relevant particles is found in ground-based collider experiments~\cite{ATLAS:2024fub, Ghosh:2024joj, Constantin:2025mex}, and, to the best of our knowledge, the most important constraint comes from the baryon-density-isocurvature (BDI) bound inferred from Cosmic Microwave Background (CMB) observations \cite{Planck:2018jri} (see also \cite{Enqvist:1987wi,Ellis:1987rw, Charng:2008ke, Kasuya:2008xp} for early developments of BDI perturbations in AD baryogenesis).

Inflation generates primordial fluctuations that seed the large-scale structure of the universe we observe today. In recent decades, it has been proposed that the fundamental physics operative during inflation can be probed by studying properties of the correlation functions of these fluctuations, particularly through primordial features and primordial non-Gaussianities. This is because particles and interactions involved during inflation, such as those responsible for baryogenesis, may leave characteristic imprints in these correlation functions. See \cite{Chen:2010xka,Chluba:2015bqa,Slosar:2019gvt, Achucarro:2022qrl} for reviews. In this paper, we investigate the possibility of using primordial features to probe the flat-direction baryogenesis model.

Primordial features are strongly scale-dependent deviations from the otherwise approximately scale-invariant density perturbations, arising from fundamental physics in the primordial universe, such as during inflation. Their characteristics can encode a wide range of fundamental physics --- from the detailed dynamics of the inflationary model, to the properties of elementary particles present during inflation, and to signatures that can model-independently distinguish inflation from alternative primordial universe scenarios. 

Observationally, several candidate signals of large-scale and small-scale primordial features have been identified in the WMAP and Planck data \cite{Peiris:2003ff,Akrami:2018odb, Adams:2001vc, Bean:2008na, Mortonson:2009qv, Hazra:2010ve, Hazra:2014goa, Miranda:2014fwa, Hazra:2014jwa, Canas-Herrera:2020mme, Braglia:2021ckn, Braglia:2021sun, Braglia:2021rej, Braglia:2022ftm}. Although feature models can improve the fit to the observed power spectrum, these improvements are currently statistically insignificant due to the additional model parameters they introduce. Future observations from CMB polarization measurements \cite{Miranda:2014fwa,Braglia:2022ftm, Petretti:2024mjy} and galaxy surveys \cite{Huang:2012mr,Hazra:2012vs,Chen:2016vvw,Ballardini:2016hpi,Palma:2017wxu,LHuillier:2017lgm,Ballardini:2017qwq,Vasudevan:2019ewf,Beutler:2019ojk,Ballardini:2019tuc,Debono:2020emh,Chen:2020ckc,Li:2021jvz} are expected to enhance precision, potentially confirming or ruling out these feature candidates. If a detection can be confirmed, such observations could even discriminate between different feature models \cite{Braglia:2022ftm}.

As an application of primordial feature physics, we explore how a sharp feature in inflationary dynamics can probe properties of flat-direction baryogenesis models that are otherwise very difficult to access. Since the baryogenesis sector is coupled to the inflaton, either through minimal gravitational interactions or via direct couplings that naturally arise in effective field theories, a sharp change in inflaton dynamics can suddenly inject a surge of energy into the coupled sector. This process excites massive fields that were initially sitting at their minima, kick-starts their classical oscillations, and generates ``clock signals'' as direct observational signatures of these massive fields \cite{Chen:2011zf,Chen:2014cwa,Quintin:2024boj} (in addition to ``sharp feature signals'' that encode the nature of the sharp change but are not themselves signatures of massive fields). Therefore, primordial features offer a rare opportunity to probe UV physics associated with baryogenesis.

This paper is organized as follows. In Section~\ref{Sec. Model}, we provide a brief review of the flat-direction model \cite{Dine:1995kz} from supersymmetry breaking in the early universe, and perform numerical computations of the final baryon asymmetry in Section~\ref{Sec_baryogenesis} under the assumption of instantaneous reheating. For the baryon asymmetry obtained from given model parameters, we work out the corresponding BDI perturbations in Section~\ref{Section_BDI} based on (1) the separated universe approach (Section~\ref{Sec_Separated_Universe}) and (2) the linear perturbation theory (Section~\ref{Sec_Linear_Perturbation}). The general existence of direct couplings between the inflaton and the AD field arising from kinetic terms of the superfield Lagrangian is shown in Section~\ref{Sec_SUSY_breaking}. Using a step feature as an illustrative example in Section~\ref{Sec_primordial_features}, we demonstrate how a sharp feature in the inflaton potential can excite classical oscillations of the heavy mode of the AD field, leaving potentially observable sharp feature signals and clock signals in both the curvature power spectrum and the BDI power spectrum, with all such signals being correlated between the two spectra. We summarize and discuss future directions in Section~\ref{Sec. conclusion}.

\section{Flat directions from supersymmetry}\label{Sec. Model}
\begin{figure}[]
	\begin{center}
		\includegraphics[width=12 cm]{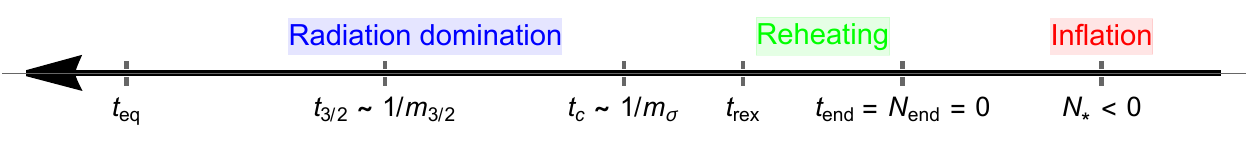}
	\end{center}
	\caption{\label{fig.timeline}  The timeline of the baryogenesis scenario considered in this work. $N$ is the $e$-folding number of inflation up to a constant shift. $t$ is the physical time. We define $N_{\rm end} = t_{\rm end} = 0$ at the end of inflation where reheating of the universe starts due to the decay of inflaton $\phi$ into radiation. Reheating is assumed to be completed by the onset of the relaxation of the complex scalar $\sigma$ at $t = t_{\rm rex}$. $t =t_c \sim 1/m_\sigma$ is the time scale for which $\sigma$ enters the phase of harmonic oscillation with a constant angular velocity. $t=t_{3/2}\sim 1/m_{3/2}$ is the time scale for the $A$ term in the potential \eqref{def_UFD} dominates over the $c_A$ term. $m_\sigma\sim m_{3/2}$ for the standard gravity mediation and $m_\sigma\gg m_{3/2}$ for split supersymmetry. $t_{\rm eq}$ denotes the time at matter-radiation equality around the temperature $T_{\rm eq} \approx 9.8\times 10^{-10}$ GeV. A primordial feature in the inflaton potential is present at $N = N_\ast$.}
\end{figure}
\begin{figure}[]
	\begin{center}
		\includegraphics[width=10.3 cm]{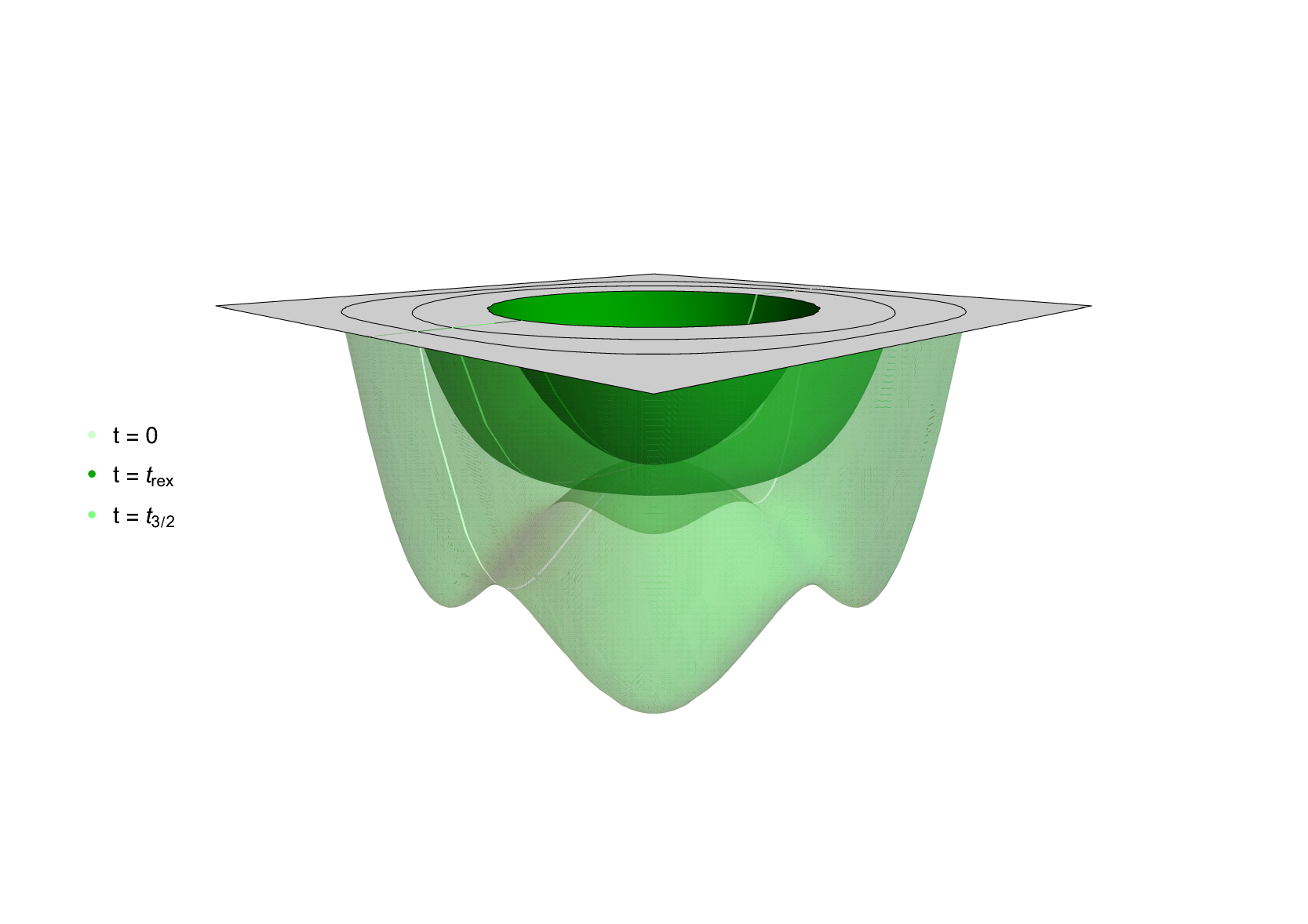}
        \hfill
        \includegraphics[width=5 cm]{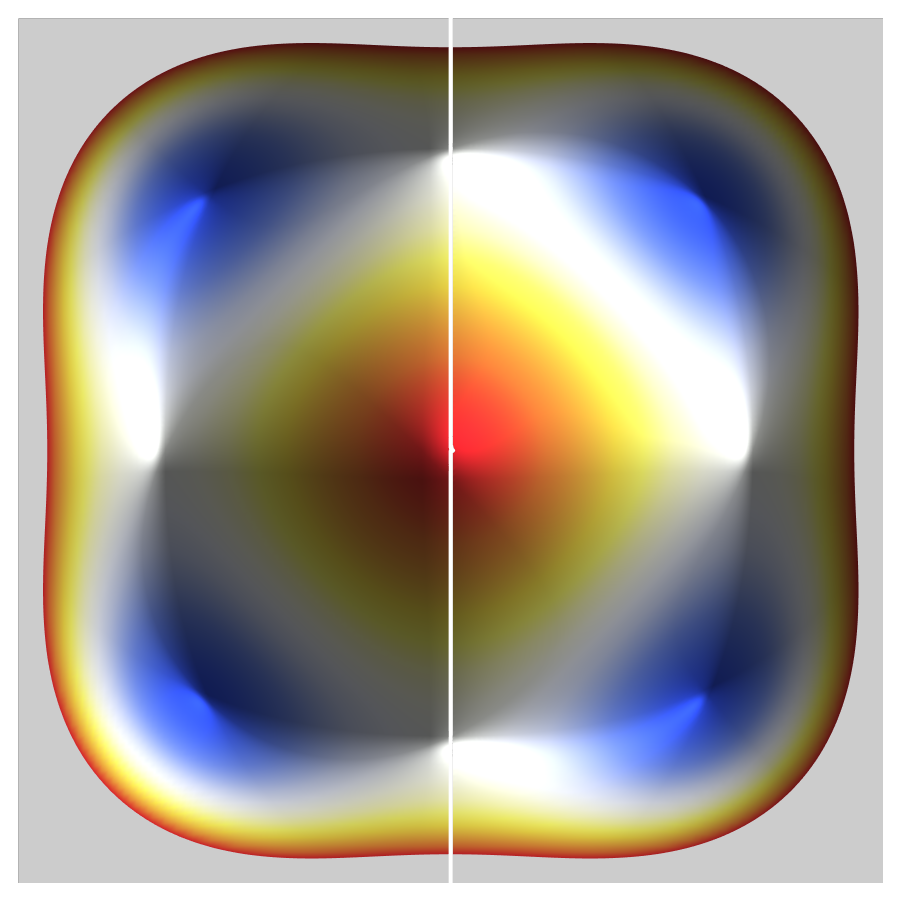}
	\end{center}
    \caption{\label{fig.UFD_snapshots} 
    [Left Panel] Snapshots of the flat-direction potential $U_{\rm FD}$ at $t = 0$, $t = t_{\rm rex}$ and $t = t_{3/2}$ with a thermal mass term. [Right Panel] A bird-eye view of the potential at $t = 0$. Parameters with $n = 4$, $\xi = 2$, $m_\sigma/H_I = 10^{-3}$, $\lambda = c_A = A = 1$, $\Lambda/H_I = 100$, $m_{3/2}/H_I = 10^{-6}$ and $\delta = \pi/2$ are used in these plots. }
\end{figure}

\noindent
To demonstrate how primordial features can be used to probe the physics of baryogenesis models, we begin by reviewing the supersymmetric flat-direction model and then use a simple example to compute the BDI spectrum it generates.
For illustration, it suffices to consider global supersymmetry in absence of gravity, where the Lagrangian of chiral and vector superfields, denoted by $\Phi_i$ and $v^a$, is specified by a Kähler potential term with internal gauge interactions, a gauge kinetic term, and a term led by a holomorphic function, $W(\Phi_i)$, known as the superpotential. 

We focus on terms that contribute as potentials of the scalar components $\phi_i$ of the chiral superfields $\Phi_i$ and a canonical Kähler potential $K(\Phi_i,\Phi_i^\dagger) = \sum_i K\Phi_i^\dagger \Phi_i$. In the case with a canonical Kähler potential $K(\Phi_i,\Phi_i^\dagger) = \sum_i \Pi_a\Phi_i^\dagger e^{2g_a v^aT^a}\Phi_i$, we collect terms up to the first order in $v_a$ for the purpose of our discussion. $T^a$ are generators of internal gauge symmetries and $g_a$ are gauge coupling constants. 
After integrating over the superspace, the superpotential becomes an ordinary function of the scalar components: $W = W(\phi_i)$. Eliminating the non-dynamical auxiliary fields using algebraic equations of motion, we obtain the scalar potential for $\phi_i$ given by:
\begin{align}
	V &= V_F + V_D
    = g_{i\bar{j}}F^i \bar{F}^{\bar{j}} + \frac{1}{2} \sum_a \left(D^aD^a\right),
	\label{MSSM_scalar_potential}
\end{align}
where, the auxiliary ``F'' and ``D'' fields are solved as:
\begin{align}
	F^i = - g^{i\bar{j}} \frac{\partial\bar{W}}{\partial\phi_j^\ast} \equiv -g^{i\bar{j}} \bar{W}_{\bar{j}} ,\quad
	D^a = g_a \sum_i K_i (T^a \phi)^i,
\end{align}
with $K_i \equiv \partial_{\phi_i}K$ and $W_j \equiv \partial_{\phi_j}W$.
Here $g_{i\bar{j}} = \partial_{\phi_i}\partial_{\bar{\phi}_j}K$ and $g^{i\bar{j}} = g_{i\bar{j}}^{-1}$ are the so-called Kähler metric and $K = K(\phi_i,\phi_i^\ast)$ is now a scalar function.
The bosonic part of the field Lagrangian relevant to our discussion is then given by
\begin{align}
    \mathcal{L}_{\rm bosonic} = -g_{i\bar{j}}\partial_\mu\phi^i\partial^\mu\phi^{\ast \bar{j}} -V_F - V_D.
\end{align}
For the canonical Kähler potential with $ \sum_i\vert\phi_i\vert^2 $ as the leading term, $g_{i\bar{j}} = \delta_{i\bar{j}}$ and $K_i \sim \phi_i^\ast$.
At the classical level, a non-zero VEV of any one of the $F$ or $D$ fields (leading to a non-zero potential energy) breaks the supersymmetry.

One of the distinguishing features of supersymmetric theories with a large number of fields is that they often have accidental degeneracies along which the classical potential vanishes. These are called "flat directions" and occur along directions in the field space where both F-term and D-term VEVs are zero. When supersymmetry is unbroken, a single flat direction, referring to the scalar component $\phi_i$ of the full chiral superfield $\Phi_i$, necessarily carries a global $U(1)$ number. As a result, a scalar condensate developed in the flat direction can carry a net particle number of $U(1)$, which will facilitate AD baryogenesis.

In the early universe, flat directions can acquire classical potentials and be lifted due to the effects of (1) non-renormalizable terms in the superpotential $W(\Phi_i)$, (2) finite-energy supersymmetric breaking terms, and (3) hidden sector supersymmetric breaking terms.

First, non-renormalizable, higher-dimensional interactions with more fields in the superpotential can lift these flat directions. These terms arise from new physics above a high-energy scale $\Lambda$ (typically Planck or string scale) and take the form \cite{Dine:1995kz}:
\begin{align}
	W \sim \lambda_n \frac{\Phi_\sigma^n}{n \Lambda^{n-3}} \sim \lambda_n \frac{\sigma^n}{n \Lambda^{n-3}},
\end{align}
where $n \geq 4$, $\Phi_\sigma$ is the superfield contains a flat direction $\phi_\sigma = \sigma$ and its fermionic partner, and $\lambda_n \sim \mathcal{O}(1)$. Note that these terms also break the $U(1)$ symmetry associated with the flat direction explicitly, indicating their potential relation with (quantum) gravity. The term also gives rise to an $F$-term potential $V_F \sim \vert\sigma\vert^{2(n-1)}/\Lambda^{2(n-3)}$, 
which contributes to the total effective potential \eqref{def_UFD} as the $|\lambda|^2$ term.

Second, a cosmological epoch with a finite but non-zero background energy density implies a vacuum state that breaks supersymmetry. Inflation is a prime example. During inflation, we must include gravitational effects and hence the local supersymmetry (supergravity) effects. In this case, the partial derivative is replaced with the covariant Kähler derivative $D_i W\equiv \partial_i W + K_iW/M_P^2$ in the $F$-term potentials. Inflation can be driven by a scalar mode within a complex scalar field $I$ as the scalar component of a superfield $\Phi_I$. If the inflationary background density is governed by the $F$-term potential as $\rho_I = 3M_P^2 H_I^2 \sim V_F(I)$, then one finds $D_I W(I) \sim M_PH_I$ and therefore $W(I) \sim \vert I\vert M_PH_I \lesssim M_P^2 H_I$. As a result, all terms that couple to $W(I)$ or $D_IW(I)$ gain a linear dependence on the Hubble parameter $H$, while terms that couple to $V_F(I)$, $\vert W(I)\vert^2$, $\bar{W}(I)D_IW(I)$ get the $H^2$ dependence. 

Third, hidden sectors provide model-independent mechanisms for soft supersymmetric breaking. Their effects are usually transmitted at some mass scales much lower than the Hubble scale during inflation (which is approximately a constant denotes as $H_I$),  such as the bare mass $m_\sigma$ of the flat direction and the $m_{3/2}$ identified as the gravitino mass in \cite{Dine:1995kz}. These masses could be as low as the electroweak scale if supersymmetric breaking is mediated by gravity. In this work, we intend to perform a general survey in the parameter space, allowing $m_\sigma$ and $m_{3/2}$ to have split mediation scales \cite{Arkani-Hamed:2004ymt,Giudice:2004tc,Arkani-Hamed:2004zhs,Kasuya:2005sb}.
Therefore, $m_{3/2}$ can be smaller than $m_\sigma$ by many orders of magnitude.

Following the above discussion (a more detailed derivation can be found in \cite{Dine:1995kz}), the effective potential for a flat direction can be parametrized by a complex scalar field $\sigma$ in the form of:
\begin{align}\label{def_UFD}
	U_{\rm FD} = \left(m_\sigma^2 + c_H H^2\right) \vert\sigma\vert^2 + \vert\lambda\vert^2 \frac{\vert\sigma\vert^{2(n-1)}}{\Lambda^{2(n-3)}} 
	+ \frac{c_A}{n}\lambda \frac{H}{\Lambda^{n-3}}\sigma^n + \frac{A}{n}\lambda\frac{m_{3/2}}{\Lambda^{n-3}}\sigma^n + h.c. ,
\end{align}
where $m_\sigma$ describes a soft mass term from supersymmetry breaking of some hidden sector and is responsible for creating the asymptotic minimum to stabilize the final baryon asymmetry. $\Lambda \gg H_I$ is the cutoff scale of the theory. $c_H < 0$ is necessary for baryogenesis. Since the $c_A$ term decays with the Hubble parameter $H$, it is the $A$ term which controls the total amount of baryon asymmetry generated in the post-inflationary epochs. For the rest of this work, we will refer to the explicit $H$ dependence in $U_{\rm FD}$ as the ``gravitational'' coupling, because so far the inflaton acts as a spectator field to these fields and is only responsible for creating the inflationary background. 
We provide the possible origin of the Hubble-induced coefficients $c_H$ and $c_A$ in Appendix~\ref{Appendix_Hubble_coefficients}, which is closely related to the presence of non-canonical kinetic terms considered in Section~\ref{Sec_SUSY_breaking}. 
If the $A$ term also arises from hidden sector supersymmetry breaking, then $m_{3/2} \lesssim m_\sigma$ is, in general, far below the Hubble scale of inflation $H_I$ \cite{Dine:1995kz}.
In polar coordinates, we have the parameterization
\begin{align}\label{def_FD_parametrizations}
	\sigma \equiv \frac{R}{\sqrt{2}}e^{i\theta}, \quad \xi \equiv \vert c_H\vert, \quad A \equiv \vert A \vert e^{i \delta},
\end{align}
with $\lambda = \vert \lambda\vert$ being a real constant. We can rewrite \eqref{def_UFD} as
\begin{align}
	\label{def_UFD_polar}
	U_{\rm FD}(R, \theta) = \left(m_\sigma^2 -\xi H^2\right)\frac{R^2}{2} 
	+  \frac{\lambda^2}{2^{n-1}} \frac{R^{2(n-1)}}{\Lambda^{2(n-3)}} 
	+\frac{\alpha}{n} \frac{\lambda}{2^{n/2-1}} \frac{H}{\Lambda^{n-3}} R^n, 
\end{align}
where we have defined the composite parameter;
\begin{align}\label{def_alpha}
	\alpha H \equiv c_A H \cos\left(n\theta\right) + A m_{3/2} \cos\left(n\theta + \delta\right),
\end{align}
that collects the angular dependence in the potential due to the $U(1)$ symmetry breaking terms. Note that we have henceforth replaced $\vert A \vert$ by $A$ for convenience.

The angular minima solved by $\partial_\theta U_{\rm FD} =0$ are equivalent to solving $\partial_\theta \alpha = 0$. After some manipulations, we obtain
\begin{align}\label{def_theta_min}
	\theta_{\rm min} = -\frac{i}{n} \ln\left[-\frac{\alpha_x -i\alpha_y}{\sqrt{\alpha_x^2 +\alpha_y^2}}\right],
\end{align}
where the dimensionless parameters are $\alpha_x  \equiv c_A H/H_I + A M_{3/2}\cos\delta$, $\alpha_y \equiv A M_{3/2}\sin\delta$, and they satisfy $\alpha H/H_I = \alpha_x \cos(n\theta) -\alpha_y\sin(n\theta)$.
$M_{3/2} \equiv m_{3/2}/H_I$ is in the unit of $H_I$.
If $A$ is small enough, the terms proportional to $A$ are negligible during inflation, 
and we find $\theta_{\rm min}\approx \pi/n$ with $\alpha_y \approx 0$.
On the other hand, with $c_A \rightarrow 0$ during reheating after the inflation ends (see Appendix~\ref{Appendix_Hubble_coefficients}), we find $\alpha_x \rightarrow A M_{3/2}\cos\delta$ and so that $\theta_{\rm min} \rightarrow (\pi -\delta)/n$. The change in the angular minima in the post-inflationary epoch is the key to generate the rotation of the $\sigma$ field, which is directly associated with the baryon number density.

To compute the radial minimum of the flat-direction potential, it is convenient to use the dimensionless variable
\begin{align}\label{def_X}
	X \equiv \frac{\lambda}{2^{n/2- 1}} \frac{R^{n-2}}{\Lambda^{n-3} H}.
\end{align}
The equation $\partial_R U_{\rm FD} =0$ for the radial minimum is now translated to $(n-1)X^2 + \alpha X = \xi - m_\sigma^2/H^2$ for the $X$ variable. The solution is
\begin{align}\label{def_Xmin}
	X_{\rm min} = \left[\frac{\xi - m_\sigma^2/H^2}{n -1} + \frac{\alpha^2}{4(n-1)^2}\right]^{1/2} - \frac{\alpha}{2(n-1)}.
\end{align}
Note that if $\xi \gg m_\sigma^2/H^2$, the value $X_{\rm min}$ is nearly independent of the background Hubble parameter.
The location of the radial minimum can be obtained from \eqref{def_X} and \eqref{def_Xmin}:
\begin{align}\label{def_Rmin}
	R_{\rm min} = \left(\frac{2^{n/2-1}}{\lambda} \Lambda^{n-3} X_{\rm min} H\right)^{\frac{1}{n-2}}.
\end{align}
A useful feature for our analysis is that $X_{\rm min} \sim \mathcal{O}(1)$ in all the cases considered in this work. 

In terms of $\alpha_{\rm min}\equiv \alpha(\theta_{\rm min})$ and $X_{\rm min}$, we can define the radial and angular masses at one of the potential minima as
\begin{align}
	m_R^2 &\equiv \partial_R^2 U_{\rm FD} = (2n-4)\left(\xi H^2 - m_\sigma^2\right) +(2-n) \alpha_{\rm min} X_{\rm min} H^2, \label{def_m_R}
	\\\label{def_m_theta}
	m_\theta^2 &\equiv \frac{1}{R_{\rm min}^2}\partial_\theta^2 U_{\rm FD} = -n \alpha_{\rm min} X_{\rm min} H^2.
\end{align}
Note that $\alpha_{\rm min} = -c_A$ if $A = 0$ where $\theta_{\rm min} = \pi/n$. The radial mass $m_R$ defined with $\xi$ is used during inflation only, while the angular mass is generally true even in post-inflationary epoches with arbitrary $c_A$ and $A$.

\section{Baryogenesis from flat directions}\label{Sec_baryogenesis}
In this section, we take the flat-direction model \eqref{def_UFD} to realize baryogenesis based on the AD mechanism~\cite{Affleck:1984fy}. We proceed with initial conditions given by instantaneous reheating for the post-inflationary epoch and compute the background evolution of the AD field. The field carries a global quantum number, which could be recognized as the baryon number $B$, lepton number $L$ or their linear combinations.\footnote{The Noether current can also be a combination of baryon numbers from both the ordinary and dark matter sectors, see \cite{Petraki:2013wwa} for a review.} The final baryon asymmetry in radiation domination from instantaneous reheating is in general different from the scenario with a long duration of the preheating phase led by coherent inflaton oscillations considered in~\cite{Dine:1995kz}.

Before the $U(1)$ breaking terms are introduced, the scalar condensate of $\sigma$ has a conserved Noether current:
\begin{align}
	j^\mu = i \left(\sigma^\ast \partial^\mu\sigma - \sigma\partial^\mu\sigma^\ast\right).
\end{align}
With the relaxation dynamics, which will be detailed below, the rotating $\sigma$ field introduces a non-zero $U(1)$ charge density $j^0 \propto R^2 \dot{\theta}$. In this work, we assign the Noether current of the unbroken $U(1)$ symmetry with the unitary baryon number $B$ for simplicity, namely $j^\mu = j_B^\mu$. Other choices of charge assignment changes the ratio between $j^\mu$ and $j^\mu_B$ by a model-dependent constant.\footnote{There are examples of flat directions that carry a pure baryon number ($B\neq 0$ but $L = 0$). In the Minimal Supersymmetric Standard Model (MSSM), a typical example is the $\mathcal{O}_{udd} = u^cd^cd^c$ combination which is gauge invariant, with $u^c$ and $d^c$ being right-handed squark superfields. The $D$ terms vanishes because the combination is a gauge singlet.
The F-flatness in this case is exact since VEVs of all other superfields vanish in the MSSM superpotential. See \cite{Dine:2003ax} for more details.}

For AD baryogenesis, the out-of-equilibrium condition is manifest in the post-inflationary relaxation of the scalar condensate from the large initial VEV developed during inflation. For the flat-direction model, the baryon number violation is mainly created by the $A$ term in the late stage evolution. The $C$ or $CP$ violation is measured by the phase difference between the $c_A$ and $A$ terms, or namely, a non-zero $\delta$ in our parametrization \eqref{def_FD_parametrizations}.

\subsection{Initial conditions at the end of inflation}\label{Sec. Initial_conditions}
Let us define $H_I$ as the Hubble parameter during inflation, which is taken as a constant in this work. As shown in Figure~\ref{fig.timeline}, $N \leq 0$ describes the epoch during inflation and $t \geq 0$ is the physical time for the post-inflationary phases. Initial conditions for the relaxation of $\sigma$ at the end of inflation ($t = t_{\rm end}$) are given by \eqref{def_theta_min} and \eqref{def_Rmin} with $H = H_I$, namely
\begin{align}\label{def_Rend_theta_end}
	\theta_{\rm end} = \theta_{\rm min}(H_I), \qquad R_{\rm end} = R_{\rm min}(H_I, X_{\rm end}),
\end{align} 
where $X_{\rm end} = X_{\rm min}(H_I, \alpha_{\rm end})$ is given by \eqref{def_Xmin} with $\alpha_{\rm end} = \alpha(H_I, \theta_{\rm end})$ from \eqref{def_alpha}. $R_{\rm end}$ and $\theta_{\rm end}$ are constants during inflation.

The post-inflationary evolution of the homogeneous background field values, $R_0(t)$ and $\theta_0(t)$, are governed by the equations
\begin{align}\label{eom_R0}
	\ddot{R}_0 + 3H\dot{R}_0 + \partial_R U_{\rm FD} - \dot{\theta}_0^2 R_0 &= 0, \\
	\label{eom_theta0}
	\ddot{\theta}_0 + \left(3H + 2\frac{\dot{R}_0}{R_0}\right) \dot{\theta}_0 + \frac{1}{R_0^2} \partial_\theta U_{\rm FD} & =0,
\end{align}
where we need to specify the time evolution in $a(t)$ or $H(t)$ for the background expansion. 
Taking the static assumptions $\dot{R}_0 = 0$, $\dot{\theta}_0 = 0$ with $R_0 = R_{\rm end}$, $\theta_0 = \theta_{\rm end}$ at the initial time ($t = 0$), we are now ready to evaluate the post-inflationary dynamics of the complex scalar field to the background level.

Let us consider one of the simplest scenarios where the universe is instantly reheated after the termination of slow-roll inflation. In this scenario, the inflaton field decays into radiation (or a collection of relativistic degrees of freedom) in a very short time and the energy density of the post-inflationary universe is dominated by radiation with $3M_P^2 H^2 = \rho_r$. This gives a transparent relation between temperature and the Hubble parameter as:
\begin{align}\label{temperature_instantR}
	T^4(t) = \frac{30}{\pi^2 g_\ast} 3M_P^2 H^2(t),
\end{align} 
where $g_\ast = 106.75$ is nearly a constant in the Standard Model for $T > 300$ GeV, which corresponds to $H_I \gtrsim 1$ GeV for instantaneous reheating. 
If the equation-of-state of the universe $w = -1 -2\dot{H}/3H^2$ is a constant, then the Hubble parameter $H = \dot{a}/a$ and the scale factor $a(t)$ can be solved as
\begin{align}
	H = \frac{H_I}{1+\frac{3}{2}(1+w)(\tau - \tau_{\rm end})}, \quad \frac{a}{a_{\rm end}} = \left[1+\frac{3}{2}(1+w) (\tau - \tau_{\rm end})\right]^{\frac{2}{3(1+w)}},
\end{align}
where $\tau \equiv H_I t$ is the dimensionless time parameter. Note that we have defined $\tau_{\rm end} = 0$ at the end of inflation. For a radiation-dominated universe led by instantaneous reheating, we take $w = 1/3$ so that $H = H_I/(1+2\tau)$ is applied to \eqref{eom_R0} and \eqref{eom_theta0}. 

\begin{figure}[]
	\begin{center}
		\includegraphics[width=8 cm]{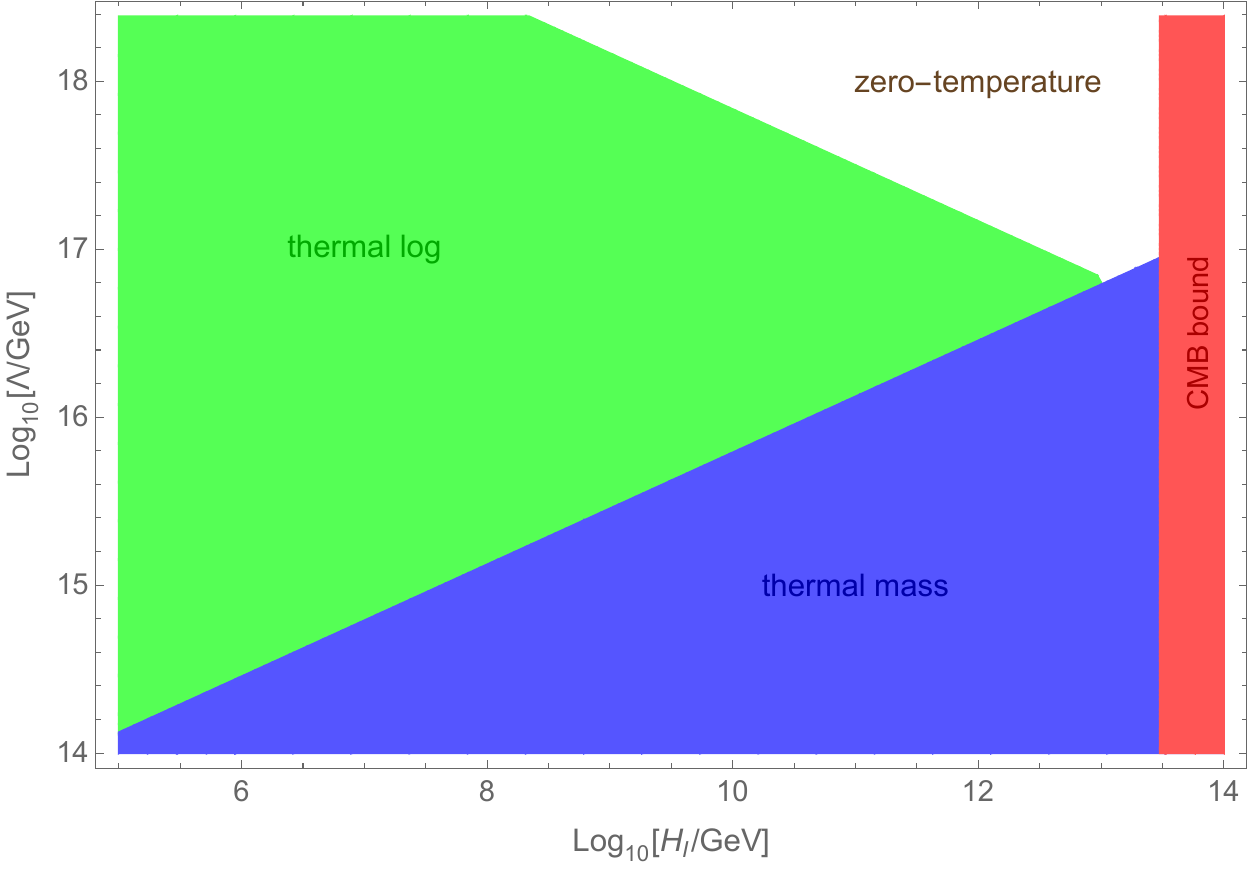}
	\end{center}
	\caption{\label{fig.finite_T_scan} The parameter space for finite-temperature effects at the end of inflation with $g = 0.33$. Model parameters with $n = 6$, $m_\sigma/H_I = m_{3/2}/H_I = 10^{-3}$, $\xi = \lambda = A = 1$, $c_A = 0.5$ are fixed in this plot. The finite-temperature corrections are negligible in the ``zero-temperature'' zone. }
\end{figure}

The general expression of the effective potential of the AD field in the post-inflationary epoch takes the form of
\begin{align}\label{def_UFD_RD}
	U_{\rm FD}(R,\theta, T) =&\;  c_H\frac{H^2}{2} R^2   + \frac{c_A}{n}\frac{\lambda}{2^{n/2-1}}\frac{H}{\Lambda^{n-3}}R^n\cos(n\theta )+ \frac{\lambda^2}{2^{n-1}}\frac{R^{2(n-1)}}{\Lambda^{2(n-3)}} \nonumber\\
	&+\frac{m_\sigma^2}{2} R^2 + \frac{A}{n}\frac{\lambda}{2^{n/2-1}}\frac{m_{3/2}}{\Lambda^{n-3}}R^n\cos(n\theta +\delta)   + \Delta U(R, T),
\end{align}
Since the origin of the $c_H$ and $c_A$ terms is the inflaton sector, for the instantaneous reheating scenario that we assume in this work, we will assume $c_H = c_A = 0$ in the above equation as soon as the inflation ends.  (See Appendix~\ref{Appendix_Hubble_coefficients} for discussions.) 

$\Delta U$ in \eqref{def_UFD_RD} accounts for the finite-temperature effects led by a dominant coupling $g$ between the AD field and Standard Model particles. The finite-temperature corrections mainly come from two types of effects as \cite{Allahverdi:2000zd, Anisimov:2000wx}:
\begin{align}
	\Delta U_{\rm log} &= \alpha_s^2 T^4 \ln\left(\frac{\vert\sigma\vert^2}{T^2}\right) = \alpha_s^2 T^4 \ln\left(\frac{R^2}{2T^2}\right), \nonumber\\
	\Delta U_{\rm mass} &= g^2 T^2 \vert\sigma\vert^2 = \frac{1}{2} g^2 T^2 R^2,
\end{align}
where $\alpha_s = g^2/4\pi$. $\Delta U_{\rm log}$ is called the thermal log effect which dominates over the thermal mass correction $\Delta U_{\rm mass}$ in the large field regime when $g R > T$. The finite-temperature term becomes operative when the effective thermal mass $m_T > H$ is larger than the background Hubble expansion, and $m_T = \sqrt{2}\alpha_s T^2/R$ for the thermal log effect while $m_T = gT$ for the thermal mass correction, respectively. We show the dominant effect in the parameter space with respect to $H_I$ and $\Lambda$ in Figure~\ref{fig.finite_T_scan}. The ``zero-temperature'' region denotes the parameter space where the finite-temperature corrections are subdominant so that the potential $U_{\rm FD}(R,\theta, T=0)$ can be used.   For the typical $udd$ flat direction in MSSM, the $SU(3)$ gauge coupling $g = g_3\sim \mathcal{O}(1)$ dominates so that the QCD thermal log correction plays an important role in the large $R_{\rm end}$ regime for baryogenesis.

\begin{figure}[]
	\begin{center}
		\includegraphics[width=7.5 cm]{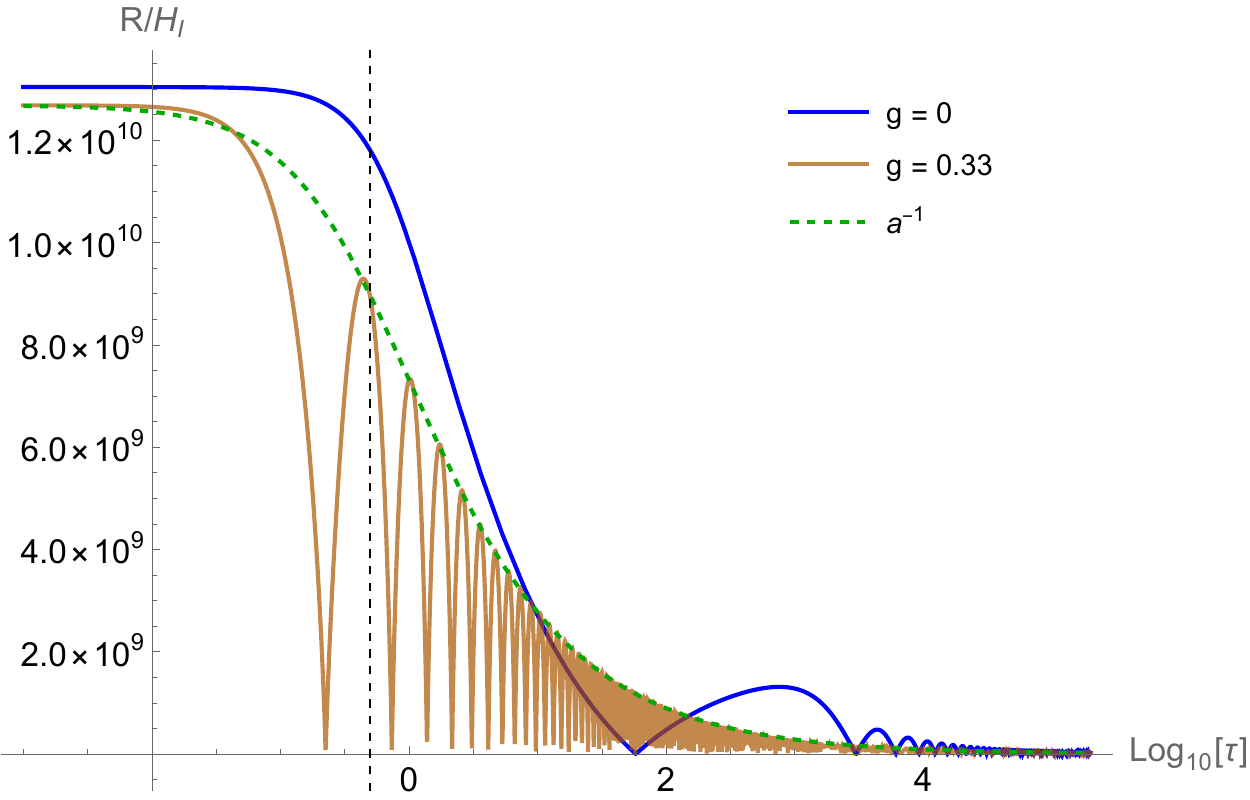}
		\hfill
		\includegraphics[width=7.5 cm]{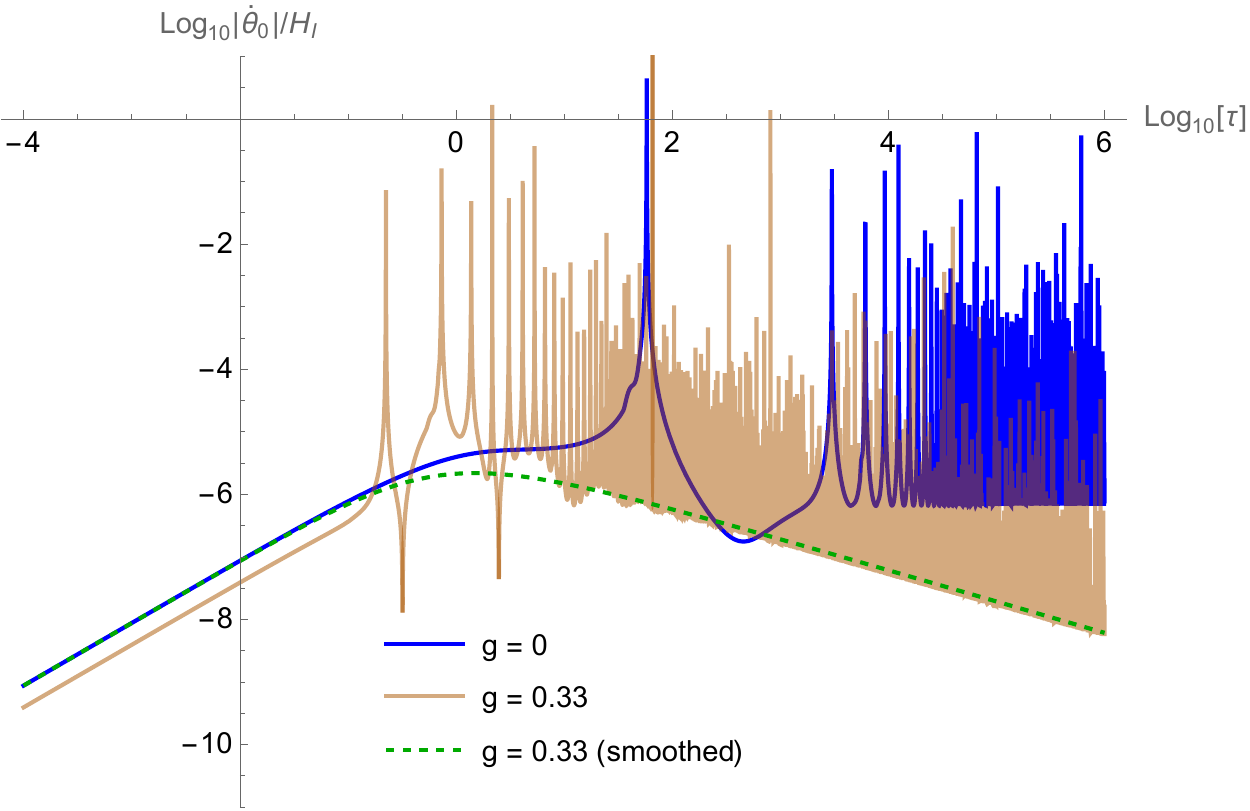}
	\end{center}
	\caption{\label{fig.finite_T_effects} The evolution of $R_0$ (left panel) and $\dot{\theta}_0$ (right panel) with $g = 0.33$ in the thermal log domination region. The $g = 0$ result turns off the finite-temperature correction in the effective potential. Model parameters with $n = 6$, $\Lambda = M_{P}$, $m_\sigma/H_I = m_{3/2}/H_I = 0.01$, $\xi = \lambda = A = 1$, $c_A = 0.5$ during inflation, $\delta = \pi/2$ and $H_I = 10^{5}$ GeV are used in these plots. The vertical dashed line in the left panel is $\tau = \tau_{\rm rex}$ given by \eqref{def_tau_rex}.}
\end{figure}
\begin{figure}[]
	\begin{center}
		\includegraphics[width=12 cm]{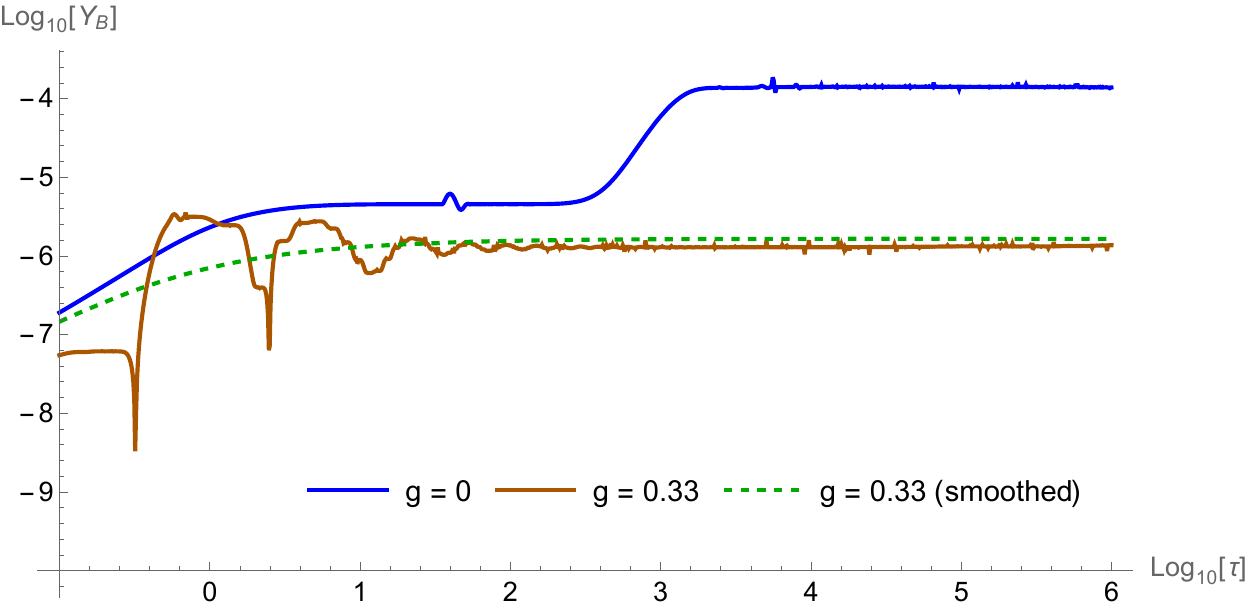}
	\end{center}
	\caption{\label{fig.finite_T_YB} The evolution of $Y_B$  corresponding to the background motion given in Figure~\ref{fig.finite_T_effects}.}
\end{figure}

\subsection{The final baryon asymmetry}\label{Sec. final_YB}
The baryon-to-entropy ratio  is defined as
\begin{align}\label{def_Y_B}
	Y_B(t) = \frac{n_B(t)}{s(t)} = \frac{45}{2\pi^2 g_\ast}\frac{n_B(t)}{T^3(t)},
\end{align} 
where $n_B = j^0_B = R^2\dot{\theta}$ is the baryon number density and $s = 2\pi^2 g_\ast T^3/45$ is the entropy density. In the scenario of instantaneous reheating, $T(t)$ is given by \eqref{temperature_instantR}.\footnote{It is possible that the decay of supersymmetric particles or string moduli at lower temperatures leads to a significant injection of entropy, resulting in the dilution of the ratio \eqref{def_Y_B} based on the Standard Model entropy density \cite{Dine:2003ax}.} 
In this work, we report constraints on the parameter space based on the benchmark value $Y_B = 10^{-10}$ obtained from the standard thermal history of radiation domination after reheating. We take the $g_\ast$ to be of the Standard Model value during the relaxation of $\sigma$.
Our results can be easily modified by the relation $Y_B = (g_{\ast}^{\rm new}/g_\ast) \times 10^{-10}$ in the presence of new relativistic degrees of freedom at high temperatures.

Let us sepcify our strategies for resolving the asymptotic value of $Y_B$ in the regions with (1) thermal log domination, (2) thermal mass domination, and (3) zero-temperature potential at the end of inflation. 

\bigbreak\noindent
$\bullet$\textit{Thermal log domination.--}
With $c_H = c_A = 0$ led by instantaneous reheating, the effective potential $U_{\rm FD}$ becomes very flat after inflation ends where the finite-temperature correction $\Delta U$ easily dominates the potential and governs the background radial evolution \eqref{eom_R0}. When $\Delta U_{\rm log}$ dominates the potential, it gives rise to a time-vary mass $m_T \sim T^2/R$. The radial mode $R_0$ can be viewed as a harmonic oscillator with a slowly varying mass $m_T$ which satisfies $n_\sigma \sim \rho_\sigma/m_T \sim a^{-3}$. Since $\rho_\sigma \sim \Delta U_{\rm log}\sim T^4$ and $T\sim a^{-1}$ in radiation domination, we get $n_\sigma \sim T^2R \sim a^{-3}$ with $R \sim a^{-1}$. The radial motion smoothed over the rapid oscillations can be captured by the solution $R_0 = R_{\rm end}(a/a_{\rm end})^{-1}$, which is shown as the dashed line in the left panel of Figure~\ref{fig.finite_T_effects}. Note that $R_{\rm end}$ is defined in \eqref{def_Rend_theta_end} and we choose $a_{\rm end} = 1$. This smoothed radial solution gives $\dot{R}_0/R_0 = -H$ so that the angular equation \eqref{eom_theta0} can be reduced as
\begin{align}\label{eom_theta0_smoothed}
	\ddot{\theta}_0 + H \dot{\theta}_0 = A m_{3/2} X_{\rm end} H_I a^{2-n} \sin(n\theta_0 + \delta),
\end{align}
where $a = (1+2\tau)^{1/2}$ in radiation domination and $X_{\rm end}$ defined in \eqref{def_Rend_theta_end} is a constant determined by parameters during inflations. The numerical solution of \eqref{eom_theta0_smoothed} is shown as the dashed line in the right panel of Figure~\ref{fig.finite_T_effects}, which is a smoothed result over rapid oscillations. Combining the analytic solution $R_0 = R_{\rm end}/a$ and the numerical solution from \eqref{eom_theta0_smoothed}, we obtain a hybrid result for the baryon asymmetry $Y_B$ to be discussed in the next section. As shown in Figure~\ref{fig.finite_T_YB}, this hybrid method is a good approximation for the full numerical solution and it is stable against the high frequency oscillations driven by the large thermal mass. We therefore adopt this hybrid method to study the baryon isocurvature perturbation with finite-temperature effects.

\bigbreak\noindent
$\bullet$\textit{Thermal mass domination.--}
If the thermal mass term $\Delta U_{\rm mass}$ dominates the AD field potential (see Figure~\ref{fig.UFD_snapshots}), the radial mode gains a time-varying mass $m_T \sim T$. The number density $n_\sigma \sim \rho_\sigma/m_T \sim TR^2$ since $\rho_\sigma \sim \Delta U_{\rm mass}\sim T^2R^2$. The adiabatic scaling for the number density of a coherent oscillating field is $n_\sigma \sim a^{-3}$, and thus in radiation domination with $T\sim a^{-1}$ we find $R\sim a^{-1}$. The $B$-violating $A$ term thus decays as $R^n\sim a^{-n}$. The radial motion smoothed over fast oscillations is again $R_0 = R_{\rm end}a^{-1}$, and therefore the smoothed equation \eqref{eom_theta0_smoothed} for $\theta_0$ can be used. The hybrid approximation for $Y_B$ is essentially the same as that for the thermal log domination until sufficient long time when $m_T = gT < m_\sigma$. At this very late stage with $T < m_\sigma/g$ (where $t > g^2M_P/m_\sigma^2 \gg 1/m_\sigma$), the AD field oscillates with a constant mass $m_\sigma$ so that $R\sim a^{-3/2}$. The baryon asymmetry $Y_B$ is conserved during this late-time transition since the charge-violating source has gone.
\footnote{Thermal A-terms are only important when $y R\lesssim T$, where $y$ is a Yukawa coupling to thermal plasma fields. Compared with thermal mass terms, thermal A-terms are further suppressed by a ratio of $(R/\Lambda)^{n-2}$ \cite{Allahverdi:2000zd}, and therefore we can neglect their contribution in our discussion.}

\bigbreak\noindent
$\bullet$\textit{Zero-temperature potential.--}
If the finite-temperature terms are small or negligible at the end of inflation, the AD field is under damping from the beginning at $t = t_{\rm end}$ and it is gradually relaxed from the initial VEV with the decay of the Hubble friction. In this case, the AD field does not enter the phase of coherent oscillation for $t_{\rm end} \leq t < t_c$ so that we need to solve the background evolution by full numerical approach, as the $g=0$ results in Figure~\ref{fig.finite_T_effects}. When $t > t_c \sim 1/m_\sigma$, the low-scale mass term dominates the AD field potential and it drives the harmonic oscillation. For split supersymmetry breaking \cite{Arkani-Hamed:2004ymt,Giudice:2004tc,Arkani-Hamed:2004zhs,Kasuya:2005sb} we focus on the cases with $m_\sigma \gg m_{3/2}$. The $B$-violating $A$ term decays faster than the $m_\sigma$ mass term since $n\geq 4$, and thus when $t > t_{3/2} = 1/m_{3/2}$ the angular equation \eqref{eom_theta0} reduces to $\partial_t(a^3R_0^2\dot{\theta}_0) \approx 0$. However, this time $m_\sigma$ is a constant mass so that the adiabatic condition $n_\sigma \sim n_B \sim a^{-3}$ for the oscillating AD field gives $R_0\sim a^{-3/2}$ with $\dot{\theta}_0$ modulating around a certain level from a constant value each time when $R_0$ passes the origin, as shown in Figure~\ref{fig.finite_T_effects} (see also~\cite{Wu:2020pej} for more detailed analysis).
For the zero-temperature potential with $g = 0$ cases, evolution of the AD field is not sensitive to a finite duration of the preheating process led by a coherent oscillation of the inflaton, as long as reheating completes by the onset of the relaxation of the radial mode, $t = t_{\rm rex}$, from the initial value at $R_0  =R_{\rm end}$. Since $R_0 \approx R_{\rm end}$ remains frozen by $t = t_{\rm rex}$, the onset of relaxation of the $\sigma$ field occurs around
\begin{align}
	\frac{\lambda^2}{2^{n-1}} \frac{R_{\rm end}^{2(n-1)}}{\Lambda^{2(n-3)}} = \frac{1}{2}X_{\rm end}^2 R_{\rm end}^2 H_I^2 \approx \frac{1}{2} \xi H^2(t_{\rm rex}) R_{\rm end}^2.
\end{align}
With $H = H_I/(1+2\tau)$, we find
\begin{align}\label{def_tau_rex}
	\tau_{\rm rex} \approx \frac{\sqrt{\xi}}{2X_{\rm end}} - \frac{1}{2},
\end{align}
if $\sqrt{\xi}/X_{\rm end} > 1$, and $\tau_{\rm rex} = 0$ if $\sqrt{\xi}/X_{\rm end} \leq 1$. 
(Note that $\tau_{\rm rex}\rightarrow 0$ in the cases where the finite-temperature term $\Delta U$ dominates the field potential.)

In summary, we will use the hybrid semi-analytic method, $R_0 = R_{\rm end}/a$ and $\theta_0$ solved by \eqref{eom_theta0_smoothed}, to compute $Y_B$ with finite-temperature effects, and we will use the full numerical approach to solve the AD field motion with zero-temperature potential.
It is possible to make an analytic estimation of the final $Y_B$ in the case with a finite duration of preheating led by a heavy field oscillation until its decay at $t > t_{3/2}$. This heavy field is part of the inflaton sector that dominates the energy density during inflaiton.
If the preheating phase lasts long, the universe enters a matter domination era with $H = H_I/(1+3\tau/2)$. For $n \geq 4$, the radial VEV $R_0$ stays at the instantaneous minimum \eqref{def_Rmin}, which decreases as $R_{\rm min}\sim H^{1/(n-2)}$. This is the scenario considered in \cite{Dine:1995kz}.

In any viable case, the baryon number per comoving volume is eventually frozen and the final baryon asymmetry given by \eqref{def_Y_B} approaches a constant value. The scalar condensate is expected to decay into Standard Model particles through $B-L$ conserving processes \cite{Dine:1995kz}. The asymptotic constant $Y_B$ is insensitive to the details of the decay.
As the $\sigma$ field oscillates and eventually decays, the baryon number carried by the condensate will be converted to asymmetries in the lighter species, such as quarks. If the radiation temperature at $t_{\rm rex}$ is above the electroweak phase transition, the sphaleron process will be active as the $\sigma$ field decays. The $B\!-\!L$ conserving sphaleron process will convert part of $B$ into $L$ as the chemical equilibrium is reached. In our case with initial $L=0$, the net baryon asymmetry is suppressed by a factor of $28/79$ assuming the thermal bath is dominated by the Standard Model radiation around $t_{\rm rex}$~\cite{Cline:2006ts}. In other models, such as the case where the $\sigma$ condensate carries $L$ only, the ratio flips sign~\cite{Luty:1992un}. Since this sphaleron-induced reprocessing amounts only to an overall $\mathcal{O}(1)$ factor, the benchmark value of $Y_B$ (already $\ll 1$) is unchanged at the level of logarithmic precision. For simplicity, we therefore keep using the previous numerical benchmark for $Y_B$ in what follows.

Despite that supersymmetry breaking from gravitational mediation suggests $m_\sigma \sim m_{3/2}$, the parameter space for baryogenesis is more stable in the following mass range:
\begin{align}\label{bare_mass_condition}
	\sqrt{\xi} H_I \gg m_\sigma > \frac{A}{2\sqrt{n-1}} m_{3/2}.
\end{align}
The condition $\sqrt{\xi} H_I \gg m_\sigma$ ensures that $X_{\rm min}$ given by \eqref{def_Xmin} is a constant during inflation so that the radial mass $m_R \sim H_I$. The inequality $m_\sigma > A m_{3/2}/(2\sqrt{n-1})$ is to prevent the unstable growth of the final baryon asymmetry led by the presence of non-trivial minima in the late stage of evolution, as discussed in more detail in Appendix~\ref{Appendix_unwanted_minima}.

\section{Baryon density isocurvature perturbations}\label{Section_BDI}

In this section, we compute the perturbations of the AD field and the corresponding baryon density perturbations of this model.
The density perturbation from the simplest single-field inflation models is adiabatic as the curvature perturbation is the only source of perturbations to all components of the universe during reheating. The situation can be different for more complicated models. Perturbations in the number density of a particle species $n_i$ is said to be an isocurvature mode if the gauge invariant quantity $\mathcal{I}_i \equiv \delta(n_i/n_\gamma)$ is not a constant, where $n_\gamma$ is the number density of radiation.  
Neglecting the contribution from the neutrino sector, the (total) matter density isocurvature (MDI) perturbation is given by \cite{Planck:2018jri}:
\begin{align}\label{def_iso_matter_perturbation}
	\mathcal{I}_{\rm MDI} = \left(\frac{\Omega_{\rm CDM0}}{\Omega_{m0}}\right) \mathcal{I}_{\rm CDI} + \left(\frac{\Omega_{b0}}{\Omega_{m0}}\right) \mathcal{I}_{\rm BDI},
\end{align}
where $\mathcal{I}_{\rm CDI}$ is the cold-dark-matter (CDM) density isocurvature perturbation and $\mathcal{I}_{\rm BDI}$ is the perturbation of the baryon density.
$\Omega_{\rm CDM0} \approx 0.265$ ($\Omega_{m0} = \Omega_{\rm CDM0} + \Omega_{b0} \approx 0.315$) is the CDM (total matter) density today \cite{Planck:2018vyg}.
The observational constraints on the fraction of the MDI power spectrum in the total amount is defined as
\begin{align}\label{def_beta_iso}
	\beta_{\rm iso} \equiv \frac{P_{II}(k_{\rm pivot})}{P_{II}(k_{\rm pivot}) + P_{\mathcal{RR}}(k_{\rm pivot})},
\end{align}
which depends on the pivot scale $k_{\rm pivot}$. The choices of $k_{\rm pivot}$ used in \cite{Planck:2018jri} are $k_{\rm low} = 0.002$ Mpc$^{-1}$, $k_{\rm mid} = 0.050$ Mpc$^{-1}$ and $k_{\rm high} = 0.100$ Mpc$^{-1}$, respectively. Note that $P_{II}(k) = (k^3/2\pi^2)\langle \mathcal{I}_{\rm MDI}^2(k)\rangle^\prime$ uses the dimensionless definition. $\mathcal{R}$ stands for the comoving curvature perturbation, which coincides with the gauge invariant definition of the curvature perturbation $\zeta$ on superhorizon scales. 

Assuming that there is no correlation between the perturbations of the dark matter and baryons (namely $\langle \mathcal{I}_{\rm CDI}\mathcal{I}_{\rm BDI}\rangle = 0$), we have
\begin{align}
	P_{II} = \left(\frac{\Omega_{\rm CDM0}}{\Omega_{m0}}\right)^2 P_{\rm CDI} + \left(\frac{\Omega_{b0}}{\Omega_{m0}}\right)^2 P_{\rm BDI}.
\end{align}
Taking $\beta_{\rm iso} P_{\mathcal{RR}} = (1-\beta_{\rm iso}) P_{II}$ from \eqref{def_beta_iso}, we therefore obtain
\begin{align}
	P_{\rm BDI} = 
	\frac{\beta_{\rm iso}}{1-\beta_{\rm iso}} \left(\frac{\Omega_{m0}}{\Omega_{b0}}\right)^2 P_{\mathcal{RR}} 
	- \left(\frac{\Omega_{\rm CDM0}}{\Omega_{b0}}\right)^2 P_{\rm CDI}.
\end{align}
If we take $P_{\rm CDI} = 0$, $P_{\mathcal{RR}} = 2.2\times 10^{-9}$ and a conservative bound $\beta_{\rm iso} \lesssim 0.02$ from the generally correlated models (see Section 9.2 in \cite{Planck:2018jri}) with three isocurvature parameters, we get $P_{\rm BDI} \lesssim 1.8 \times 10^{-9}$.

Note that $Y_B \sim n_B/T^3$ gives $\delta Y_B/Y_B = \delta n_B/n_B -3\delta T/T$, and the energy density of radiation $\rho_\gamma \sim T^4$ indicates $\delta \rho_\gamma/\rho_\gamma = 4\delta T/T$. As a result, the gauge invariant definition of the baryon density isocurvature (BDI) perturbation reads
\begin{align}\label{def_BDI}
	\mathcal{I}_{\rm BDI} \equiv \delta \ln\left(\frac{n_B}{n_\gamma}\right) = \frac{\delta n_B}{n_B} - \frac{\delta n_\gamma}{n_\gamma}
	= \frac{\delta \rho_B}{\rho_B} - \frac{3}{4} \frac{\delta \rho_\gamma}{\rho_\gamma} =  \frac{\delta Y_B}{Y_B}, 
\end{align}
where the effect of radiation density perturbation (assumed to be sourced by the adiabatic curvature perturbation) drops out naturally.

\subsection{A separated Universe approach}\label{Sec_Separated_Universe}
Due to the exponential expansion of the spatial distance during inflation, the physical wavelength ($\lambda_p \sim a/k$) of quantum fluctuations, such as $\delta R_k$ or $\delta \theta_k$, can be stretched larger than the horizon size ($\sim 1/H_I$) so that they behave very similar to the zero-mode ($k = 0$) background components $R_0$ or $\theta_0$. If these superhorizon fluctuations survive until the end of inflation, they can be viewed as slightly different initial conditions \eqref{def_Rend_theta_end} for the post-inflationary relaxation of the complex scalar $\sigma$ in different patches of the universe. In a local patch of the universe, the $\sigma$ field is approximately homogeneous, and its dynamics is governed by equations for the homogeneous background, namely \eqref{eom_R0} and \eqref{eom_theta0}.

The radius of each separated patch of the universe may be different from the horizon scale depending on our choices of model parameters. To be specific, let us focus on cases with a sufficiently large $\xi$ such that the radial mass \eqref{def_m_R} during inflation satisfies $M_R = m_R/H_I >  3/2$. This condition ensures that no radial perturbation $\delta R_k$ on scales relevant to the targeting cosmological observations can survive till the end of inflation. As a result, we have a homogeneous radial value $R = R_0$ for the entire universe, and a local  patch is specified by the correlation length of the angular perturbation $\delta\theta_k$.

The separated universe approach has been well-established for the study of curvature perturbations from multi-field inflation scenarios \cite{Starobinsky:1985ibc, Sasaki:1995aw, Sasaki:1998ug, Lyth:2004gb}. Here we adopt the basic idea for computing the BDI perturbations from the initial AD field fluctuations at the end of inflation.  
To make the discussion more concrete, let us consider two separated patches of the universe with the initial conditions at the end of inflation as:
\begin{itemize}
	\item Patch 1: $\theta_1 = \theta_{\rm end}$, $R_1 = R_{\rm end}$ as given by \eqref{def_Rend_theta_end}.
	
	\item  Patch 2: $\theta_2 = \theta_{\rm end} + \Delta\theta$, $R_2 = R_{\rm end}$, where $- P_\theta^{1/2} \leq \Delta\theta \leq P_\theta^{1/2}$ and $P_\theta$ is the spectral amplitude of the power spectrum of $\delta\theta$.
\end{itemize}  
Here we use $\Delta\theta$ to denote a deviation in background values of $\theta_0$, which is not the same as the linear perturbations $\delta\theta$.

In each universe, the final baryon asymmetry $Y_f = Y_B(t_f)$ is solved by \eqref{eom_R0} and \eqref{eom_theta0} for the homogeneous background values $R_0$ and $\theta_0$. Results for the two separated universe are denoted as
\begin{align}
	Y_{f1} = \frac{R_f^2 \dot{\theta}_{f1}}{s_f}, \qquad Y_{f2} =  \frac{R_f^2 \dot{\theta}_{f2}}{s_f}.
\end{align}
The BDI perturbation \eqref{def_BDI} is thus mimicked by the quantity
\begin{align}
	\mathcal{I}_{\rm BDI} \approx 	\left\vert \frac{Y_{f2}- Y_{f1}}{Y_{f1}}\right\vert,
\end{align}
with $\Delta\theta = \pm P_\theta^{1/2}$ chosen at the averaged value.\footnote{For the computation of curvature perturbations, we often take the spatially flat slicing at the initial time so that inflaton takes different values in different patches of the universe. As we take the uniform-field gauge of inflaton at the end of inflation, the curvature perturbation is thus revealed as the difference in the number of $e$-folds among different patches of the universe. For the computation of the BDI perturbation, we can still take the spatially flat slicing at the initial time and solve $Y_B$ for different patches of the universe by the same background equations, since the adiabatic perturbation is not involved in the gauge invariant definition \eqref{def_BDI}. We terminate all computations with a uniform numerical time so that the BDI perturbation is reflected as the difference of $Y_B$ led by the initial difference of $\theta_{\rm end}$. }

We now determine the largest $\Delta\theta$ that originated from quantum fluctuations in the angular mode $\theta$ during inflation. Treating $\theta$ as a massive scalar decoupled from the heavy radial mode during inflation, we estimate the size of the fluctuation $\delta\theta$ as
\begin{align}\label{def_theta_k}
	\delta\theta_k(\eta) = \frac{H_I}{R_{\rm min}} u_k(\eta), \qquad u_k(\eta) \equiv c_\theta (-k\eta)^{3/2} H_{\nu_\theta}^{(1)}(-k\eta),
\end{align}
where $-\infty < \eta < \eta_{\rm end}$ is the conformal time defined for the epoch of inflation, $u_k$ is the standard mode function of a massive scalar field during inflation, and
\begin{align}
	c_\theta = -\frac{i}{2} \sqrt{\frac{\pi}{k^3}} e^{i\pi(\nu_\theta +1/2)/2}, \qquad \nu_\theta = \sqrt{\frac{9}{4} - M_\theta^2} =  \sqrt{\frac{9}{4} - n\alpha_{\rm end} X_{\rm end}},
\end{align} 
where $M_\theta \equiv m_\theta/H$.
The power spectrum can be computed via the two-point function according to the definition
\begin{align}
	\left\langle \delta\theta_{\vec{k}}\delta\theta_{\vec{p}} \right\rangle = \left(2\pi\right)^3 \delta^{(3)} \left(\vec{k}+\vec{p}\right) P_\theta(k;\eta) \frac{2\pi^2}{k^3}.
\end{align}
Taking the mode function \eqref{def_theta_k}, we find
\begin{align}\label{def_P_theta_0}
	P_\theta(k, \eta) = \frac{k^3}{2\pi^2} \frac{H_I^2}{R_{\rm min}^2} \vert u_k(\eta)\vert^2.
\end{align} 
Note that $R_{\rm min} = R_{\rm end}$ is a constant during inflation unless we consider the presence of primordial features in Section~\ref{Sec_primordial_features}.
In practice, inflation must end within a finite number of $e$-folds and we are interested in the spectral amplitude at the end of inflation, denoted by the conformal time $\eta_{\rm end}$, which can be also written in terms of the horizon crossing mode $k_{\rm end}$ at $\eta=\eta_{\rm end}$, $\eta_{\rm end}=-1/k_{\rm end}$.
Note that in our example we choose $k_{\rm end}=e^{55} k_{\rm pivot}$.
For $m_\theta/H_I < 3/2$, $P_\theta(k, \eta_{\rm end})$ scales as
\begin{equation}
    P_\theta(k, \eta_{\rm end}) \propto \bigg(\frac{k}{k_{\rm end}}\bigg)^{3-2\nu_\theta}.
\end{equation}
The scale-dependence of $P_{\theta}(k,\eta_{\rm end})$ depends on the value of $m_\theta/H_I$. It is blue-tilted if $m_\theta/H_I>0$ and approximately scale-invariant if $m_\theta/H_I \approx 0$. We will consider both cases.

\begin{figure}[]
	\begin{center}
		\includegraphics[width=12 cm]{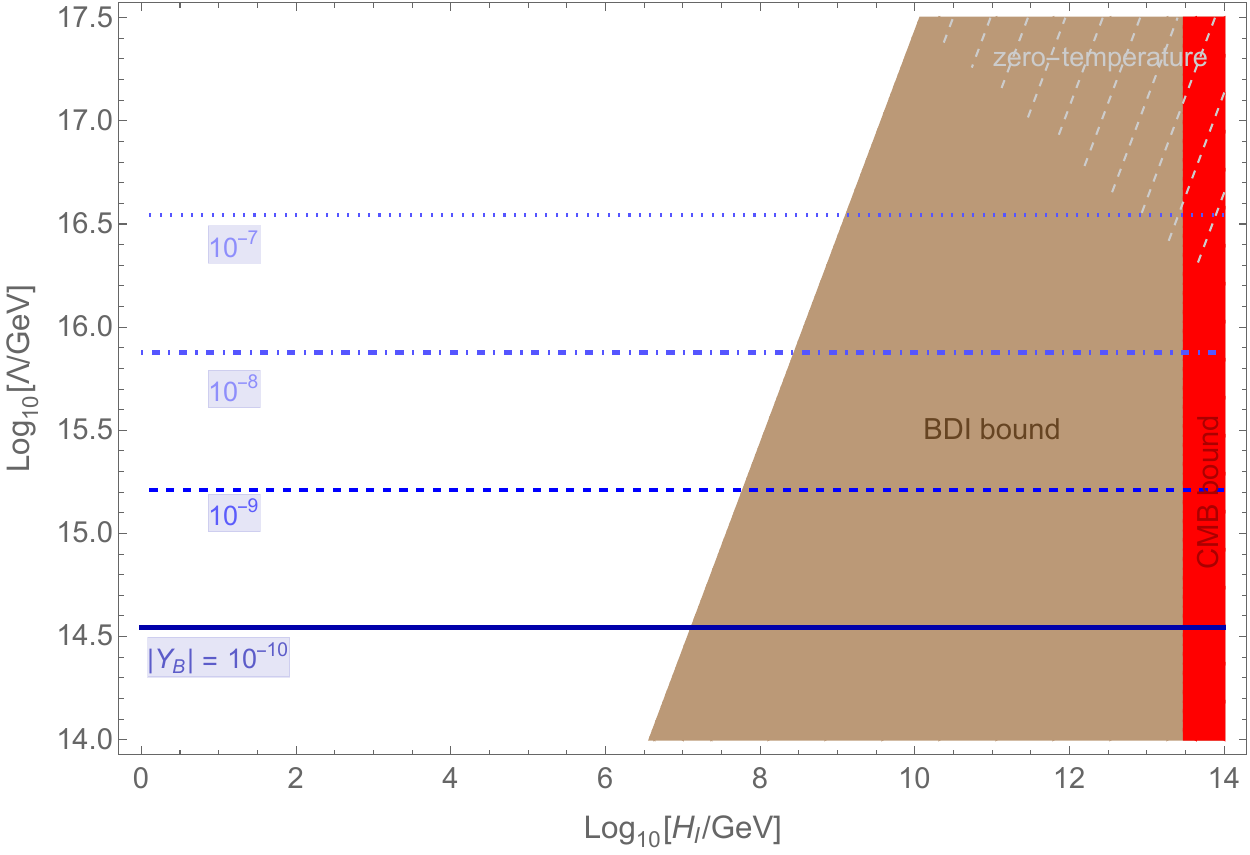}
	\end{center}
	\caption{\label{fig.scan_HI_M} A scan of the $Y_B$ contours in the $H_I$ to $\Lambda$ parameter space with $n=6$, $\xi = 100$, $m_\sigma/H_I = m_{3/2}/H_I = 10^{-3}$, $\delta = \pi/60$, and $\lambda = A = 1$. The colored regions are excluded by the BDI bound based on the hybrid semi-analytic results with finite-temperature effects and the CMB bound for single-field inflation $H_I < 3\times 10^{13}$ GeV \cite{BICEP:2021xfz}. Finite-temperature effects are subdominant in the meshed zero-temperature region where perturbations of the AD field are unstable.}
\end{figure}

\bigbreak\noindent
$\bullet$\textit{The BDI bound.--} 
To investigate the BDI constraints with finite-temperature effects, we adopt the hybrid semi-analytic solutions, $R_0 = R_{\rm end}/a$ and $\theta_0$ solved by \eqref{eom_theta0_smoothed}, to the separated universe approach. For instantaneous reheating, we use $c_H = c_A = 0$ for the effective potential \eqref{def_UFD_RD} in radiation domination. However, the hybrid method has two important features as (1) it does not care about the values of Hubble-induced coefficients $c_H$ and $c_A$ since the thermal corrections dominate the evolution, and, (2) it is independent of the $\mathcal{O}(1)$ gauge coupling $g$. 

We reduce the number of free parameters in our survey by fixing $\lambda = A = 1$. $\xi = 1$ is the lower bound for having a sufficiently large radial mass \eqref{def_m_R} to stabilize the model. The final $Y_B$ is not very sensitive to an arbitrary choice of $\xi \geq 1$. $\lambda = 1$ can be achieved by rescaling the ratio $\Lambda/H_I$ and $A = 1$ can also be achieved by rescaling the ratio $m_{3/2}/H_I$. $m_\sigma \ll H_I$ shall satisfy the condition \eqref{condition_unwanted_VEV}, and the ratio $m_\sigma/H_I$ determines the required numerical time to reach a constant $Y_B$ in radiation domination, but it does not play an important role in the production of the net baryon asymmetry.

Following the setup, $R_{\rm end}$ and $m_{3/2}$ are the two most important parameters that control the final $Y_B$, and $R_{\rm end}$ given in \eqref{def_Rend_theta_end} is mainly determined by $\Lambda$ and $H_I$. An example of the viable parameter space allowed by the current BDI bound is shown in Figure~\ref{fig.scan_HI_M} with $m_{3/2} = m_\sigma$, as motivated by gravity mediation.
We find that the final BDI perturbation $I_{\rm BDI}$ is, at leading order, linear to the initial angular perturbation $\delta\theta$ once other parameters like $H_I$ or $\Lambda$ are fixed. In this case the perturbation mode $\delta R$ is vanishing at large scales due to the heavy radial mass. We therefore denote the relation as $ \vert\mathcal{I}_{\rm BDI}\vert \simeq \chi \vert\delta\theta\vert$. Since $\delta\theta$ is given by \eqref{def_theta_k}, we only need to find the ratio $\chi$ by numerical computations.
Since $T\sim H^{1/2}$ and $R_{\rm end}/H_I \sim (\Lambda/H_I)^{(n-3)/(n-2)}$ as given by \eqref{def_Rmin}, the $n = 6$ scenario, as motivated by the $udd$ flat direction in MSSM, is a special case where $Y_B$ becomes independent of $H_I$.
As shown in Figure~\ref{fig.scan_HI_M}, the highest Hubble scale of inflation $H_I$ is set by the BDI bound. We report the $Y_B = 10^{-10}$ contours based on the Standard Model entropy density in \eqref{def_Y_B}. (More precisely, $8.7\times 10^{-11}$ is the reference value for the baryon asymmetry observed today.) The parameter space with $Y_B > 10^{-10}$ can become valid if one considers a model of particle physics with a dilution driven by entropy production at lower temperatures.

\subsection{Numerical results of linear perturbations}\label{Sec_Linear_Perturbation}
With $c_H = c_A =0$ in radiation domination following instantaneous reheating, the zero-temperature potential is very flat so that small scale perturbations of the AD field can develop spinoidal instability and collapse into Q-balls \cite{Kusenko:1997si}. This is an interesting topic for the flat-direction model but it is beyond the scope of this work. In this section, we numerically solve the linear perturbations in the zero-temperature zone to verify the stability of the system. 
As explained at the end of this section, we also design consistency checks for BDI perturbations solved by the linear perturbation theory with those obtained from the seperated universe approach in the zero-temperature limit. 

We adopt the method used in \cite{Kusenko:1997si} to study linear perturbations of the AD field. As a first step, let us include the spatial dependence into the radial and angular modes as
\begin{align}
	R = R(t, \vec{x}), \qquad \theta = \theta(t, \vec{x}),
\end{align}
where the full equations of motion with spatial derivatives read:
\begin{align}\label{eom_full_R_RD}
	\ddot{R} + 3H \dot{R} +\partial_R U_{\rm FD}+ \left[-\frac{\Delta}{a^2} -\dot{\theta}^2 +\frac{\partial^i\theta\partial_i\theta}{a^2}\right] R =0, 
	\\\label{eom_full_thtea_RD}
	\ddot{\theta} + \left(3H + 2 \frac{\dot{R}}{R}\right) \dot{\theta} -\frac{\Delta}{a^2}\theta -\frac{2}{R}\frac{\partial^i R\partial_i \theta}{a^2} 
	+ \frac{1}{R^2}\partial_\theta U_{\rm FD}= 0.
\end{align}
Here $\Delta \equiv \delta^{ij}\partial_i\partial_j$ and $U_{\rm FD}$ is given by \eqref{def_UFD_polar}. 

Next, we can perform the decomposition of the two modes into
\begin{align}
	R(t ,\vec{x}) = R_0(t) + \delta R(t, \vec{x}), \qquad \theta(t, \vec{x}) = \theta_0(t) + \delta\theta(t, \vec{x}),
\end{align}
where $R_0$ and $\theta_0$ are the homogeneous part (namely the zero-mode) of the fields and their evolution are governed by \eqref{eom_R0} and \eqref{eom_theta0}.
Equations for the linear perturbations are 
\begin{align}
	\label{eom_dR}
	\delta\ddot{R}_k + 3H\delta\dot{R}_k + \left[\frac{k^2}{a^2}+\partial_{RR} U_{\rm FD} -\dot{\theta}_0^2 \right] \delta R_k =&
	2\dot{\theta}_0 R_0 \delta\dot{\theta}_k - \partial_{R\theta} U_{\rm FD} \delta\theta_k,
	\\\label{eom_dtheta}
	\delta\ddot{\theta}_k + \left(3H + 2\frac{\dot{R}_0}{R_0}\right) \delta\dot{\theta}_k 
	+\frac{k^2}{a^2} \delta\theta_k +\frac{1}{R_0^2}\partial_{\theta\theta} U_{\rm FD}\delta\theta_k =&
	\\\nonumber
	2 \frac{\dot{R}_0 \dot{\theta}_0}{R_0^2} \delta R_k - 2 \frac{\dot{\theta}_0}{R_0} \delta \dot{R}_k
	+2\frac{\delta R_k}{R_0^3}\partial_\theta U_{\rm FD} &- \frac{1}{R_0^2}\partial_{\theta R} U_{\rm FD}\delta R_k,
\end{align}
where $\delta R_k$ and $\delta\theta_k$ are Fourier modes associated with the wavenumber $k$, and $\partial_R\partial_{\theta} \equiv\partial_{R\theta} = \partial_{\theta R}$.
During inflation, $\delta R_k$ and $\delta\theta_k$ are governed by the canonical equation of motion for massive scalar perturbations with $\dot{R}_0 = \dot{\theta}_0 = 0$ and $\theta_0 = \theta_{\rm min}$, $R_{0} = R_{\rm min}$ so that $\partial_\theta U_{\rm FD} = \partial_{R\theta}U_{\rm FD} =0$. The scalar masses are defined as \eqref{def_m_R} and \eqref{def_m_theta}. 
The initial value of $\delta R$ is therefore given by
\begin{align}\label{def_R_k}
	\delta R_k(\eta) =  c_R H_I (-k\eta)^{3/2} H_{\nu_R}^{(1)}(-k\eta), \qquad \nu_R =  \sqrt{\frac{9}{4} - M_R^2}
\end{align} 
where $M_R^2 \equiv m_R^2/H_I^2$ is given by \eqref{def_m_R} and
\begin{align}
	c_R = -\frac{i}{2} \sqrt{\frac{\pi}{k^3}} e^{i\pi(\nu_R +1/2)/2},
\end{align}
is determined by the Bunch-Davis vacuum condition.
The initial value of $\delta\theta$ can be found in \eqref{def_theta_k}.

\begin{figure}[]
	\begin{center}
		\includegraphics[width=12 cm]{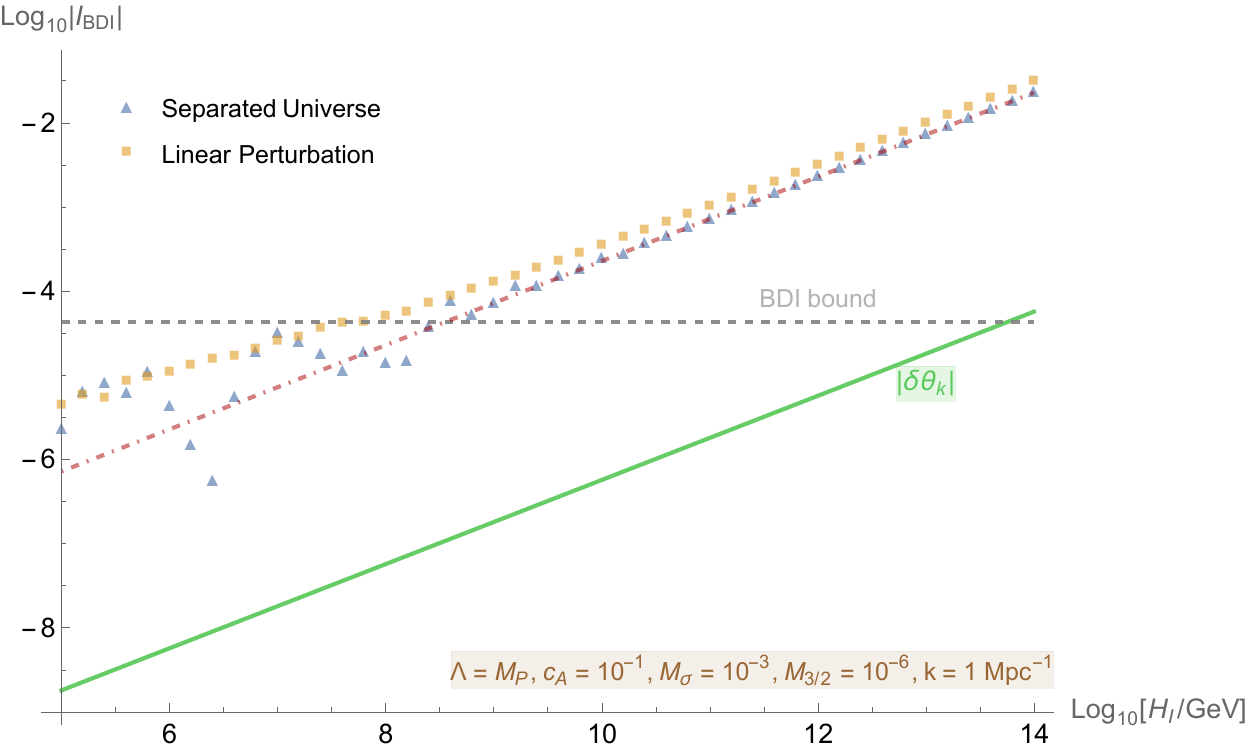}
	\end{center}
	\caption{\label{fig.IBDI} A consistency check of the baryon density isocurvature (BDI) perturbation with respect to the Hubble parameter of inflation $H_I$ based on the separated universe approach (Section~\ref{Sec_Separated_Universe}) and the linear perturbation theory (Section~\ref{Sec_Linear_Perturbation}). The dashed line is the upper bound of the BDI perturbation from the generally correlated adiabatic and isocurvature models in \cite{Planck:2018jri}. The solid line for $\vert\delta\theta\vert$ is the initial value of the angular perturbation used in both methods, which differ from $I_{\rm BDI}$ by a constant ratio. The dot-dashed line is the estimation of BDI perturbation based on  \eqref{Delta_theta_formalism}. $\xi = \lambda = A = 1$ are used in the computations of these results. We use the dimensionless notations $M_\sigma\equiv m_\sigma/H_I$ and $M_{3/2} \equiv m_{3/2}/H_I$ in this plot. The corresponding angular mass is $m_\theta/H_I = 0.488$. }
\end{figure}

By solving the perturbations from \eqref{eom_dR} and \eqref{eom_dtheta}, we can obtain the BDI perturbation \eqref{def_BDI} for each $k$ mode as
\begin{align}\label{I_BDI_linear_perturbation}
	\mathcal{I}_{\rm BDI}(k) = \frac{\delta Y_B(k)}{Y_B} =  \frac{\delta n_B(k)}{n_B} \approx 2\frac{\delta R_k}{R_0} + \frac{\delta\dot{\theta}_k}{\dot{\theta}_0}.
\end{align}
We show a numerical example in Figure~\ref{fig.IBDI}, and compared with the results from the separated universe approach.

\bigbreak\noindent
$\bullet$\textit{Consistency checks.--} 
Let us consider a consistency check for the separated universe approach and the linear perturbation theory of the AD field. We perform such a consistency check with the zero-temperature potential ($g = 0$) where the full numerical computation of the background evolution is available. In particular, we use the full numerical background evolution to obtain the BDI perturbations from the two approaches. With $c_H = c_A = 0$ set by instantaneous reheating, the zero-temperature potential is too flat so that the late-time evolution generally encounters with the spinoidal instability, $\dot{\theta}_0^2 > \partial_{RR}U_{\rm FD} \sim m_\sigma^2$, which drives the unstable growth of linear perturbations and it can lead to Q-ball formation \cite{Kusenko:1997si,Dine:2003ax}. Such an issue can exist in AD baryogenesis but is not the concern of consistency checks for the two approahes. To avoid this issue, we impose $c_H = -\xi$ and $c_A$ as non-vanished constant values at the end of inflation with $H_I = H$ to Hubble-induced terms in the zero-temperature potential. This introduces a sufficently large mass to avoid the spinoidal instability for linear perturbations while keeping the background motion under critical damping without rapid oscillations. Such a special treatment goes beyond the instantaneous reheating but it ensures that the full numerical solutions in our consistency checks are stable. Here, our purpose is to see if results obtained from the two approaches agree with each other provided with the same background evolution. 

Let us see how the separated universe approach works in such a designed setup for consistency checks.
In the late-stage of evolution, $t>t_c \sim 1/m_\sigma$, the conservation of the angular momentum $\partial_t(a^3 R_0^2\dot{\theta}_0) \approx 0$ is reached. 
At this stage, $R_0 \sim a^{-3/2}$ is governed by the harmonic oscillation led by the $m_\sigma$ mass term, and we approximate $\dot{\theta}_0 \approx \dot{\theta}_c$ as the constant background value for $t > t_c$, neglecting the rapid spiky modulation led by the $R_0$ oscillations. Given that the $c_A$ term decays with the Hubble parameter, we have $\partial_\theta U_{\rm FD} \approx -A X H m_{3/2} R_0^2 \sin(n\theta_0+\delta)$ at this stage, where $X$ is defined in \eqref{def_X}. This is the source term of the angular rotation of the AD field in \eqref{eom_theta0}, and from which we can estimate the value of $\dot{\theta}_c$ as
\begin{align}\label{theta_c}
	\vert\dot{\theta}_c\vert \approx \frac{\partial_\theta U_{\rm FD}}{3HR_0^2} \approx \frac{A}{3} m_{3/2} X|_{t=t_c},
\end{align}
ignoring the rapid oscillations. 

If there is a small fluctuation $\Delta\theta/\theta_{\rm end} \ll 1$ of the initial condition in different patches of the universe, one can check that the angular gradient in the flat-direction potential at the end of inflation is
\begin{align}
	\left.\frac{1}{R_0^2}\partial_{\theta}U_{\rm FD} \right\vert_{\theta_{\rm end} + \Delta\theta} 
	\approx 
	\left.\frac{1}{R_0^2}\partial_{\theta}U_{\rm FD} \right\vert_{\theta_{\rm end}}
	+ \left.\frac{1}{R_0^2}\partial_{\theta}^2U_{\rm FD} \right\vert_{\theta_{\rm end}} \Delta\theta,
\end{align}
up to linear order in $\Delta\theta$, where $\partial^2_\theta U_{\rm FD} = -n\alpha H^2X R^2$ and $\alpha$ is defined in \eqref{def_alpha}.
Note that $\partial_{\theta}U_{\rm FD}\vert_{\theta_{\rm end}} =\partial_\theta\alpha\vert_{\theta_{\rm end}} = 0$ by our definition of $\theta_{\rm end}$ given by \eqref{def_theta_min}. Therefore the initial angular fluctuation $\Delta\theta$ leads to a small difference in $\dot{\theta}_c$ for $t \geq t_c$, which is estimated as
\begin{align}\label{Delta_theta_c}
	\Delta\dot{\theta}_c \approx \frac{c_A}{3}H_c X_c n \Delta\theta,
\end{align}
where $H_c \equiv H(t_c) = 1/(1+2H_I/m_\sigma)$.
Here, the approximation holds only if $A m_{3/2} \ll c_A H_c \sim c_A m_\sigma$.
One can check that $\alpha(\theta_{\rm end} + \Delta\theta) = \alpha(\theta_{\rm end})$ up to the linear correction in $\Delta\theta$, and therefore $X_{\rm min}(\theta_{\rm end} + \Delta\theta) \approx X_{\rm min}(\theta_{\rm end})$ as given by \eqref{def_Xmin}. 

We can estimate the BDI perturbation \eqref{def_BDI} for the separated universe approach by using \eqref{theta_c} and \eqref{Delta_theta_c} as
\begin{align}\label{Delta_theta_formalism}
	\mathcal{I}_{\rm BDI} \approx \Delta(\ln Y_B) = \Delta (\ln n_B) \approx \frac{\Delta\dot{\theta}_c}{\dot{\theta}_c} \sim \frac{c_A m_\sigma}{A m_{3/2}} n \Delta\theta.  
\end{align}  
This estimation is given in Figure~\ref{fig.IBDI} as the dot-dashed red line.  
Our consistency checks show that BDI perturbations obtained from the separated universe approach agree with the full numerical solutions from the linear perturbation theory, provided with the same effective potential of the AD field in radiation domination.

\subsection{Baryogenesis with a light angular mass}
So far, we have focused on the generation of baryon asymmetry in cases with a sizable $c_A$ or $\xi \sim \mathcal{O}(10^2)$ during inflation that explicitly breaks the conservation of the global quantum number, $B$, of the theory with $m_\theta/H_I > 0.1$. If $m_\theta/H_I > 1$, we will still have successful baryogenesis, but unfortunately, the resulting BDI perturbations at large scales will be too small for us to probe the underlying physics. 

On the other hand, it is possible that the $B$-violating $c_A$ term to be vanishingly small during and after inflation \cite{Dine:1995kz}. If $c_A < 10^{-2}$ and $\xi \sim\mathcal{O}(1)$, the angular mass given by \eqref{def_m_theta} is much smaller than the Hubble parameter of inflation ($m_\theta/H_I < 0.1$). In this case, the potential barrier in the angular direction is too shallow to trap the AD scalar so that the angular mode $\theta$ can be treated approximately as massless. The cases with $m_\theta/H_I \ll 1$ are still interesting, since we can have detectable BDI perturbations with a nearly scale-invariant spectrum (in contrast to the blue spectrum index in the $m_\theta\sim H_I$ case). 

\begin{figure}[]
	\begin{center}
		\includegraphics[width=10 cm]{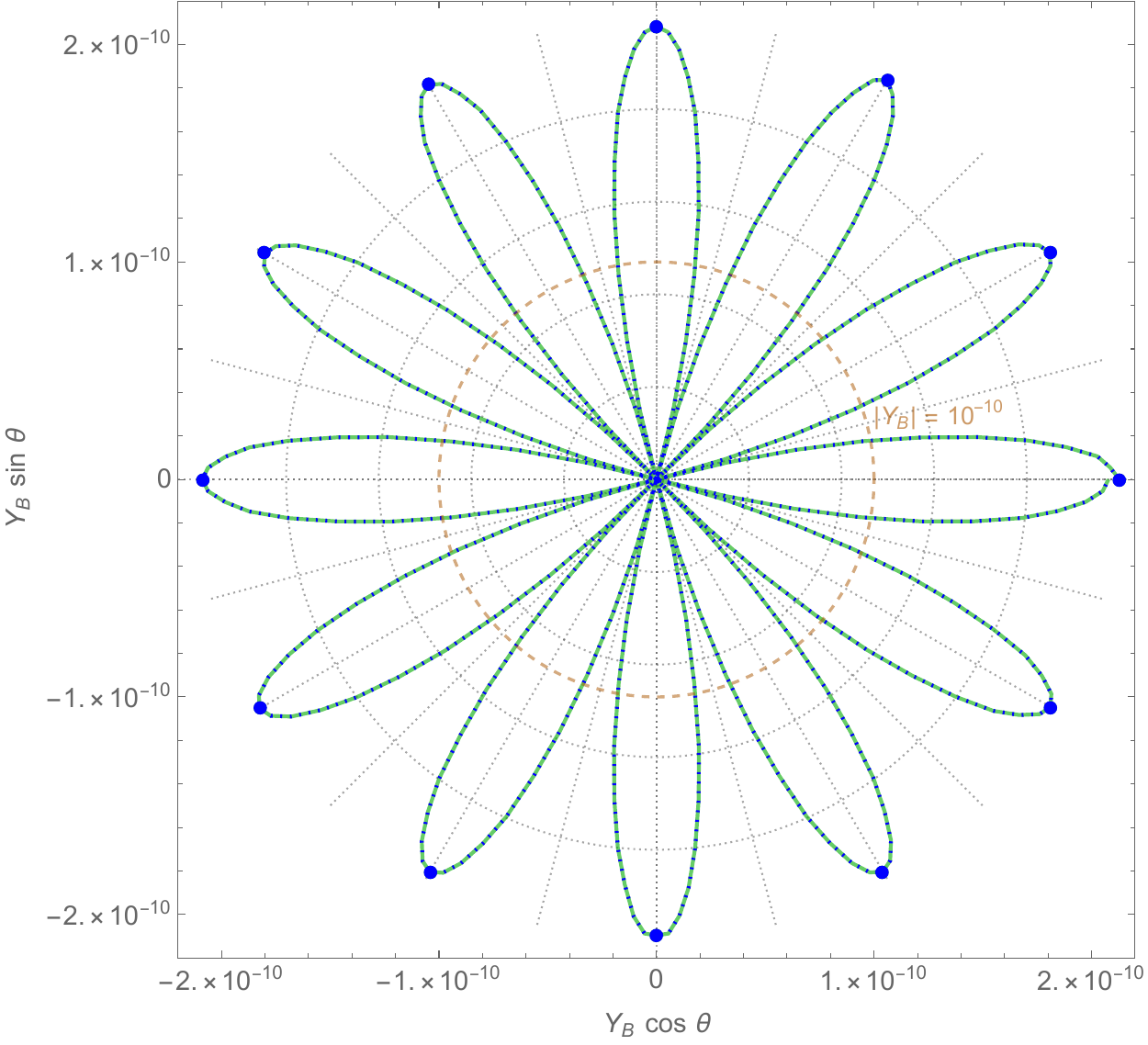}
	\end{center}
	\caption{\label{fig.YB_scan_AM} The scan of final baryon asymmetry with respect to the initial misalignment angle $\theta \in [0,2\pi)$  with $H_I = 10^{12}$ GeV, $n = 6$, $\xi = \lambda = A = 1$, $\delta = \pi/2$, $\Lambda/H_I = 10^4$, $M_\sigma = 10^{-3}$ and $M_{3/2} = 10^{-5}$. The blue-dotted line result uses $\theta = \theta_I$ given by \eqref{def_theta_I} and the green-solid line result uses $\theta = \theta_{\rm end}$ as initial conditions. The dots denote data points at $\theta = n\pi/6$.}
\end{figure}

By neglecting the small $c_A$ term and the low-energy $A$ term during inflation, the baryon number $B$ conservation is recovered in the theory. In such a case, baryon asymmetry can still be generated in a local universe from an initial angular VEV randomly picked up at the end of inflation. This initial VEV must be misaligned with the low-energy potential minima nor maxima caused by the $B$-violating $A$ term so that the AD field relaxation can start from a $CP$ violating state. Remarkably, in this scenario the net baryon number averaged over all inflationary domains vanishes \cite{Hook:2015foa, Wu:2020pej}, as can be seen in Figure~\ref{fig.YB_scan_AM}.

Without $B$ violating terms, the radial minimum \eqref{def_Rmin} is simplified as
\begin{align}\label{def_Rend_angular_misalignment}
	R_{\rm end} = \left(\frac{2^{n/2-1}}{\lambda}\Lambda^{n-3}X_{\rm min}H_I\right)^{\frac{1}{n-2}}, \quad X_{\rm min} = \left(\frac{\xi- m_\sigma^2/H_I^2}{n-1}\right)^{1/2}.
\end{align} 
The angular VEV during inflation is now realized by the spontaneous symmetry breaking of the $U(1)$ so that $\theta_{\rm end}$ becomes a free parameter.\footnote{The current upper bound $H_I \leq 3\times 10^{13}$ GeV for the Hubble parameter in single-field inflation indicates that $R_{\rm end} > H_I/(2\pi)$ is required for the generation of $Y_B \sim 10^{-10}$, and thus $U(1)$ symmetry is never restored by quantum fluctuations.}
Taking into account the stochastic corrections to $\theta_{\rm end}$ led by the angular perturbation $\delta\theta = H_I/(2\pi R_{\rm end})$, the angular VEV at the end of inflation is given by
\begin{align}\label{def_theta_I}
	\langle \theta_I^2\rangle = \theta_{\rm end}^2 + \left( \frac{H_I}{2\pi R_{\rm end}}\right)^2.
\end{align}

Since the light angular mode does not decay after the horizon exit during inflation, it randomly walks along the angular direction in the potential. At the end of inflation, different values of $\theta_{I}$ in different patches of the universe can result in very different numbers of baryons or even antibaryons. 
To ensure that we can have a local universe with only small fluctuations around the central value $Y_B = 10^{-10}$ of the baryon asymmetry, we require the overall size of this random walk accumulated during inflation to be much smaller than the circumference of the orbit of the radial minimum  in the potential. Such a condition reads
\begin{equation}
\frac{H_I}{2\pi} \sqrt{N_{\rm tot}} \ll 2\pi R_{\rm end} \sim 2\pi 
\left(\frac{\Lambda}{H_I}\right)^{\frac{n-3}{n-2}} H_I,
\label{eq:masslesscondition}
\end{equation}
where $N_{\rm tot} = \vert N_{\rm end} -N_\ast\vert$, and the factor $(2^{n/2-1}X_{\rm min}/\lambda)^{1/(n-2)}\sim \mathcal{O}(1)$ in $R_{\rm end}$ is given by \eqref{def_Rend_angular_misalignment}. Since $\sqrt{N_{\rm tot}}/4\pi^2\approx0.2$ with $N_{\rm tot} = 55$, we can see that the condition is clearly satisfied with $\Lambda/H_I \gg 1$.

\begin{figure}[]
	\begin{center}
		\includegraphics[width=7.7 cm]{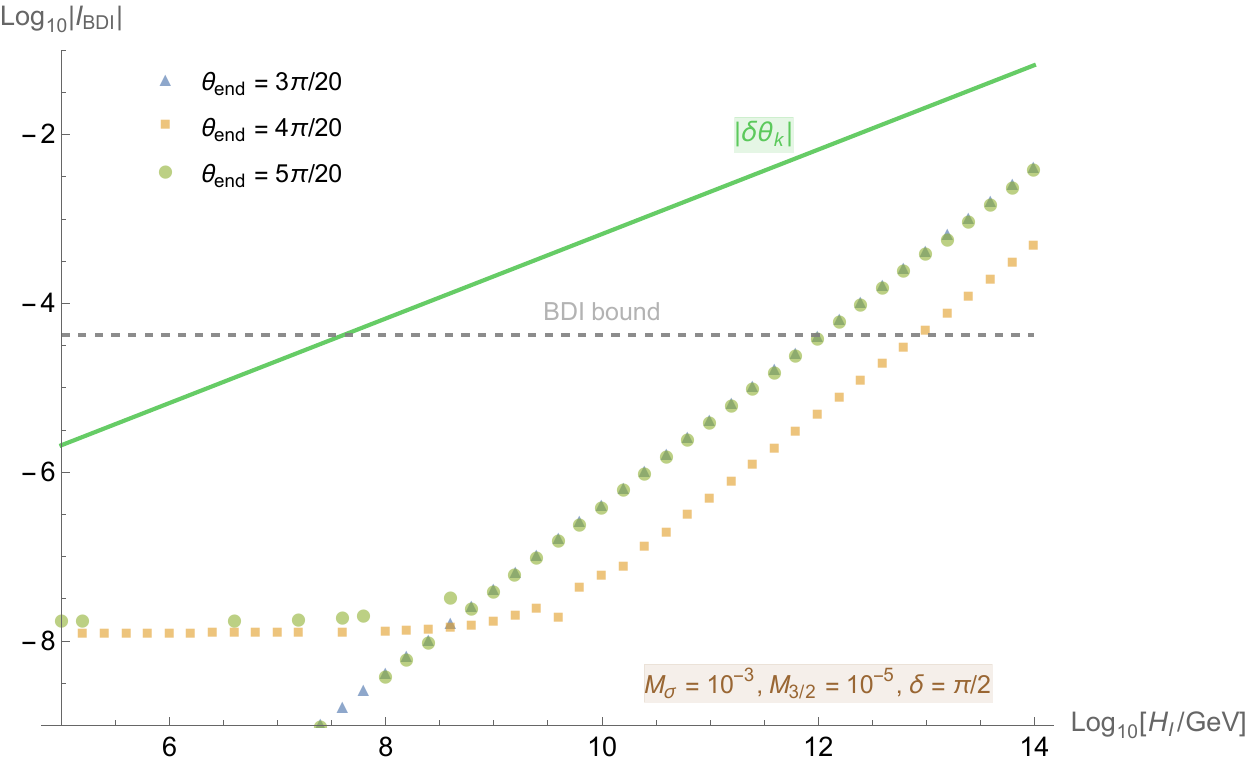}
			\hfill
		\includegraphics[width=7.5 cm]{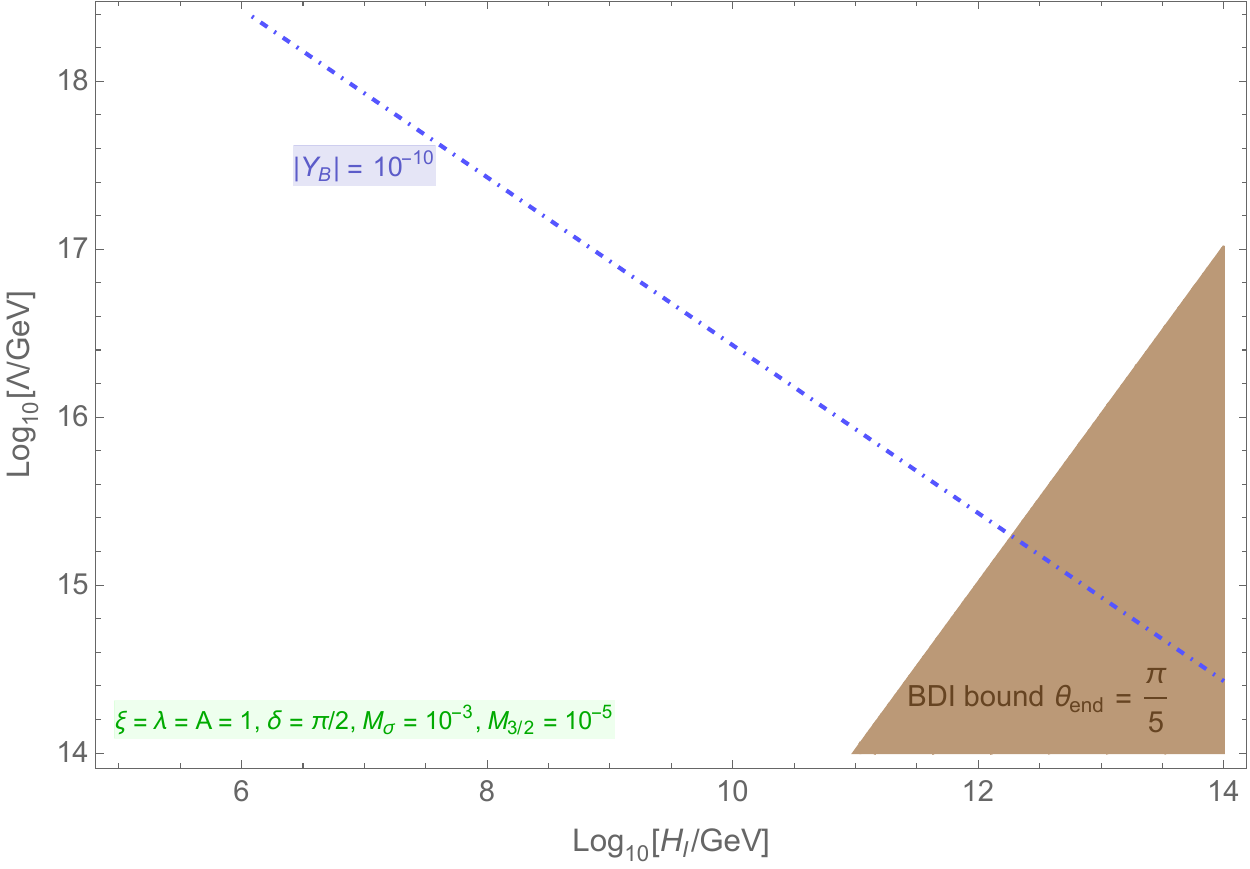}
	\end{center}
	\caption{\label{fig.IBDI_AM}[Left Panel] The baryon density isocurvature (BDI) perturbation with respect to the Hubble parameter of inflation $H_I$ based on the separated universe approach (Section~\ref{Sec_Separated_Universe}) with $c_H = -\xi$, $c_A = 0$ during inflation and $c_H = c_A = 0$ in radiation domination. The dashed line is the upper bound of the BDI perturbation from the generally correlated adiabatic and isocurvature models in \cite{Planck:2018jri}. The solid line for $\vert\delta\theta\vert$ is the initial value of the massless angular perturbation. $n= 4$, $\Lambda = 10^{16}$ GeV, and $\xi = \lambda = A = 1$ are used in the computations of these results. 
	[Right panel] The $Y_B = 10^{-10}$ contour for the initial misalignment angle $\theta_{\rm end} =\pi/5$ in the left panel. The colored regions are excluded by the BDI bound based on hybrid semianalytic background evolution for the separated universe approach.}
\end{figure}

In Figure~\ref{fig.IBDI_AM}, we solve the BDI perturbation with $c_H = -\xi$, $c_A = 0$ during inflation and $c_H = c_A = 0$ in radiation domination via the separated universe approach based on the hybrid semi-analytic background evolution with finite-temperature effects. The initial angular perturbation, $\delta\theta_k$, is just the massless limit ($m_\theta = 0, \nu_\theta = 3/2$) of the standard mode functions \eqref{def_theta_k}. An interesting finding is that the size of the BDI perturbation is sensitive to the initial misalignment angle $\theta_{\rm end}$. In general, the final BDI perturbation is smaller than the initial $\delta\theta_k$ during inflation. 

However, the results with small   in the left panel of Figure~\ref{fig.IBDI_AM} do not indicate that high-scale inflation with $H_I \lesssim 10^{13}$ GeV can pass the current BDI bound. This is because $R_{\rm end}$ given by \eqref{def_Rend_angular_misalignment} depends on $H_I$, and thus both the baryon number density $n_B = R_0^2\dot{\theta}_0$ and the initial angular perturbations $\delta\theta \sim H_I/R_{\rm end}$ vary with $H_I$. In the right panel of Figure~\ref{fig.IBDI_AM}, we scan the allowed parameter space for $H_I$ and $\Lambda$ with respect to the BDI bound from different initial angular misalignment angles $\theta_{\rm end}$. We can see that in the case with $\delta = \pi/2$ and $\theta_{\rm end} = \pi/5$, the contour of $Y_B = 10^{-10}$ implies that $H_I < 10^{12}$ GeV.

\section{Non-canonical kinetic couplings} \label{Sec_SUSY_breaking}
In order to study the effects of primordial features on the isocurvature perturbations, in this section we discuss possible couplings between the inflaton and the AD field. Although kinetic terms are not explicitly specified in the discussion of \cite{Dine:1995kz}, a general Kähler potential is suggested in the construction of the effective potential \eqref{def_UFD} from the finite-energy supersymmetry breaking during inflation. In fact, a non-canonical Kähler potential, the effective field theory after integrated out heavy mediators, loop threshold corrections, and normalization of moduli from geometric compactifications of ultraviolet physics can all give rise to non-canonical kinetic couplings between the inflaton and the AD field. 

To see this, let us focus on the lowest components of the superfields $\Phi_I \equiv I$, $\Phi_\sigma \equiv \sigma$ for simplicity. We denote $I \equiv I_R e^{i \tilde{\phi} /f_I}/\sqrt{2}$ as a complex scalar which includes the inflaton $\tilde{\phi}$ as a  pseudoscalar component up to a rescaling of the field values. (Of course, it is just one of the possible ways for us to assign the inflaton in $I$.) $I_R$ is the radial mode of $I$ and it has a mass scale much higher than that of inflation. Therefore we treat $I_R$ as a non-dynamical constant. Let us first consider a small mixing in the Kähler potential suppressed by the Planck scale as:
\begin{align}
	K = I\bar{I} + \sigma \bar{\sigma} + \frac{c_I}{M_P^2}  I\bar{I}  \sigma \bar{\sigma}, 
\end{align}  
where $\sigma = \tilde{R}e^{i\theta}/\sqrt{2}$ is the AD field for baryogenesis with a rescaled radial mode $\tilde{R}$. The corresponding Kähler metric reads
\begin{align}
	g_{I\bar{I}} = 1 + \frac{c_I}{M_P^2} \vert\sigma\vert^2, \quad	g_{\sigma\bar{\sigma}} = 1 + \frac{c_I}{M_P^2} \vert I\vert^2, \quad
	g_{I\bar{\sigma}} = \bar{g}_{\sigma \bar{I}} = \frac{c_I}{M_P^2} \bar{I} \sigma.
\end{align}
A canonical Kähler metric can be reproduced in the limit of $c_I = 0$.
With the rescaling of $\phi = I_R \tilde{\phi}/f_I$, we can obtain the kinetic terms of the bosonic part of the Lagrangian as 
\begin{align}
    \label{eq:Kählerexpansion}
	\mathcal{L}_{\rm kin} 
    \supset \frac{1}{2}\left[1 + c_I\frac{\vert\sigma\vert^2}{M_P^2}\right] (\partial\phi)^2 + \left[1+c_I\frac{I_R^2}{2M_P^2}\right]\vert\partial\sigma\vert^2 + i c_I\frac{ I_R}{2 M_P^2}\partial \phi \left(\sigma \partial \bar{\sigma}- \bar{\sigma}\partial \sigma\right).
\end{align}
This gives rise to non-canonical kinetic couplings of $\sigma$ with the inflaton. Both kinetic couplings in \eqref{eq:Kählerexpansion} are investigated in axion dark matter models \cite{Chen:2023txq}. 
In such a simple construction, the Hubble-induced mass \eqref{def_cH} for the AD field and the non-canonical kinetic terms are all sourced by the quartic $c_I$ coupling. As a result, the effective operators of those kinetic couplings during inflation set by the cutoff scale $\Lambda$ is controlled by the radial mass \eqref{def_m_R}. 

Next, let us explore the scenario where the non-canonical kinetic terms arise from an origin independent of the Hubble-induced mass \eqref{def_cH} for the AD field. It is interesting to consider the multi-field realization of inflation in supergravity led by the Kähler potential:
\begin{align}
	K = \frac{1}{2}\left(I+\bar{I}\right)^2 + S\bar{S} + \sigma \bar{\sigma} + \frac{c_I}{2M_P^2} \left(I+\bar{I}\right)^2   \sigma \bar{\sigma}+ \frac{c_S}{M_P^2}  S\bar{S}  \sigma \bar{\sigma},
\end{align}
where the inflaton sector $I \rightarrow I + i C$ is protected by the shift symmetry \cite{Brax:2005jv,Antusch:2009ef}, and $S$ is a stabilizer field that is the heavy direction supporting the slow-roll trajectory of inflation. In this scenario, the superpotential $W = W_0(S, I) + W_{\rm AD}(S,I,\sigma)$ with $W_{\rm AD} \ll W_0$ and we adopt a simple factorization $W_0 = S f(I)$.

Along the slow-roll trajectory, the background values of these superfields are given by $S = 0$, $I = i \phi/\sqrt{2}$, and $\sigma = \sigma_0$, where $\phi$ is the slow-roll inflaton. This background gives $W_0 = D_I W_0 = 0$. In the small-field limit with $\vert I\vert/M_P\ll 1$, the supergravity factor $e^{K/M_P^2} \approx 1$ and the background F-term potential $V_{F0} \approx g^{S\bar{S}}\vert D_SW_0\vert^2 = \vert f(I)\vert^2 = 3M_P^2H^2$. One can check that all contributions from the $c_I$ coupling to the Hubble-induced mass listed in \eqref{Hubble_mass_cI} vanish at the leading order. Instead, the Hubble-induced mass is led by the stabilizer F-term potential via $g^{S\bar{S}}\vert D_S W_0\vert^2 \supset -c_S\vert f(I)\vert^2\vert\sigma\vert^2/M_P^2 = -3c_S H^2\vert\sigma\vert^2$. Note that this implies that the radial mass \eqref{def_m_R} is mainly governed by $\xi \approx 3c_S$, which is independent of the coupling with the inflaton sector $c_I$.

Meanwhile, non-canonical kinetic terms are only generated by 
\begin{align}
g^{I\bar{I}}\vert\partial I\vert^2 = \frac{1}{2} \left( 1- c_I \frac{\vert\sigma\vert^2}{M_P^2} \right) (\partial\phi)^2,
\end{align}
with all other terms vanish along the slow-roll trajectory. This non-canonical kinetic term is sourced by the $c_I$ coupling and its contribution to the Hubble-induced mass of the AD field is suppressed by the slow-roll parameter. Overall, as a low-energy effective operator during inflation, we denote the non-canonical kinetic interaction between the inflaton and the AD field $\sigma$ as:
\begin{align}\label{def_phi_sigma_couplings}
	\mathcal{L}_{\phi\sigma} \supset - c_6 \frac{\vert\sigma\vert^2}{\Lambda^2} \left(\partial\phi\right)^2, \qquad \frac{c_6}{\Lambda^2} = \frac{c_I}{M_P^2} .
\end{align} 
Therefore, it is possible to have $c_I \sim M_P^2/\Lambda^2 \gg 1$ with $c_6 \sim \mathcal{O}(1)$ as long as $c_I \vert\sigma\vert^2/M_P^2\sim \vert\sigma\vert^2/\Lambda^2 \ll 1$. This converts to the constraint on the VEV $\vert\sigma\vert = R_{\rm end}/\sqrt{2} \lesssim \Lambda$ for us to explore the large $c_I$ limit during inflation in the next section. It is remarkable that in the limit with $c_I \sim M_P^2/\Lambda^2 \gg 1$, we can still have the regular radial mass controlled by $\xi \approx 3c_S$ independently. This is the important difference from the scenario given in Appendix~\ref{Appendix_Hubble_coefficients} in which $\xi \approx 3c_I$ so that a large kinetic coupling will result in a large radial mass.

During inflation where $\dot{\phi_0} \equiv\partial_t \phi$ obtains a large background value, $\mathcal{L}_{\phi\sigma}$ introduces an effective mass to the radial mode of $\sigma$,
	\begin{align}\label{def_mc6}
		m_{c_6}^2 = -c_6 \frac{\dot{\phi}_0^2}{\Lambda^2} 
		\approx -2c_6 \epsilon_{H0} \frac{M_P^2}{\Lambda^2}  H_I^2,
	\end{align}
where $\phi_0$ is the background inflaton value and the slow-roll parameter $\epsilon_{H 0} = \dot{\phi}_0^2/(2M_P^2H_I^2)$. For $c_6 \sim \mathcal{O}(1)$, we can see that $m_{c_6} \ll H_I$ in all cases even if $\Lambda \simeq M_P$ so that the coupling induced masses can be small corrections to the flat-direction potential $U_{\rm FD}$. We also restrict $m_{c_6}/H \ll \xi$ so that our background analysis in Section~\ref{Sec. final_YB} for the final baryon asymmetry will not be affected by the couplings with inflaton.

In our following study we focus on the dynamics of the inflaton $\phi$ and the AD field $\sigma = \frac{R}{\sqrt{2}}e^{i\theta}$ during the epoch of inflation ($N \leq N_{\rm end}$). The effective Lagrangian is given by
\begin{align}
	{\cal L} &= -\frac{1}{2} \left( \partial \phi \right)^2  - V(\phi) -\vert\partial
    \sigma\vert^2 - U_{\rm FD}(\sigma) - c_6 \frac{\vert\sigma\vert^2}{\Lambda^2} \left(\partial\phi\right)^2, 
    \\ \label{Lagrangian_phi_R_theta_kinetic_coupling}
    &\approx-\frac{1}{2}\left(1+c_6\frac{R^2}{\Lambda^2}\right)\left( \partial \phi \right)^2 
    \\\nonumber &\qquad\qquad-\frac{1}{2}\left( \partial R \right)^2 - \frac{1}{2}m_R^2  (R-R_{\rm end})^2
    -\frac{R^2}{2}\left( \partial \theta \right)^2
    - \frac{R^2}{2} m_\theta^2  (\theta-\theta_{\rm end})^2 
    ,
\end{align}
where we concentrate on the potential \eqref{def_UFD_polar} around the particular choices of VEVs $R_{\rm end}$ and $\theta_{\rm end}$ given by \eqref{def_Rend_theta_end} that can generate the correct amount of baryon asymmetry today. The radial and angular masses defined at those VEVs are found as \eqref{def_m_R} and \eqref{def_m_theta}.
The exact form of the inflaton potential $V(\phi)$ does not play a significant role in our purpose. However, to generate primordial feature signals, one may require $V(\phi)$ to contain a small component which breaks the scale-invariance of inflation explicitly. The properties of $V(\phi)$ will be discussed with details in Section~\ref{Sec_primordial_features}. 

Let us specify the field notations in our following study in the presence of non-canonical kinetic couplings in the theory. We define the decomposition:
\begin{align}
    \phi &= \phi_b(t) +\delta\phi(t,\vec{x}), &\phi_b(t) &= \phi_0(t) + \phi_1(t), \\
    R &= R_b(t) + \delta R(t,\vec{x}), &R_b(t) &= R_{\rm end} + R_1(t),\label{Eq:R_decomp}\\
     \theta &= \theta_b(t) + \delta \theta (t,\vec{x}), &\theta_b(t) &= \theta_{\rm end} + \theta_1(t),
\end{align}
where $\phi_b$, $R_b$, $\theta_b$ are homogeneous background components and $\delta\phi$, $\delta R$, $\delta\theta$ are quantum fluctuations. The background components are further decomposed into their VEVs during inflation in absence of any primordial features, $\phi_0(t)$, $R_{\rm end}$ $\theta_{\rm end}$, and the modulations induced by primordial features in the inflaton potential, $\phi_1(t)$, $R_1(t)$, $\theta_1(t)$. We will specify  these modulations with explicit examples of primordial features in Section~\ref{Sec_primordial_features}.

A brief comment on our model assumptions is in order. Effective couplings between the inflaton and the AD scalar can arise in several possible ways as long as they are no longer prohibited by the symmetries of UV physics. In principle, these couplings can play an important role during or after inflation and lead to drastic modification of the AD field dynamics. 
In this work, we assume that these inflaton–AD couplings do not affect the post-inflationary evolution or the baryon asymmetry computed above. This is primarily ensured by assuming instantaneous reheating at the end of inflation, although we do not specify its details.

\section{Baryogenesis with primordial features}\label{Sec_primordial_features}

We have so far considered the vanilla single-field slow-roll model because it agrees well with the recent observations \cite{Planck:2018jri}.
In this section, we consider a class of inflation models beyond the simplest one, namely, models with primordial features. As we briefly reviewed in Sec.~\ref{Sec. introduction},
primordial features could arise naturally in a generic inflationary landscape. Observationally, some $2$ -- $3\sigma$ CMB anomalies might hint the existence of such features that could be realized through a variety of inflation models, see e.g.~\cite{Braglia:2022ftm} for a summary. Such features modify the dynamics of $\phi$, leading to scale-dependent corrections to the adiabatic power spectrum. The feature can also affect the background evolution or quantum fluctuations of isocurvature fields during inflation, and therefore be used as probes of properties of those fields.
In this section, our aim is to study the imprints of such features on the background evolution, as well as on field perturbations of the inflaton field and the complex scalar AD field, $\sigma$, during inflation and, eventually, on observables such as the curvature power spectrum and the BDI power spectrum.

\subsection{Inflationary background with a step in potential}
\label{Sec:step_background}
In this paper, we consider a simple example of primordial feature model in which the slow-roll potential is modified by a sudden change, $\Delta V$, in the potential energy of the inflaton sector,
\begin{align}
	V(\phi) = V_0(\phi) + \Delta V(\phi), 
\end{align}
where $V_0$ dominates the energy density of the inflationary universe and $\vert\Delta V\vert/V_0 \ll 1$ is constrained by CMB observations. $\phi$ denotes the slow-roll scalar along the adiabatic trajectory of inflation, and its perturbation $\delta\phi$ is fully responsible for the generation of the curvature perturbation.

The feature term introduces some modulations to the background dynamics as
\begin{align}\label{def_modulation_ansatz}
	\phi_b(t) = \phi_0(t) + \phi_1(t), \qquad H(t) = H_I + \Delta H(t),
\end{align}
with $\vert\phi_1/\phi_0\vert \ll 1$ and $\vert\Delta H/H_I\vert \ll 1$. 
Such a feature also modifies the background Hubble parameter through the background Friedman equation of \eqref{Lagrangian_phi_R_theta_kinetic_coupling}: 
\begin{align}
    3M_P^2 H^2 &= \left(1+c_6\frac{R_b^2}{\Lambda^2}\right)\frac{\dot{\phi}_b^2}{2}+V_0 + \Delta V + \cdots,\\
    -2M_P^2\dot{H} &= \left(1+c_6\frac{R_b^2}{\Lambda^2}\right) \dot{\phi}_b^2+\cdots,
\end{align}
where all contributions from the AD field are negligible.
To stay closely in the background of the standard single field inflation, we require $c_6R_b^2/\Lambda^2 \approx c_6R_{\rm end}^2/\Lambda^2 \ll 1$. 

Taking the ansatz \eqref{def_modulation_ansatz} into the background energy density, at the leading order we find the $\Delta H$ induced by the modulated inflaton density as
\begin{align}\label{def_Delta_H}
	2\frac{\Delta H}{H_I} = \left(1+c_6\frac{R_b^2}{\Lambda^2}\right)\frac{\dot{\phi}_0\dot{\phi}_1}{V_0} + \frac{\Delta V}{V_0},
\end{align} 
where $V_0 = (3-\epsilon_H)M_P^2H_I^2$. We denote
\begin{align}\label{def_epsilon_H0}
	\epsilon_{H} \equiv - \frac{\dot{H}}{H^2} = \left(1+c_6\frac{R_b^2}{\Lambda^2}\right)\frac{\dot{\phi}_b^2}{2M_P^2H^2} +\cdots, 
\end{align}
where the contributions from the AD field are higher-order corrections.
We denote $\epsilon_{H 0} = \dot{\phi}_0^2/(2M_P^2H_I^2)$ as the (first) slow-roll parameter in the absence of primordial features. $\epsilon_{H 0} < 0.0063$  \cite{Planck:2018jri} can be treated as a constant throughout the relevant scales in our discussion.

For a AD scalar this results in a modulated radial minimum \eqref{def_Rmin} during inflation in terms of $\Delta H$ as
\begin{align}\label{radial_modulation}
	R_{\rm min}(H_I + \Delta H) 
    \equiv R_{\rm end} + \Delta R_{\rm end}(t)
    \approx R_{\rm end} + \frac{R_{\rm end}}{n-2} \frac{\Delta H}{H_I},
\end{align} 
where $R_{\rm end}$ is the VEV at the radial minimum during inflation defined in \eqref{def_Rend_theta_end} with respect to $H_I$, and $\Delta R_{\rm end}$ is the modulation induced by $\Delta H$. Note that $\Delta R_{\rm end}$ is a component of $R_1$, as defined in \eqref{Eq:R_decomp}, which is the source term of the radial oscillation to be considered in Section~\ref{Sec:bkgd_osci_radial_mode}. 

To provide some simple estimates and obtain some analytical understanding of the models, let us first specify the sharp feature in the potential as an infinitely-sharp step-function, modeled by the Heaviside function,
\begin{align}\label{def_step_feature}
	\Delta V(\phi) = B \; V_0 \Theta(\phi - \phi_\ast),
\end{align}
where $B < 0$ ($B > 0$) corresponds to a downward (upward) step in the inflaton potential. $\vert B \vert \ll 1$ is generally required by CMB constraints \cite{Braglia:2022ftm} and $V_0 \approx 3M_P^2 H_I^2$ parametrizes the background energy density of inflation. The Heaviside $\Theta$ function $\Theta(\phi - \phi_\ast) = 1$ for $\phi > \phi_\ast$ and $\Theta(\phi - \phi_\ast)  = 0$ otherwise.  We will study the more realistic smoothed steps later in the paper.

With a step feature given by \eqref{def_step_feature}, the background modulation \eqref{def_modulation_ansatz} at leading order satisfies
\begin{align}\label{eom_phi1_t}
	\ddot{\phi}_1 + 3H_I\dot{\phi}_1 +3\Delta H \dot{\phi}_0 + B\; V_0\; \delta(\phi_b - \phi_\ast) = 0.
\end{align}
The third term is small when comparing with others, so we can neglect its contribution in this equation. This will be checked once we have explicit solutions of $\dot{\phi}_1$ and $\Delta H$. 

In terms of the $e$-folding number $N \equiv \ln a$, we define $\partial_N \phi \equiv \phi^{\prime}$. The modulated inflaton motion has an analytic solution given by \cite{Chen:2023txq}:
\begin{align}\label{sol_phi1_step}
	\phi_1(N) &= -B \frac{V_0}{3H_I\dot{\phi}_0}\left(1+c_6\frac{R_{\rm end}^2}{\Lambda^2}\right)^{-1}\left[1- e^{-3(N-N_\ast)}\right]  \Theta(N-N_\ast), \nonumber\\
	\phi_1^\prime (N) &=  -B \frac{V_0}{H_I\dot{\phi}_0}\left(1+c_6\frac{R_{\rm end}^2}{\Lambda^2}\right)^{-1} e^{-3(N-N_\ast)}  \Theta(N-N_\ast),
\end{align}
where $\phi_\ast \equiv \phi_b(N_\ast)$. 
Using \eqref{sol_phi1_step} and \eqref{def_step_feature} in \eqref{def_Delta_H}, we can obtain the change in the Hubble parameter induced by the step feature as
\begin{align}\label{Delta_H_step}
	\frac{\Delta H}{H_I} = \frac{B}{2}\left(1-e^{-3(N - N_\ast)}\right) \Theta(N - N_\ast).
\end{align}
This means that, in terms of the physical time, the Hubble parameter reads
\begin{align}\label{Hubble_step}
	H &= H_I, && t\leq t_\ast, \\\nonumber
	&= H_I + \frac{B}{2}H_I \left(1-\frac{a_\ast^3}{a^3}\right), && t > t_\ast,
\end{align}
where the step feature occurs at $t = t_\ast$. In the second phase, one can solve the scale factor to obtain 
\begin{align}\label{a_time_varying_Hubble}
	\frac{a(t)}{a_\ast} = \left[\frac{1}{1+B/2} \left(\frac{B}{2}+ e^{3(1+B/2)H_I(t-t_\ast)}\right)\right]^{1/3}.
\end{align}
Taking the above solutions from the step feature, we get
\begin{align}
    \epsilon_H = -\frac{3}{2}Be^{-3(N-N_\ast)}\Theta(N-N_\ast), 
\end{align}
up to the leading order in the step size $B$.

For the step feature to be small perturbation, we need $B \lesssim \epsilon_{H0}\ll 1$.

We also check if the third term in \eqref{eom_phi1_t} is neglected properly. We can see that $\Delta H/H_I \sim B$ from \eqref{Delta_H_step}, and therefore the condition for neglecting the $\Delta H$ correction in \eqref{eom_phi1_t} is $B \dot{\phi}_0 \ll \dot{\phi}_1$. Given that $\dot{\phi}_1 \sim B V_0 /\dot{\phi}_0 \sim 3B M_P^2 H_I^2/\dot{\phi}_0$ from \eqref{sol_phi1_step}, and that the adiabatic spectral amplitude $P_{\mathcal{RR}} \approx H_I^4/(4\pi^2\dot{\phi}_0^2) \approx 2.2\times 10^{-9}$, the condition $B \dot{\phi}_0 \ll \dot{\phi}_1$ can be translated into $H_I \ll 2\pi \sqrt{3P_{\mathcal{RR}} }M_P \approx 1.2 \times 10^{15}$ GeV. This condition is clearly satisfied due to the upper bound $H_I \leq 3\times10^{13}$ from the tensor-to-scalar ratio bound \cite{BICEP:2021xfz}.

\subsection{Background oscillation of radial mode}
\label{Sec:bkgd_osci_radial_mode}

The changes in the inflationary background induced by sharp features can trigger oscillations of the heavy radial mode $R$ in the effective potential $U_{\rm FD}$. In the flat-direction model, there are two possible sources that can lead to the radial oscillations during inflation, which we investigate in this subsection.

For numerical examinations, we solve the background equations for the radial and angular modes of the AD field given by
\begin{align}
	\ddot{R}_b + 3H\dot{R}_b + \partial_R U_{\rm FD}+ \left(m_{c_6}^2 - \dot{\theta}_b^2\right) R_0 &= 0, \\
	\ddot{\theta}_b + \left(3H + 2\frac{\dot{R}_b}{R_b}\right) \dot{\theta}_b + \frac{1}{R_b^2} \partial_\theta U_{\rm FD} & =0,
\end{align}
where $m_{c_6}$ is the effective mass introduced by the kinetic coupling defined as \eqref{def_mc6}. We refer to the oscillation of $R_b$ driven by the feature-induced modulation of the radial VEV $\Delta R_{\rm end}$ (given by \eqref{radial_modulation}) as the ``gravitational'' effect, which can be realized without the presence of the kinetic coupling ($c_6 = m_{c_6} = 0$). This is to be distinguished from the oscillation of $R_b$ driven by the $m_{c_6}$ term, which we referred as the ``kinetic coupling'' effect. Note that in both cases, one can check that $\theta_0 = \theta_{\rm end}$ keeps a constant value (namely $\theta_1 =0$) throughout inflation. This is because $\partial_\theta U_{\rm FD} = 0$ at $\theta_b = \theta_{\rm end}$ and the background radial motion only contribute to the $\theta_b$ equation  as additional friction terms.

These cases represent two different classes of examples in which a massive field can be excited and start oscillating classically due to a sharp feature in the inflaton sector. In the first case, although the massive field and the inflaton are not coupled non-gravitationally, the VEV of the massive field (which in this example is the radial field $R$) depends on the Hubble parameter $H$. A sharp feature in the inflaton sector gravitationally induces a sharp change in the background value of $H$, abruptly shifting the VEV of the massive field and triggering its oscillation around the new VEV. In the second case, a sudden change in the inflaton velocity, through a generic type of direct coupling between the massive field and the inflaton, kicks the massive field off the bottom of the potential and triggers its classical oscillation.\footnote{A third kind of mechanism involves tachyonic falling; see \cite{Chen:2014cwa} for an example in which a direct coupling is also used.} In both cases, the observational signature of the oscillating massive field is imprinted in the curvature perturbation and/or isocurvature perturbation, providing a range of valuable information about the underlying model and the inflationary background \cite{Chen:2011zf, Chen:2014cwa, Quintin:2024boj}. As we will also see in the subsequent analyses, the direct coupling case is much more flexible in generating clock signals with larger amplitudes.

To separate different effects, let us decompose the background radial field as 
\begin{align}\label{def_background_decomposition}
	 R_b(t) = R_{\rm end} + R_{\rm 1g}(t) + R_{\rm 1k}(t) ,
\end{align} 
where $R_{\rm 1g}$ ($R_{\rm 1k}$) denote the gravitational (kinetic) coupling induced background modulations, respectively. During inflation, we can ignore the low-energy terms from hidden sector supersymmetry breaking by taking $m_\sigma = m_{3/2} = 0$. This gives $\theta_{\rm end} = \pi/n$ and 
\begin{align}
    R_{\rm end} = \left(\frac{2^{n/2-1}}{\lambda}\Lambda^{n-3}X_{\rm min}H_I\right)^{\frac{1}{n-2}},\; X_{\rm min} = \left[\frac{\xi}{n-1}+\frac{c_A^2}{4(n-1)^2}\right]^{1/2}+\frac{c_A}{2(n-1)}. 
\end{align}
These are the initial conditions for the AD field sitting at the potential minimum in the presence of the sharp feature with $t = t_\ast$, where $\dot{R}_b = \dot{\theta}_b = 0$.  

\subsubsection{Oscillating massive field induced by gravitational couplings}
Let us first solve the gravitational coupling effect by taking $c_6 = m_{c_6} =0$. In this case, we have $R_b = R_{\rm end} + R_{\rm 1g}$, where $R_{\rm 1g} \sim \mathcal{O}(B)$. We drop the angular equation since $\theta_b =\dot{\theta}_b = 0$ is not dynamically excited. At the leading order in the step size $B$, the radial equation in terms of the $e$-folding number $\partial_t = H\partial_N$ reads
\begin{align}
    R_{\rm 1g}^{\prime\prime} + 3 R_{\rm 1g}^\prime + \Delta\left(\frac{1}{H^2} \partial_RU_{\rm FD}\right)_{R = R_{\rm end}}=0,
\end{align}
where the prime derivative stands for $\partial_N$. $\Delta(\cdot)$ denotes the change of $\cdot$ induced by the feature-induced changes in $R$ and $H$.    

The background modulation induced by the sharp feature only shifts the potential minimum in the $R$ direction. Note that $\partial_RU_{\rm FD}/H^2 = (n-1)XR^2-\xi R + c_A XR\cos(n\theta)$, where $X$ is given by \eqref{def_X}. Given that $\Delta X = (n-2)X R_{\rm 1g}/R_{\rm end} - X \Delta H/H_I$, with $\Delta H$ given by \eqref{def_Delta_H}, the force term in the equation of $R_{\rm 1g}$ is therefore
\begin{align}
    \Delta\left(\frac{1}{H^2}\partial_RU_{\rm FD}\right)_{R = R_{\rm end}}  = M_R^2 R_{\rm 1g} - M_R^2 \Delta R_{\rm end},
\end{align}
where $M_R^2 = m_R^2/H^2 = (2n-4)\xi + (n-2)c_AX_{\rm min}$ is given by \eqref{def_m_R} and $\Delta R_{\rm end} \equiv  \partial_H R_{\rm min}\Delta H$ is the radial modulation generated by $\Delta H$ as given by \eqref{radial_modulation}, which is part of the gravitational effect in $R_{\rm 1g}$. The condition $(n-1)X_{\rm min}^2 -c_AX_{\rm min} = \xi$ for the radial minimum is used in this derivation. As a result, we find that the radial modulation induced by the gravitational coupling satisfies
\begin{align}\label{eom_R1g}
    R_{\rm 1g}^{\prime\prime} +3R_{\rm 1g}^\prime + M_R^2R_{\rm 1g} = M_R^2 \Delta R_{\rm end} = M_R^2\frac{R_{\rm end}}{n-2}\frac{B}{2}\left[1-e^{-3(N-N_\ast)}\right]\Theta(N-N_\ast).
\end{align}
We can further separate $R_{\rm 1g} = \Delta R_{\rm end} + R_{\rm 1g}^{\rm osc}$ for convenience. The solution is obtained as
\begin{align}\label{R1_gravitational}
    \frac{R_{\rm 1g}^{\rm osc}}{R_{\rm end}} = -\frac{3B}{2(n-2)\mu_R} e^{-\frac{3}{2}(N-N_\ast)}\sin\left[\mu_R(N-N_\ast)\right],
\end{align}
where $\mu_R \equiv(M_R^2-9/4)^{1/2} \approx M_R$ if $M_R \gg 1$. 

\begin{figure}
    \centering
    \includegraphics[width=0.47\linewidth]{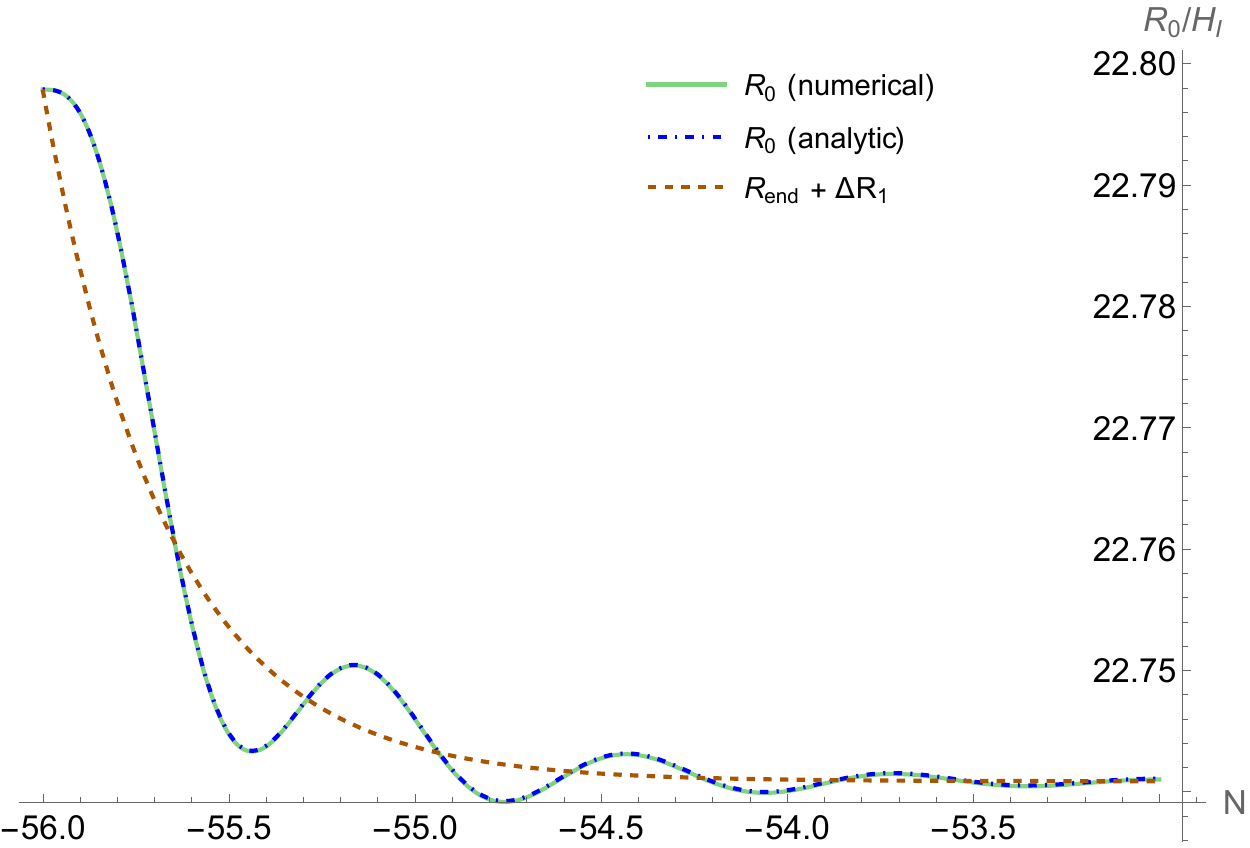}
    \hfill
    \includegraphics[width=0.47\linewidth]{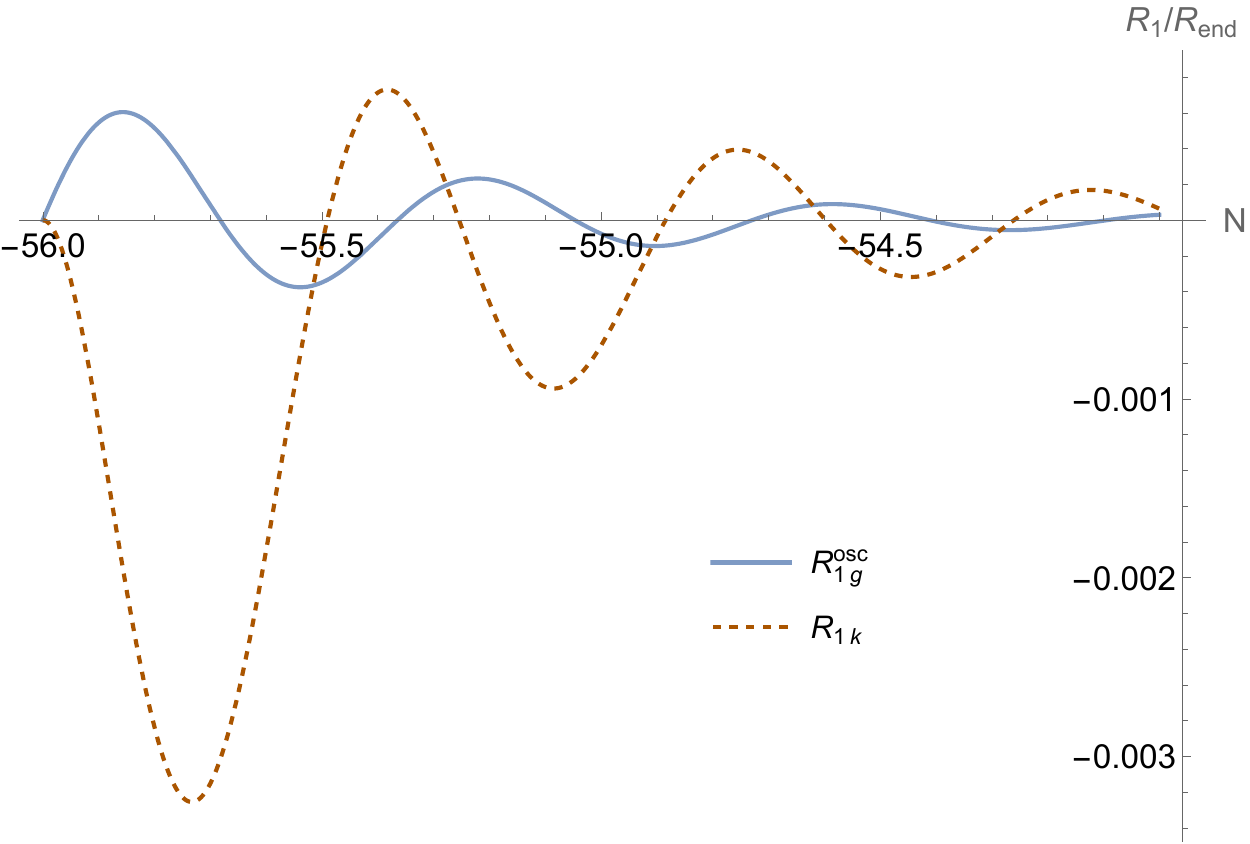}
    \caption{[Left Panel] Radial oscillations on top of the time-varying background induced by the step feature through the gravitational coupling in the flat-direction potential with $n = 4$, $\xi = 100$, $c_A = 0.1$, $B = -0.01$, $N_\ast = -56$. The corresponding radial mass is $M_R=20$. The analytic solution is given by \eqref{R1_gravitational}. 
    [Right Panel] Radial oscillations induced by the step feature through the gravitational coupling ($R_{\rm 1g}^{\rm osc}$) or the kinetic coupling ($R_{\rm 1k}$) with $n = 4$, $\xi = 10$, $B = -0.01$, $N_\ast = -56$. The corresponding radial mass is $M_R=6.3$. We use $M_P/\Lambda = \sqrt{5}$ with $c_6 = 1$ in this plot.
    \label{fig:R1g}
    }
\end{figure}

We compare the analytical solution with the numerical results in Figure~\ref{fig:R1g}. Although the amplitude of the oscillation $R_{\rm 1g}^{\rm osc}/R_{\rm end}\sim B/M_R$ is small compared to the overall shift of the potential minimum $\Delta R_{\rm end}/R_{\rm end}\sim B$ when $M_R\gg 1$, this oscillatory component decays slower than the non-oscillatory component. As we will see in Section~\ref{Sec_observation_signals}, this oscillation can result in the so-called clock signal \cite{Chen:2011zf,Chen:2014cwa} in the isocurvature power spectrum via the time derivative friction term $\sim R_b^\prime/R_b$ in the equation of motion for the angular mode perturbations.
On the other hand, due to the lack of a direct coupling between the radial field and the inflaton, the clock signal in the curvature perturbation power spectrum is only produced through the feedback of background fields. Consequently, we expect the clock signal in the curvature spectrum to be relatively much smaller than, and swamped by, the sharp feature signal, in contrast to the next scenario we will consider.

\subsubsection{Oscillating massive field induced by direct and kinetic couplings}
\label{Sec:direct_couping_bkgd_oscillations}
The AD field can also couple directly to the inflaton through terms such as those in \eqref{def_phi_sigma_couplings}.
Let us focus on the case with a step feature \eqref{def_step_feature}. The solution \eqref{sol_phi1_step} remains the leading order approximation. 

We now solve the modulated radial motion $R_{\rm 1k}$ induced by the kinetic coupling \eqref{def_phi_sigma_couplings}. In the leading order in step size $B$, the radial motion in terms of the $e$-folding number $N = \ln a$ with $\partial_N R_1 \equiv R_1^\prime$ satisfies
\begin{align}\label{eom_R1}
	R_{\rm 1k}^{\prime\prime} + 3R_{\rm 1k}^\prime + M_R^2 R_{\rm 1k} = - \Delta M_{c_6}^2  R_{\rm end} - M_{c_6}^2 \Delta R_{\rm end},
\end{align}
where $M_R^2 \equiv m_R^2/H^2$ given by \eqref{def_m_R} is a constant as long as $m_\sigma/H \ll \xi$. 
Similarly we also define the dimensionless $ M_{c_6}^2 \equiv m_{c_6}^2/H^2$ and $\Delta M_{c_6}^2 \equiv \Delta m_{c_6}^2/H^2$, where $\Delta m_{c_6}^2$ denotes the background modulation in the coupling induced mass $m_{c_6}^2$ led by $\phi_1$. Note that the source term $M_R^2\Delta R_{\rm end}$ for the gravitational coupling induced oscillation is not included on the right-hand side of \eqref{eom_R1}. 

Recalling that from \eqref{def_phi_sigma_couplings}, we have  
\begin{align}
    M_{c_6}^2 = -c_6  \frac{\dot{\phi}_b^2}{\Lambda^2 H^2},
\end{align}
where the modulated masses in the linear order in $B$ with the analytic solutions \eqref{sol_phi1_step} are given by
\begin{align}
	\Delta M_{c_6}^2 &= 6B c_6  \frac{M_P^2}{\Lambda^2}\left(1+c_6\frac{R_{\rm end}^2}{\Lambda^2}\right)^{-1} e^{-3(N-N_\ast)} \Theta(N-N_\ast).
\end{align}
Since $\Delta R_{\rm end}/R_{\rm end} \sim \Delta H/H_I \sim B$, one can observe that $M_{c_6}^2 \Delta R_{\rm end}/R_{\rm end} \sim \epsilon_{H 0}  c_6 B  M_P^2/\Lambda^2$ and $  \Delta M_{c_6}^2\sim 6c_6B M_P^2/\Lambda^2$ and thus $ M_{c_6}^2 \Delta R_{\rm end} \ll \Delta M_{c_6}^2 R_{\rm end}$. Therefore, we can solve $R_{\rm 1k}$ by neglecting the second term on the right-hand side of \eqref{eom_R1}.

The solution of \eqref{eom_R1} reads
\begin{align}\label{sol_R1_step}
	\frac{R_{\rm 1k}(N)}{R_{\rm end}}  &= 6c_6B \frac{M_P^2}{\Lambda^2M_R^2}\left(1+c_6\frac{R_{\rm end}^2}{\Lambda^2}\right)^{-1}	\Theta(N-N_\ast)\times \\\nonumber
	&\quad\left\{e^{-3(N-N_\ast)} 
	- e^{-\frac{3}{2}(N-N_\ast)} \left[\cos(\mu_R(N-N_\ast)) -\frac{3}{2\mu_R} \sin(\mu_R(N-N_\ast)) \right]\right\},
\end{align}
where $\mu_R\equiv (M_R^2 - 9/4)^{1/2}$. Note that $c_6R_{\rm end}^2/\Lambda^2\ll 1$ is required by the sub-dominance of the kinetic coupling.
As in the previous case, we consider the $M_R \gg 1$ case in which the radial VEV is stable during inflation.
In this case, the primordial feature $\Delta V$ in the inflaton potential induces the background oscillation of the complex AD scalar. As we will show, this oscillation can produce clock signals in both the angular mode perturbation (i.e.~isocurvature perturbation) and the inflaton perturbation (i.e.~curvature perturbation).

\bigbreak\noindent
$\bullet$ {\it Summary.}
To summarize the results of Section~\ref{Sec:bkgd_osci_radial_mode}, a sharp feature in the inflaton sector can excite classical oscillations of the radial component of the AD field. These processes can be mediated by a direct coupling and/or by gravitational couplings. In the large radial mass $M_R \gg 1$ and infinitely sharp step limits, the oscillations in the gravitational and direct couplings case are estimated by  \eqref{R1_gravitational} and \eqref{sol_R1_step}, respectively, and their fractional amplitudes are summarized below:
\begin{align}
    \frac{R_{\rm 1g}}{R_{\rm end}} &\sim -\frac{B}{(n-2)M_R}\left(\frac{a}{a_\ast}\right)^{-3/2}\sin\left(M_R\ln\frac{a}{a_\ast}\right), \\
    \frac{R_{\rm 1k}}{R_{\rm end}} &\sim 6c_6B\frac{ M_P^2}{\Lambda^2M_R^2} \left(\frac{a}{a_\ast}\right)^{-3/2}\cos\left(M_R\ln\frac{a}{a_\ast}\right),
\end{align}
where we can also see that they have a $\pi/2$ phase difference, as shown in Figure~\ref{fig:R1g}. 
Note that, since $\vert B\vert \ll 1$ and $M_R > 1$, the relative oscillation amplitude for the gravitational case is always small, ${R_{\rm 1g}}/{R_{\rm end}} \ll 1 $.
On the other hand, it is possible to have $c_6M_P^2/\Lambda^2 \gg M_R$ while keeping $  c_6\vert B\vert M_P^2  < \Lambda^2M_R^2$. So, the kinetic-coupling-induced radial oscillation can have a much larger relative amplitude even if $\vert B \vert \ll 1$. Examples are shown in the right panel of Figure~\ref{fig:R1g} with $c_6 = 1$, $M_P^2/\Lambda^2 = 5$ and $M_R=6.3$. 

These estimates are based on the infinitely sharp step potential \eqref{def_step_feature}. More realistic smoothed step potentials are studied in Appendix \ref{Append_Gaussian_step}. As shown there, the smoothed step examples used in this paper do not alter, at the background level, the parametric order-of-magnitude estimates obtained using the infinitely sharp step approximation.

\subsection{Observational signals in isocurvature and curvature power spectrum}\label{Sec_observation_signals}
In this section, we compute the observational signals in these models by studying the angular perturbation $\delta\theta$ in the AD field. Again, we focus on the scenario investigated in Section~\ref{Section_BDI}, where the radial perturbation decays well before inflation ends and the initial angular perturbation at the end of inflation acts as the main source of the final BDI perturbation. 

During inflation,  the angular perturbation is governed by
\begin{align}
	\delta\mathcal{L}_\theta^{(2)} = \frac{a^3}{2} R_b^2 \left[(\delta\dot{\theta})^2 - \frac{1}{a^2}(\partial_i\delta\theta)^2\right] -  \frac{a^3}{2} R_b^2 m_\theta^2 \delta\theta^2,
\end{align}
where $m_\theta$ is given by \eqref{def_m_theta}. The Hamiltonian density, $\mathcal{H}_\theta = \delta\pi_\theta \delta\dot{\theta} - \delta\mathcal{L}_\theta^{(2)}$, where $\delta\pi_\theta = \partial\delta\mathcal{L}/\partial(\delta\dot{\theta})$, is
\begin{align}
	\mathcal{H}_\theta = \frac{a^3}{2} R_b^2 \left[(\delta\dot{\theta}_I)^2 + \frac{1}{a^2}(\partial_i\delta\theta_I)^2\right] -  \frac{a^3}{2} R_b^2 m_\theta^2 \delta\theta_I^2,
\end{align}  
where $\delta\theta_I$ is the angular perturbation in the interaction picture and $\delta\dot{\theta}_I\equiv \partial\mathcal{H}_0/\partial\delta\pi_\theta$. 

If the VEV of the radial field is a constant, $R_b = R_{\rm end}$, then $\mathcal{H}_\theta = \mathcal{H}_{\theta 0}$ and the mode functions of $\delta\theta_I$ takes the form of \eqref{def_theta_k}. This is the leading order behavior.
In the presence of primordial features, the radial field receives a time-dependent background correction, $R_b = R_{\rm end} + R_1(t)$, and so is the Hamiltonian density, $\mathcal{H}_\theta =  \mathcal{H}_{\theta 0} + \Delta\mathcal{H}_\theta$,
\begin{align}
	\Delta\mathcal{H}_\theta =  a^3 R_{\rm end} R_1 \left[(\delta\dot{\theta}_I)^2 + \frac{1}{a^2}(\partial_i\delta\theta_I)^2\right] -  a^3R_{\rm end}R_1 m_\theta^2 \delta\theta_I^2.
\end{align}
Following the in-in formalism, the leading correction to the two-point correlation function led by $\Delta\mathcal{H}_\theta$ can be computed as
\begin{align}\label{in_in_correlator_2pt}
	\left\langle \delta\theta^2\right\rangle = \left\langle 0 \vert \delta\theta_I^2 \vert 0 \right\rangle 
	+ 2i \int dt \left\langle 0 \vert 	\Delta\mathcal{H}_\theta(t) \delta\theta_I^2 \vert 0 \right\rangle.
\end{align}
 We are interested in the value of the correlator evaluated at the end of inflation with $N = N_{\rm end}$. 
The power spectrum associated with \eqref{in_in_correlator_2pt} can be separated as
\begin{align}
	P_\theta = P_\theta^{(0)} + \Delta P_\theta ,
\end{align}
where the leading order $P_\theta^{(0)}$ obtained from $\langle 0\vert\delta\theta_I^2\vert 0\rangle$ is the same as that computed in \eqref{def_P_theta_0}. The isocurvature (i.e.~BDI) power spectrum is $P_{\rm BDI}\approx \chi^2P_\theta$, where $\chi$ is a numerical constant obtained in Section~\ref{Section_BDI}. (We will also loosely refer to $P_\theta$ as the isocurvature power spectrum.) 

We now compute the corrections to the isocurvature power spectrum induced by the correction terms from both the Hubble parameter, $\Delta H$, and the radial field, $R_1$.
The change in the Hubble parameter drives an evolution of the radial minimum, as given in \eqref{radial_modulation}, while the angular minima remain unchanged. This gives $\theta_b = \theta_{\rm end}$ with $\dot{\theta}_b = 0$ so that the equation of motion for the angular perturbation \eqref{eom_dtheta} during inflation remains decoupled from the radial perturbation, which reads
\begin{align}\label{eom_dtheta_general}
	\delta\ddot{\theta}_k + \left(3H + 2\frac{\dot{R}_b}{R_b}\right)\delta\dot{\theta}_k + \left(\frac{k^2}{a^2} + m_\theta^2(t)\right)\delta\theta_k = 0,
\end{align}
where $H(t)$ and $a(t)$ are given by \eqref{Hubble_step} and \eqref{a_time_varying_Hubble}. In terms of the $e$-fold number of inflation, $N = \ln a$, we have
\begin{align}
	\delta\theta_k^{\prime\prime} + \left(3+\frac{H^\prime}{H} + 2\frac{R_b^\prime}{R_b}\right) \delta\theta_k^\prime + 
	\left(\frac{k^2}{a^2 H^2} + \frac{m_\theta^2}{H^2}\right) \delta\theta_k=0,
    \label{Eq:EOM_deltatheta}
\end{align} 
where $\delta\theta_k^\prime\equiv\partial_N \delta\theta_k$.
Note that, in \eqref{Eq:EOM_deltatheta}, there are a couple of properties that distinguish the angular mode from a free scalar field with constant mass. 
First, in addition to Hubble friction, there is an additional friction term led by $2R_b^\prime/R_b$. Thus, a time-dependent background radial motion can affect the angular perturbation, which is the main focus of our study.
Second, the ratio of the angular mass to the Hubble parameter, $M_\theta \equiv m_\theta/H = \sqrt{n c_A X_{\rm min}}$, at the local minima of the flat direction potential remains constant despite the feature-induced change in the Hubble parameter. These differences can be potentially observable as part of the isocurvature power spectrum which we will soon compute.

\bigbreak\noindent
$\bullet$ {\it Numerical methods.}
In the following subsections, we provide full numerical solutions to the power spectra, along with analytical estimates.
The numerical approach we use in this paper to solve scalar perturbations with a friction term that varies over time during inflation can be found in \cite{Wu:2024wti}. 
At some time before the appearance of the feature $\eta \leq \eta_i < \eta_\ast$, we approximate the evolution of the mode function by the analytic solution \eqref{def_theta_k}, where $\eta_i$ is the initial time of our numerical computations. For $\eta > \eta_i$, we numerically solve the equation of motion \eqref{Eq:EOM_deltatheta} for the mode function $u_k = (R_{\rm end}/H_I)\delta\theta_k$ with the initial conditions given by the analytic solution \eqref{def_theta_k} at $\eta_i = \eta_\ast k_\ast/k_i$. 
Defining $ z_\ast \equiv -k\eta_\ast = k/k_\ast$, where $k_\ast = -1/\eta_\ast = a(N_\ast)H_I$ is the mode that crosses the horizon when the sharp feature occurs, $\eta=\eta_\ast$, and $z_i \equiv k\eta_i = z_\ast k_\ast/k_i$, the initial conditions for the mode functions are
\begin{align}\label{mode_uk_boundary_conditions}
	u_k(z_i) &= c_\theta z_i^{3/2} H_{\nu_\theta}^{(1)}(z_i), \\
	\partial_N u_k\vert_{z = z_i} &= - c_\theta z_i
	\left[z_i^{3/2} H_{\nu_\theta -1}^{(1)}(z_i) +\left(\frac{3}{2}- \nu_\theta \right)z_i^{1/2} H_{\nu_\theta}^{(1)}(z_i)\right].
\end{align}
Using \eqref{Hubble_step}, \eqref{radial_modulation}, and \eqref{sol_R1_step} as the background evolution induced by the feature,
we can solve \eqref{Eq:EOM_deltatheta} for arbitrary $z_\ast = k/k_\ast$ by choosing an initial time $N_i < N_\ast$ for $\eta_i$, where $k_\ast/k_i = e^{N_\ast-N_i}$.
The power spectrum of the isocurvature perturbation, $\delta\theta$, at the end of inflation, $\eta=\eta_{\rm end} \equiv -1/k_{\rm end}$,
is thus
\begin{align}
	P_\theta(k,\eta_{\rm end}) = \frac{k^3}{2\pi^2} \vert\delta\theta_k(\eta_{\rm end})\vert^2 
	=  \frac{H_I^2}{R_{\rm end}^2} \frac{k^3}{2\pi^2} \left\vert u_k(-1/k_{\rm end})\right\vert^2 .
\end{align}
Note that feature-induced oscillations, such as $R_{\rm 1g}^{\rm osc}$ or $R_{\rm 1k}$, decay at the end of inflation so that $R_b(\eta_{\rm end}) = R_{\rm end}(1+B/2(n-2)) \approx R_{\rm end}$.

The Heaviside form of the step feature used in \eqref{def_step_feature} is infinitely sharp, which can easily result in unrealistic large signals in the power spectrum of the curvature perturbation ruled out by the current data. To have a more realistic construction of the step feature and implement a complete treatment in numerical calculation, we smooth out the infinitely sharp feature as follows:
\begin{align}\label{Eq:Smoothed_potential}
    \Delta V = B V_0 \Theta_b(\phi-\phi_\ast),
\end{align}
where $\Theta_b$ is the so-called Gaussian-smeared step function with the definition:
\begin{align}\label{Gaussian_smeared_step_function}
    \Theta_b(x) \equiv \frac{1}{2}\left[1+\text{erf}\left(\frac{x}{b}\right) \right], \quad \partial_x\Theta_b = \frac{1}{b\sqrt{\pi}} e^{-x^2/b^2} \equiv \delta_b(x).
\end{align}
Note that erf$(x)$ is the error function and $\delta_b$ is the Gaussian delta function. $\Theta_b$ and $\delta_b$ reproduce the Heaviside step function and the Dirac delta function in the limit of $b\rightarrow 0$, respectively.
This smoothing is applied to all numerical calculations including background evolution and density perturbations.
Equations of motions with the smoothed step potential can be found in Appendix \ref{Append_Gaussian_step}.

\subsubsection{Observational signals from gravitational couplings}
\label{Sec:gravitational_coupling}

Let us first study the minimal observable from the gravitational effect in the isocurvature and curvature power spectrum when the kinetic coupling \eqref{def_phi_sigma_couplings} vanishes (which is the case with a canonical Kähler potential).
This is equivalent to taking $c_6 =0$ in our analysis.
In this case, the AD field is classically excited by the sharp change in the inflaton field only indirectly through the change in the background Hubble parameter \eqref{Delta_H_step}.

We will first study the properties analytically using the infinitely sharp step-potential limit and obtain some estimates. We then present numerical examples with smoothed steps.

\bigbreak\noindent
$\bullet$ {\it Isocurvature power spectrum.}
For the Heaviside step feature \eqref{def_step_feature}, the background modulation in the leading order of $\mathcal{O}(B)$ is given by $H^\prime/H = \Delta H^\prime/H_I = 3B e^{3(N_\ast - N)}/2$,  $R_b^\prime/R_b = (\Delta R_{\rm end}^\prime +R_{\rm 1g}^{\rm osc\prime}) /R_{\rm end} $, where $\Delta R_{\rm end}^\prime/R_{\rm end} = \Delta H^\prime/H_I/(n-2)$ according to \eqref{radial_modulation}. The equation-of-motion for the perturbation of the isocurvature component, $\delta\theta_k$, is given by
\begin{align}\label{eom_dtheta_k_N}
	 \delta\theta_k^{\prime\prime} + \left[3+ \frac{n}{n-2}\frac{\Delta H^\prime}{H_I} + 2\frac{R_{\rm 1g}^{\rm osc\prime}}{R_{\rm end}} \right] \delta\theta_k^\prime +
	\left\{ e^{2(N_\ast - N)} \left[ \frac{k/k_\ast}{ 1+ \Delta H/H_I} \right]^2 + M_\theta^2  \right\}  \delta\theta_k =0,
\end{align}
where $\Delta H/H_I$ is given by \eqref{Delta_H_step}. Note that $e^{N_\ast} = k_\ast/H_I$, and for the Heaviside step feature \eqref{def_step_feature} we have
\begin{align}\label{Eq:Rosc/Rend}
    \frac{R_{\rm 1g}^{\rm osc\prime}}{R_{\rm end}} =& \frac{3B}{2(n-2)}e^{-\frac{3}{2}(N-N_\ast)}\Theta(N-N_\ast) \\\nonumber
    \qquad &\times\left\{ -\cos\left[\mu_R(N-N_\ast)\right]  +\frac{3}{2\mu_R}\sin\left[\mu_R(N-N_\ast)\right]  \right\},   
\end{align}
where $R_{\rm 1g}^{\rm osc}$ is the radial oscillation \eqref{R1_gravitational} triggered by the modulated potential minimum $\Delta R_{\rm end}$ through gravitational couplings.

\begin{figure}[!t]
    \centering
    \includegraphics[width=0.7\linewidth]{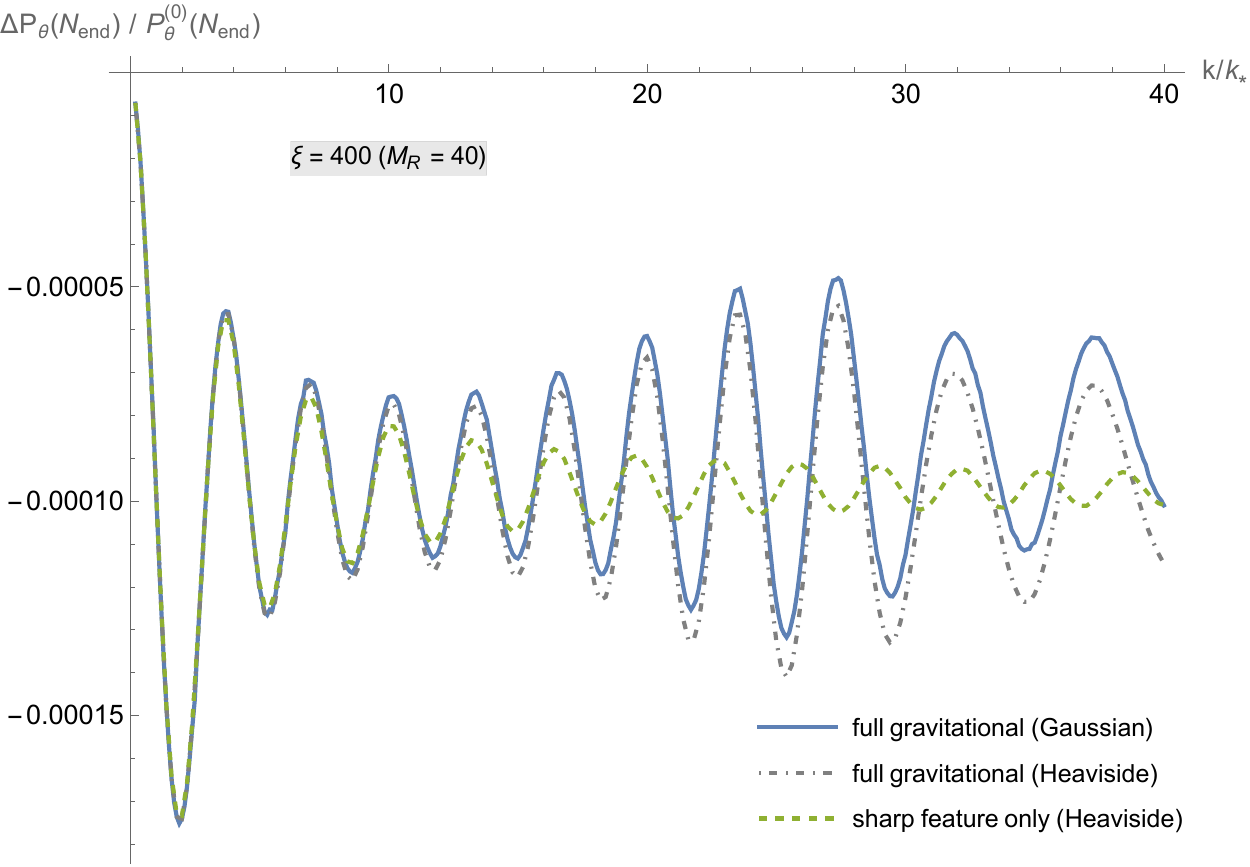}
    \caption{{\em Gravitational coupling cases.}
    The fractional corrections induced by primordial features in the isocurvature power spectra for the angular mode of the AD field, $\Delta P_\theta/P_\theta^{(0)} = P_\theta/P_\theta^{(0)} -1$, with $n = 4$, $c_A = 10^{-3}$.  We use $B = -2 \times 10^{-4}$ for demonstration. The full signal obtained from either the smoothed step feature \eqref{Eq:Smoothed_potential} or the Heaviside feature \eqref{def_step_feature} is a mixture of the sharp-feature signal, induced by the sudden change in the background, and the resonant clock signal, induced by the radial oscillations. The sharpness parameter $b = 0.01$ is used for the smoothed step case. The sharp-feature-only signal is obtained, for the purpose of comparison, by artificially ignoring the radial oscillation $R_{\rm 1g}^{\rm osc}$ in the friction term of \eqref{eom_dtheta_k_N}.
    }
    \label{fig:gravitional_clock_signal}
\end{figure}
\begin{figure}[h]
    \centering
    \includegraphics[width=0.7\linewidth]{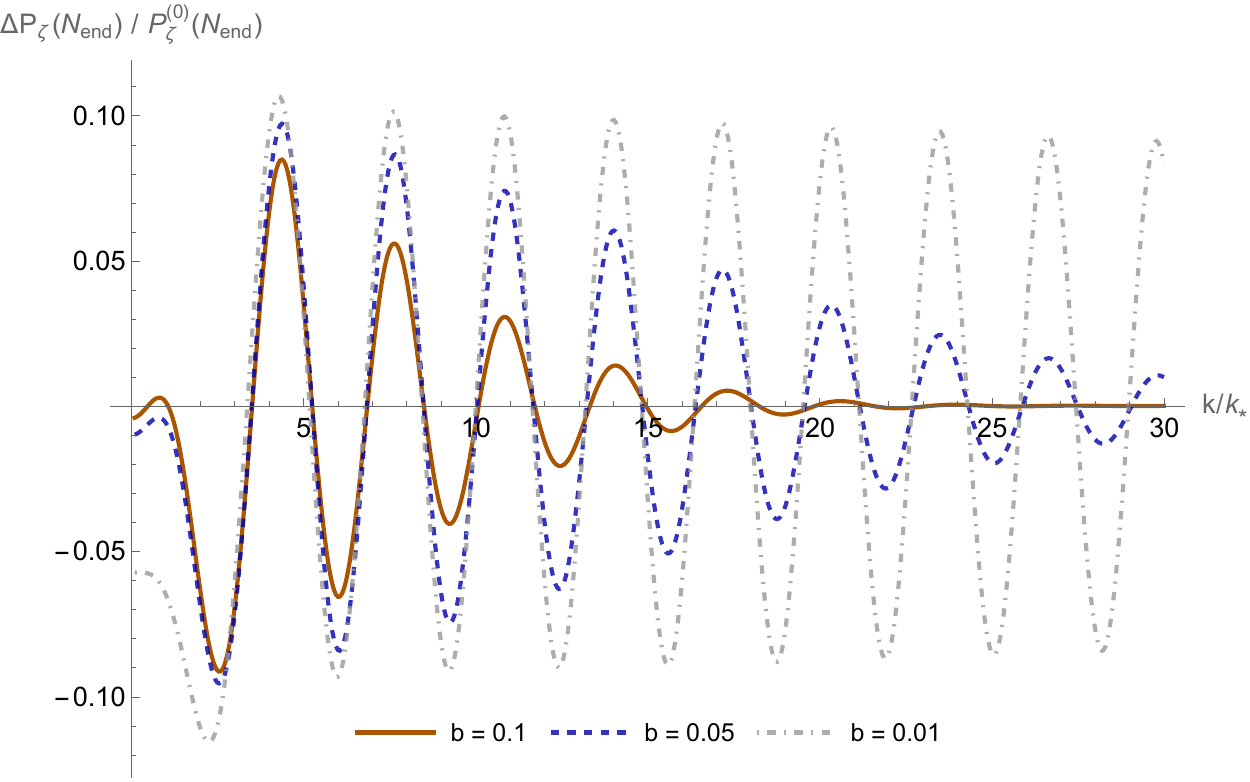}
    \caption{{\em Gravitational coupling cases.}
    The fractional corrections induced by primordial features in the curvature power spectra, $\Delta P_\zeta/P_\zeta^{(0)} = P_\zeta/P_\zeta^{(0)} -1$, with $n = 4$, $c_A = 10^{-3}$.  We use $B = -2 \times 10^{-4}$ and $\epsilon_{H0} = 0.0063$ for demonstration. $b$ is the sharpness parameter which measures the width of the smoothed step function \eqref{Eq:Smoothed_potential}.}
    \label{fig:SignalG_zeta}
\end{figure}

First, let us consider sharp feature signals. There are two sources. 

One is due to the sharp change in the friction terms 
$\sim \Delta H^\prime/H_I$ and $\sim R_{\rm 1g}^{\rm osc \prime}/R_{\rm end}$ in \eqref{eom_dtheta_k_N}. This sharp change generates a sharp-feature signal with a characteristic sinusoidal running in $k$,
\begin{align}
    S_{\rm sharp-feature}\sim  g(k/k_\ast)\cos\left(2\frac{k}{k_\ast} + \text{phase}\right).
    \label{Eq:gravitional_sharp_feature}
\end{align}
This oscillatory pattern applies to modes that are within the horizon at the time of the sharp feature, $k/k_\ast >1$.  Modes at superhorizon at the time of the sharp feature, $k/k_\ast \ll 1$, are not affected by the feature. The amplitude and envelope of the sharp feature signal, $g(k/k_\ast)$,  depend on the detail of the sharp feature, some examples are shown in Fig.~\ref{fig:gravitional_clock_signal}.

In addition to inducing an oscillatory signal, the change in 
$\sim \Delta H^\prime/H_I$ has another effect. The asymptotic change in the Hubble parameter due to the upward (downward) step also results in an enhancement (suppression) of the spectral amplitude in the small-scale limit ($k/k_\ast \gg 1$),
\begin{align}\label{sharp_feature_Amp_asyp}
    \left. \frac{\Delta P_\theta}{P_\theta^{(0)}}\right\vert_{k \ll k_\ast} \sim \frac{n}{2(n-2)} B.
\end{align}

Second, let us consider the clock signal.

The oscillation in the second friction term $\sim R_{\rm 1g}^{\rm osc \prime}/R_{\rm end}$ in \eqref{eom_dtheta_k_N} generates the so-call clock signal \cite{Chen:2011zf,Chen:2014cwa} due to the interference \cite{Chen:2008wn} of the radial oscillation $\sim e^{i m_R t}$ in $R_{\rm 1g}^{\rm osc}$ with the vacuum frequencies $\sim e^{\pm ik\eta}$ of the angular mode function $\delta\theta_k$.
The resonant clock signal induced by massive field oscillations during inflation takes the following form \cite{Chen:2014cwa}:
\begin{align}\label{Shape_resonant_clock}
    S_{\rm clock} \approx 
    \left. \frac{\Delta P}{P^{(0)}}\right\vert_{\rm clock,max}
    \;\left(\frac{k}{k_{\rm max}}\right)^{-3/2} \cos\left[ M_R\ln\left(\frac{k}{k_{\rm max}}\right) + \text{phase}\right],
\end{align}
which has an envelope that decays towards large $k$ due to the dilution of the massive field and a resonance phase directly determined by the inflationary background. Also note that this form is only valid after the first resonance occurs when $2k/a(t_\ast) \approx m_R$, which is roughly the mode that acquires the peak signal amplitude. This gives $k_{\rm max} \equiv M_Rk_\ast/2$ for the clock signal template \eqref{Shape_resonant_clock}.  Given that $M_R = m_R/H_I\gg 1$, we see that resonant-clock signals appear on scales much smaller than the onset of sharp-feature signals at $k/k_\ast = 1$.

To estimate the peak amplitude at $ k = k_{\rm max}$, we can directly use the result of Eq.~(4.16) in \cite{Chen:2014cwa} because the coupling structure is exactly the same. This is given by 
\begin{align}
	\left. \frac{\Delta P_\theta}{P_\theta^{(0)}}\right\vert_{\rm clock,max}  
    \sim  \frac{3B}{2(n-2)M_R} \sqrt{2\pi M_R} ,
    \label{Eq:Peak_Amp_theta_g}
\end{align}
where the first factor comes from the ratio of the peak oscillation amplitude to the overall radial VEV, as given in Eq.~\eqref{R1_gravitational} in the large mass limit ($M_R \gg 1$).

Overall, the clock signal \eqref{Shape_resonant_clock}, which occupies the larger $k$ region, is smoothly connected to two sharp feature signals (such as \eqref{Eq:gravitional_sharp_feature}), which occupy the smaller $k$ region. These form the full standard clocks signal. On top of this, there is also an overall shift \eqref{sharp_feature_Amp_asyp}.

\bigbreak
To demonstrate numerical examples with the smoothed step feature \eqref{Eq:Smoothed_potential}, we solve $\delta\theta$ via \eqref{Eq:EOM_deltatheta}, instead of \eqref{eom_dtheta_k_N}, with $H = H_I + \Delta H$, $R_b = R_{\rm end} + R_{\rm 1g}$, and $\Delta H$ is given by \eqref{Delta_H_smoothed}. 
$R_{\rm 1g}$ is solved from \eqref{eom_R1g} with $\Delta R_{\rm end} = R_{\rm end}\Delta H/(n-2)$ and $\Delta H$ is the smoothed Hubble modulation \eqref{Delta_H_smoothed}.
We demonstrate several examples in Figure~\ref{fig:gravitional_clock_signal}.
Consistent with the analytical estimates, in this gravitational coupling case, the full standard-clock signal, which contains both the sharp-feature signal component and the clock signal component, is present in the isocurvature power spectrum, Figure~\ref{fig:gravitional_clock_signal}. This signal is a direct probe of the radial component of the AD field during inflation. However, the amplitudes of this primordial feature signal is very small. 

\bigbreak\noindent
$\bullet$ {\it Curvature power spectrum.} 
We now consider the feature induced signals in the curvature perturbation, which is correlated with those in the isocurvature perturbation. 

We first compute the inflaton perturbation, $\delta\phi$, and then convert it to the curvature perturbation through the relation $\zeta=-H\delta\phi/\dot\phi$ in the spatially flat slicing, which should be evaluated at the end of the inflation, or, sufficient $e$-folds after the horizon exit when $\zeta$ has approached a constant. We neglect the corrections from the radial field perturbation $\delta R$. 
The inflaton perturbation is governed by the quadratic Lagrangian as
\begin{align}\label{Lagrangian_delta_phi_gravitation}
	\delta\mathcal{L}_\phi^{(2)} = \frac{a^3}{2}  \left[(\delta\dot{\phi})^2 - \frac{1}{a^2}(\partial_i\delta\phi)^2\right] -a^3 V_{\phi\phi}\delta\phi^2,
\end{align}
where we can neglect the small inflaton mass suppressed by the slow-roll parameter $\epsilon_{H0}$ from $\partial_{\phi\phi}V_0$. The sharp feature introduces a mass term given by
\begin{align}
    V_{\phi\phi} \approx \partial_{\phi\phi}\Delta V = BV_0 \frac{H_I^2}{\dot{\phi}_0^2}\partial_N\delta(N - N_\ast).
\end{align}
Adopting the smoothing strategy from \eqref{Gaussian_smeared_step_function}, the feature results in an effective mass for the inflaton perturbation in terms of the derivative of the Gaussian-smeared delta function as
\begin{align}\label{induced_inflaton mass}
    \Delta M_\phi^2 \equiv \frac{\partial_{\phi\phi}\Delta V}{H_I^2} = -3\frac{B}{\epsilon_{H0}}\frac{N-N_\ast}{b^2}\delta_b(N-N_\ast).
\end{align}
At $N = N_\ast$, $\delta_b(N-N_\ast)$ defined in \eqref{Gaussian_smeared_step_function} has a width $\sim b$ and a height $\sim 1/b$.
A sufficiently small $b \ll 1$ is used to reach the precision goal in our numerical results. The smallness of $b$ reflects the sharpness of the step feature in $\Delta V$.

The equation of motion for mode functions of the inflaton perturbation, $\delta\phi_k$, is given by
\begin{align}\label{eom_dphi_N}
	\delta\phi_k^{\prime\prime} + \left(3+  \frac{\Delta H^\prime}{H_I} \right) \delta\phi_k^\prime
	+\left(\frac{k^2}{a^2 H^2} + \Delta M_\phi^2   \right)\delta\phi_k = 0,
\end{align}
where $\Delta M_\phi$ is given in \eqref{induced_inflaton mass}. $\Delta H$ is the smoothed Hubble modulation given in \eqref{Delta_H_smoothed}. The induced mass term $\Delta M_\phi$ gives rise to a sharp feature signal with an amplitude $\sim B/\epsilon_{H0}$. This sharp-feature signal schematically has the same sinusoidal scale-dependence as in \eqref{Eq:gravitional_sharp_feature}. The envelop and phase of the signal shape function are not only model-dependent but also depend on the smoothing parameter $b$; some examples are shown in Fig.~\ref{fig:SignalG_zeta}.\footnote{ See Ref.~\cite{Ng:2021hll} for discussions on features with much larger step sizes, $\vert B\vert \gg \epsilon_{H 0}$, and their implications on primordial black hole generation.} Note that a very narrow step width $b \leq 10^{-2}$ can approximate the Heaviside step feature results, where the sharp-feature signals barely decay in the large $k/k_\ast$ limit.

In Figure~\ref{fig:SignalG_zeta}, the power spectrum of the curvature perturbation is computed as
\begin{align}
    P_\zeta = \frac{H^2}{\dot{\phi}_b^2}P_\phi \approx \frac{H_I^2}{\dot{\phi}_0^2}\left(1+2\frac{\Delta H}{H_I}\right)\left(1-2\frac{\dot{\phi}_1}{\dot{\phi}_0}\right) P_\phi^2,
\end{align}
where $P_\phi = (k^3/2\pi^2)\vert \delta\phi_k\vert^2$. Note that all the feature induced corrections (such as $\Delta H$ and $\dot{\phi}_1$) become negligibly small at the end of inflation.

So, in this gravitational coupling case, in the curvature power spectrum, only the sharp-feature signal, due to the step in the inflaton potential and correlated with the sharp-feature signal in the isocurvature power spectrum, is present. The extensiveness of the sharp-feature signal is determined by the sharpness of the step.
Unlike the isocurvature field $\theta$, here there is no direct coupling between $R$ and $\phi$. The classical oscillation of the $R$ field is not coupled to the inflaton field directly, so there is no clock signal in the curvature power spectrum. 
This is also demonstrated in numerical examples in Figure~\ref{fig:SignalG_zeta}.

\bigbreak\noindent
$\bullet$ {\it Summary.} 
To concisely summarize the main conclusions of Section~\ref{Sec:gravitational_coupling}: if the inflaton and AD field are coupled only gravitationally, the sharp feature in the inflaton sector can abruptly shift the VEV of the AD field and induce small classical oscillations in its radial component. This process generates primordial feature signals in both the curvature and isocurvature power spectra, which are correlated with each other. In the curvature power spectrum, the signal consists only of a sharp-feature signal. In contrast, the isocurvature power spectrum contains both a sharp-feature and a clock signal, which are smoothly connected and together form a full standard clock signal.
Overall, together with the tilt of the non-oscillatory component of the isocuvature power spectrum, these signals directly probe both the angular and radial components of the AD field during inflation.

On the other hand, although the amplitude of the sharp feature signal in the curvature power spectrum can be made sufficiently large to saturate the observational bound, the weak nature of the gravitational coupling implies that the correlated full standard clock signal in the isocurvature power spectrum can only appear as a tiny correction to the leading-order non-oscillatory component.

\subsubsection{Observational signals from direct couplings}
\label{Sec_inflaton_couplings}

\begin{figure}[t]
	\begin{center}
    \includegraphics[width=0.7\linewidth]{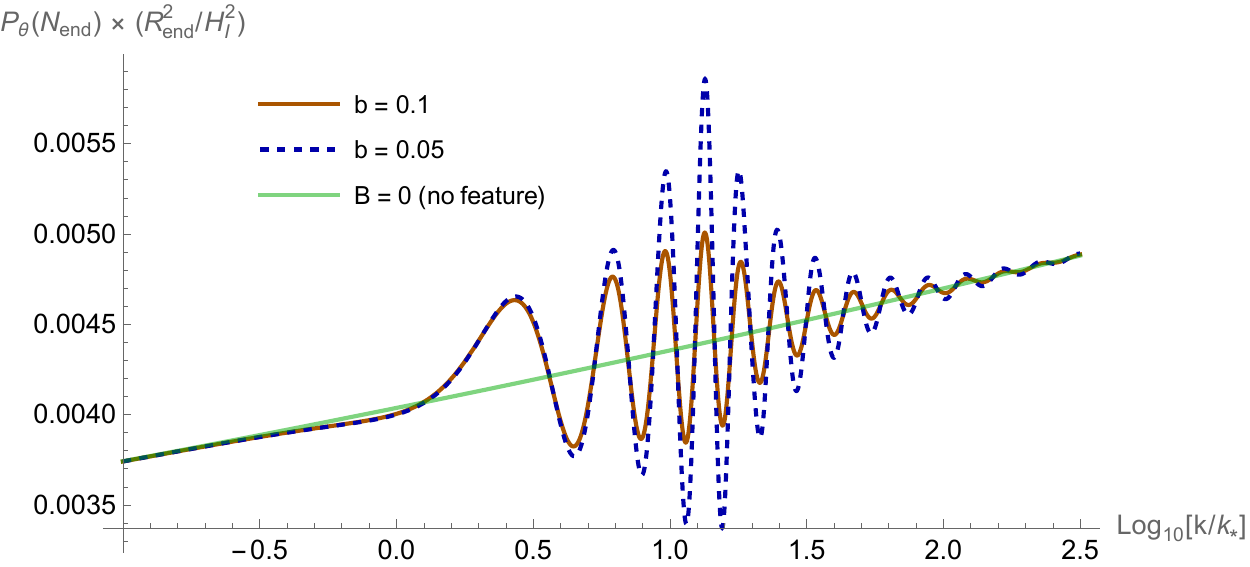}
    \hfill
	\includegraphics[width=0.7\linewidth]{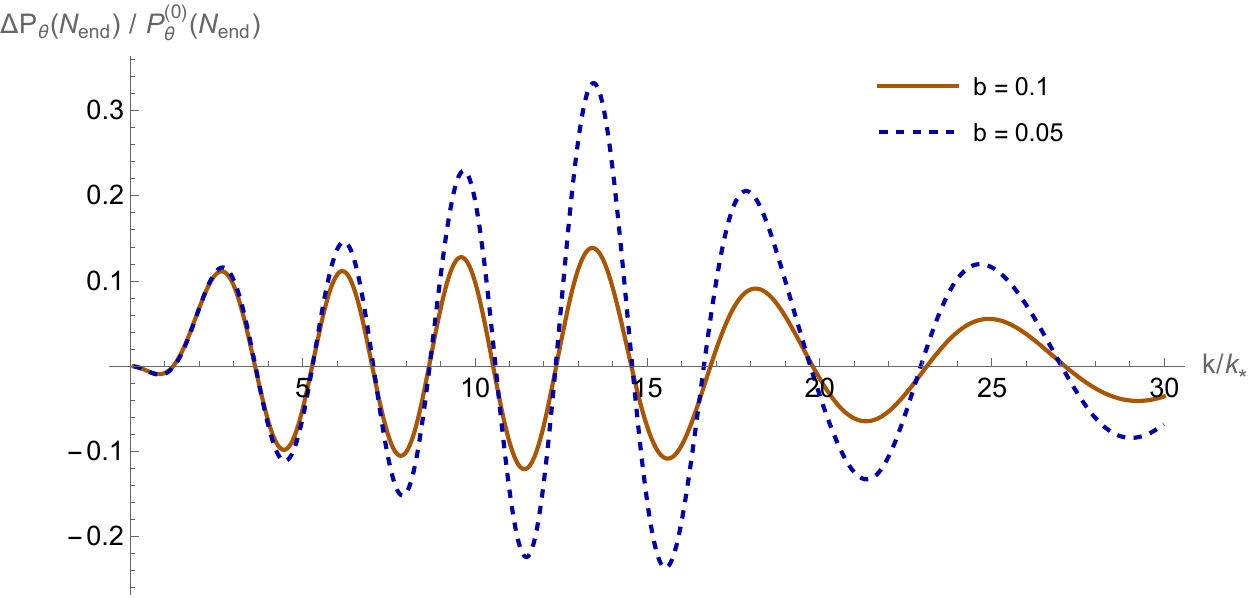}
	\end{center}
	\caption{\label{fig.spectral_correction_coupling} 
    {\em Direct coupling cases.}
    [Upper Panel] The step-feature induced signals from the kinetic coupling as seen in the isocurvature power spectrum $P_\theta$ at the end of inflation, which are initial conditions for the BDI perturbations. The flat-direction model parameters $n=4$, $\xi = 100$, $c_A = 10^{-3}$ are used, which gives $M_R = 20$, $M_\theta = 0.22$. In this plot, the smoothed step feature is used with the sharpness parameter, $b$, characterizing the width of the smoothed step feature. [Lower Panel] Fractional corrections to the power spectrum, $\Delta P_\theta/P_\theta^{(0)} = P_\theta/P_\theta^{(0)} -1$, corresponding to results in the upper panel. The step height $B = -2\times 10^{-5}$, and the parameters $M_P/\Lambda = 10^3$, $c_6 = 1/6$, are used in both panels. 
    }
\end{figure}

\begin{figure}[h]
	\begin{center}
		\includegraphics[width=10 cm]{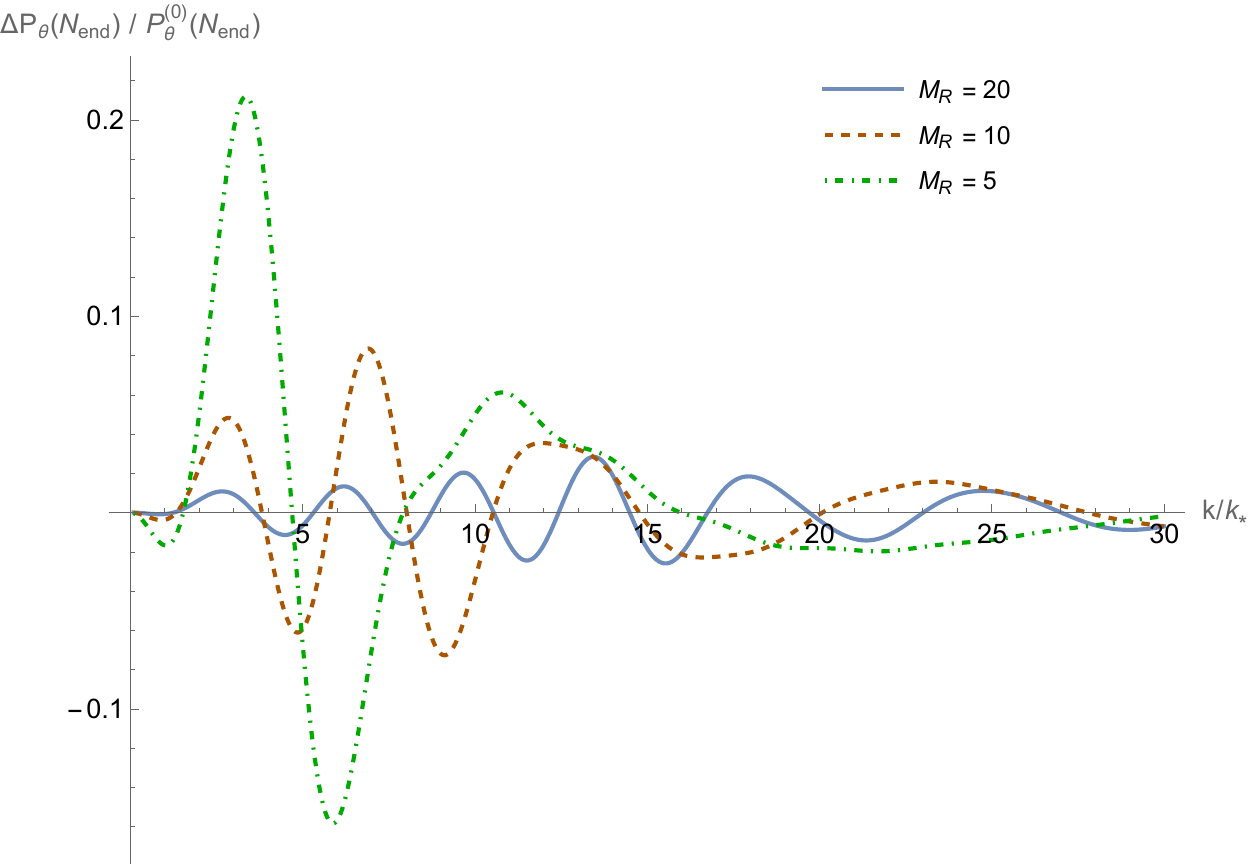}
	\end{center}
	\caption{\label{fig.spectral_correction_coupling_MR} 
    {\em Direct coupling cases.}
    The corrections to the curvature power spectra, $\Delta P_\theta/P_\theta^{(0)} = P_\theta/P_\theta^{(0)} -1$, due to oscillating massive field $R$ with different masses. Masses of the radial field $R$, $M_R$, are indicated in figure legend. We use $M_P/\Lambda = 10^{2.5}$, $c_6 = 1/6$ and a step size $B = -2\times 10^{-5}$. The sharpness parameter $b = 0.05$ is used for all results in this plot.
    }
\end{figure}

\begin{figure}[t]
    \centering
    \includegraphics[width=0.7\linewidth]{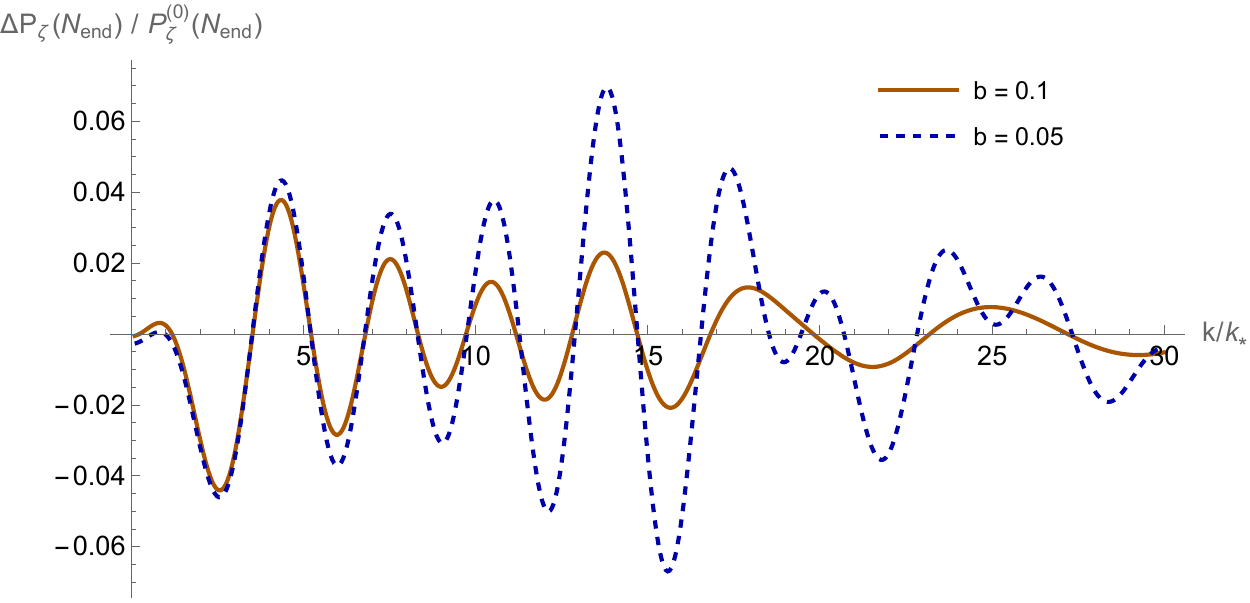}
    \hfill
    \includegraphics[width=0.7\linewidth]
    {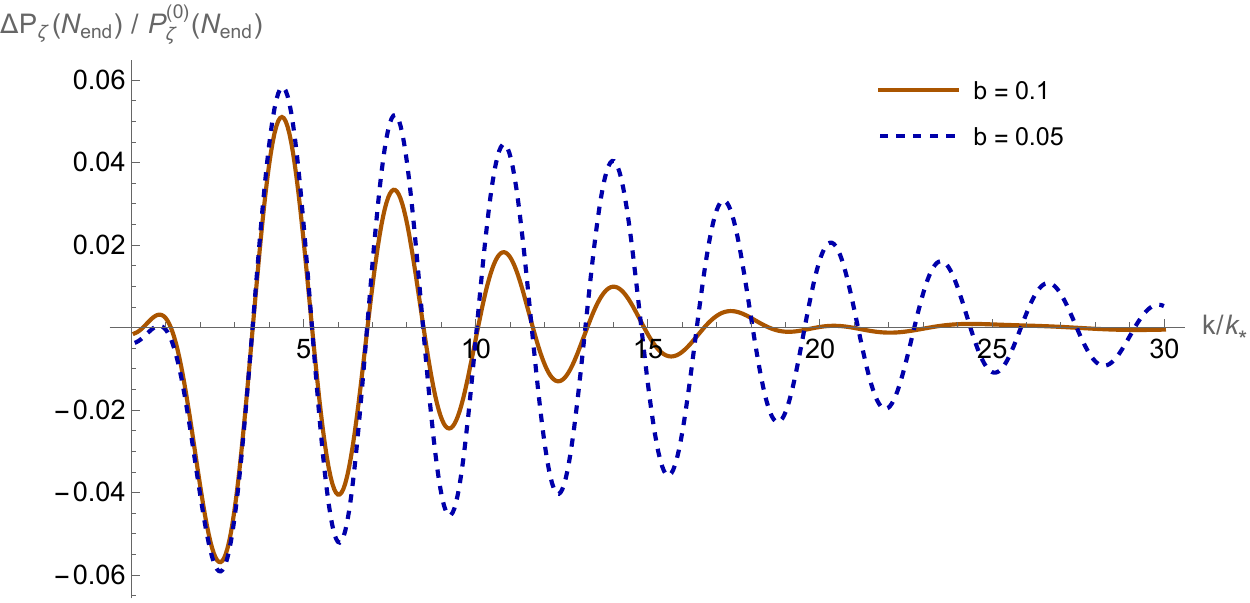}
    \caption{ {\em Direct coupling cases.}
    Fractional corrections to the curvature power spectra, $\Delta P_\zeta/P_\zeta^{(0)} = P_\zeta/P_\zeta^{(0)} -1$, due to the smoothed step feature for the curvature perturbation, correlated to the results in Figure~\ref{fig.spectral_correction_coupling}. Parameters of the flat-direction model are $n = 4$, $\Lambda/H_I = 10^2$, $\xi = 100$, $c_A = 10^{-3}$, which gives $M_R = 20$, $M_\theta = 0.22$. We use $B = -2\times 10^{-5}$, $\epsilon_{H0} =10^{-3}$, $R_{\rm end} = \Lambda$ in both panels. The upper (lower) panel uses $c_6 = 1/6$ with $M_P/\Lambda = 10^{3}$, ($M_P/\Lambda = 10^{2.5}$), respectively.}
    \label{fig:SignalK_zeta_smoothed}
\end{figure}

We now turn on the non-canonical kinetic coupling \eqref{def_phi_sigma_couplings} between the inflaton and the AD field. The classical oscillation of the radial component of the AD field is studied in Sec.~\ref{Sec:direct_couping_bkgd_oscillations}. Here, we consider its effect on the density perturbations. Again, we first qualitatively estimate the properties of the primordial feature signals using an infinitely sharp step modeled by the Heaviside function, and then present numerical examples using a smoothed step.

\bigbreak\noindent
$\bullet$ {\it Isocurvature power spectrum.}
To compute the full signals in the power spectrum induced by the radial field oscillation \eqref{R1_gravitational} and \eqref{sol_R1_step}, together with the effect from the Hubble modulation \eqref{Delta_H_step}, we solve the following angular mode function in the leading order in $B$,
\begin{align}\label{eom_dtheta_N_coupling}
	\delta\theta_k^{\prime\prime} + \left(3+ \frac{\Delta H^\prime}{H_I} + 2\frac{R_1^\prime}{R_{\rm end}}\right)\delta\theta_k^\prime
	+\left(\frac{k^2}{a^2 H^2} + M_\theta^2 \right) \delta\theta_k = 0 ~,
\end{align} 
where $R_1 = R_{\rm 1g} + R_{\rm 1k}$ includes terms induced by the gravitational effect and the direct kinetic coupling effect. The former effect, together with its effect through $\Delta H$, has been considered in the previous subsection. 

The additional effect that we are interested here is the latter effect encoded in $R_{\rm 1k}(N)$ given by \eqref{sol_R1_step} for the Heaviside step feature \eqref{def_step_feature},
\begin{align}
	\frac{R_{\rm 1k}^\prime}{R_{\rm end}} 
    &= 6c_6B\frac{M_P^2}{\Lambda^2M_R^2} \Theta(N-N_\ast)
	\left[\frac{\Delta_+^2}{2\nu_R} e^{\Delta_+(N_\ast - N)} -\frac{\Delta_-^2}{2\nu_R} e^{\Delta_-(N_\ast - N)} -3 e^{3(N_\ast - N)}\right], \\
    &\approx 6c_6B\frac{M_P^2}{\Lambda^2M_R^2} \Theta(a-a_\ast) \left(\frac{a}{a_\ast}\right)^{-3/2}\sin\left[\mu_R \ln(a/a_\ast)\right],
\end{align}
where the second line is the limit of $M_R \gg 1$ and $a/a_\ast$ is given by \eqref{a_time_varying_Hubble}.  
$\Delta_\pm\equiv 3/2\pm\nu_R$ are conformal weights of the radial mode \cite{Antoniadis:2011ib, Ng:2021hll}. $\nu_R = i \mu_R$ when $M_R > 3/2$ and $\mu_R\approx M_R$ if $M_R \gg 3/2$. For the smoothed step feature \eqref{Eq:Smoothed_potential}, we solve $R_{1k}$ numerically by \eqref{eom_r1k_smoothed}.

Both $\Delta H^\prime/H_I$ and $\Delta R_1^\prime/R_{\rm end}$ in \eqref{eom_dtheta_N_coupling} exhibit sharp changes in their evolution due to the sharp change in the inflaton sector. This induces a sharp feature signal in the isocurvature power spectrum. In addition, in $\Delta R_1^\prime/R_{\rm end}$, the sharp change is followed by classical oscillations due the oscillations of the $R$ field. This induces a clock signal. Together, they form a full standard clock signal.

As we have studied in the previous gravitational coupling case, the clock signal induced by the first term of $R_1=R_{\rm 1g}+R_{\rm 1k}$ is very small. Here, due to the second direct coupling term, the magnitude of the clock signal becomes much more adjustable and can be very large.

As in the previous case, we can estimate the amplitude of the clock signal from the kinetic coupling following \cite{Chen:2014cwa}:
\begin{align}
	\left. \frac{\Delta P_\theta}{P_\theta^{(0)}}\right\vert_{\rm clock,max}  
    \sim 6c_6 B\frac{M_P^2}{\Lambda^2M_R^2} \sqrt{2\pi M_R} .
    \label{Eq:Peak_Amp_theta}
\end{align}
This estimation agrees with our numerical results. 

Numerical examples of the combined full signals in the isocurvature power spectrum, $P_\theta$, are presented in Figure~\ref{fig.spectral_correction_coupling} and \ref{fig.spectral_correction_coupling_MR}. 
Figure~\ref{fig.spectral_correction_coupling} shows several examples of the smoothed step feature \eqref{Eq:Smoothed_potential} with several finite sharpness parameters, $b > 10^{-2}$. In Figure~\ref{fig.spectral_correction_coupling_MR} we vary the mass of the radial field, $M_R$. 

As we can see in these figures, with sufficiently large direct coupling, the full standard-clock signals can exhibit clock signals very clearly. The amplitudes of these clock signals can also be several orders of magnitude larger than those in the gravitational case --- in fact, the fractional correction they introduce to the non-oscillatory component can approach ${\cal O}(1)$ (by e.g.~choosing instead a coupling $A_{c_6} > 10^5$ for the examples in Fig.\ref{fig.spectral_correction_coupling_MR}) and become part of the leading signal in the isocurvature power spectrum.

\bigbreak\noindent
$\bullet$ {\it Curvature power spectrum.} 
As similar to the case in the gravitational coupling, we compute the curvature perturbation, $\zeta$, in terms of the inflaton perturbation, $\delta\phi$, through the relation $\zeta=-H\delta\phi/\dot\phi$ in the spatially flat slicing.  
With the presence of the kinetic coupling \eqref{def_phi_sigma_couplings}, the leading order Lagrangian is then
\begin{align}\label{Lagrangian_delta_phi}
	\delta\mathcal{L}_\phi^{(2)} = \frac{a^3}{2} \left(1+c_6\frac{R_b^2}{\Lambda^2}\right) \left[(\delta\dot{\phi})^2 - \frac{1}{a^2}(\partial_i\delta\phi)^2\right] -a^3 V_{\phi\phi}\delta\phi^2,
\end{align}
where $R_b(t)$ is given in \eqref{def_background_decomposition}. Again, we focus on the effective mass induced by the step feature under the approximation given by \eqref{induced_inflaton mass}.

The equation of motion for the inflaton mode functions, $\delta\phi_k$, is now given by
\begin{align}\label{eom_dphi_N_full}
	\delta\phi_k^{\prime\prime} + \left[3+  \frac{\Delta H^\prime}{H_I} + 2c_6\frac{R_bR_b^\prime}{\Lambda^2} \left(1+c_6\frac{R_b^2}{\Lambda^2}\right)^{-1}\right] \delta\phi_k^\prime
	+\left(\frac{k^2}{a^2 H^2} + \Delta M_\phi^2   \right)\delta\phi_k = 0.
\end{align}
We present a comparison of the power spectrum of the dimensionless inflaton perturbation $\delta\phi/H_I$ with $P_\theta$ in Figure~\ref{fig.spectral_correction_coupling} with $R_{\rm end} = \Lambda/\sqrt{6c_6}$ as an example. We take $R_{\rm end} = \Lambda/\sqrt{6c_6}$ as the largest value to keep the sub-dominance of the kinetic coupling in the inflaton kinetic energy. In a practical realization of baryogenesis, $R_{\rm end}$ given by \eqref{def_Rend_theta_end} will be determined by the flat-direction parameters.

Since now the inflaton field is directly coupled to the radial component of the AD field, the classical oscillation of $R$ can induce a large clock signal component also in the curvature power spectrum. This is in contrast to the gravitational case.
This clock signal also takes the form of \eqref{Shape_resonant_clock} and peaks
around $k = k_{\rm max} = M_Rk_\ast/2$. 
The peak amplitude can be estimated again using \cite{Chen:2014cwa} as 
\begin{align}
    \left.\frac{\Delta P_\zeta}{P_\zeta^{(0)}}\right\vert_{\rm clock,max} 
    \sim  \left. \frac{\Delta P_\theta}{P_\theta^{(0)}}\right\vert_{\rm clock,max} \times c_6\frac{R_{\rm end}^2}{\Lambda^2}  \sim 6c_6^2B \sqrt{\frac{2\pi}{M_R^3}} \;\frac{M_P^2R_{\rm end}^2}{\Lambda^4} .
    \label{Eq:Peak_Amp_phi}
\end{align}
Note that, in addition to phase correlation, the amplitudes of the clock signals in the curvature power spectrum \eqref{Eq:Peak_Amp_theta} and isocurvature power spectrum \eqref{Eq:Peak_Amp_phi} are related by a factor of $c_6R_{\rm end}^2/\Lambda^2$, as one can also see by comparing their kinetic terms in the Lagrangian \eqref{Lagrangian_phi_R_theta_kinetic_coupling}.
To ensure the perturbativeness of the inflaton non-canonical kinetic term \eqref{def_phi_sigma_couplings}, this factor is much less than one. This means that a small feature correction in the curvature spectrum can be correlated with a large feature correction in the isocurvature spectrum that can be as large as ${\Delta P_\theta}/{P_\theta^{(0)}}\sim {\cal O}(1)$.
This can also be seen by comparing Figure~\ref{fig.spectral_correction_coupling} and \ref{fig:SignalK_zeta_smoothed}.

We also note that, in the curvature spectrum, the sharp-feature signal is more affected by the sharpness of the step than that in the isocurvature spectrum, because the sharpness directly affects the sharp-feature signal in the inflaton sector, but only indirectly affect the sharp-feature signal in the $\theta$ component by exciting the $R$ oscillation. This can be seen in Figure~\ref{fig:SignalK_zeta_smoothed} by comparing results with different values of the sharpness parameter $b$. For very sharp steps, the sharp-feature signal is very extensive in the $k$-space and becomes entangled with the clock signal. 

Even with the direct coupling, the cross-correlation between $\delta\theta$ and $\delta\phi$ needs to be mediated by $\delta R$ and the $\delta\theta$-$\delta R$ bilinear coupling is proportional to the background $\dot\theta$ which vanishes. So, the curvature-isocurvature cross spectrum vanishes in this model.

\bigbreak\noindent
$\bullet$ {\it Summary.} 
To concisely summarize the main conclusions of Section~\ref{Sec_inflaton_couplings}: 
After introducing a direct, kinetic, coupling between the inflaton and the AD field allowed by effective field theories, a sharp feature in the inflaton sector can directly trigger the oscillation of the AD field in the radial direction with more flexible and larger amplitudes than those in the gravitational-coupling-only scenario. A related consequence of such a direct coupling is that the clock signal component, which carries the direct information of the massive radial field, can now show up not only in the isocurvature spectrum but also in the curvature spectrum. Moreover, the amplitudes of these primordial feature signals can be naturally much larger than those in the gravitational case. For example, the amplitude of the oscillatory feature signal can even become comparable to that of the non-oscillating component in the isocurvature spectrum, while keeping its correlated component in the curvature spectrum small and satisfy the current observational constraints.

The correlated presence of full standard clock signals in both the isocurvature power spectrum and the curvature power spectrum would be direct evidence of the structure of the AD field operating in the baryogenesis sector during inflation. They would reveal the value of the radial mass $M_R=m_R/H_I$ and, together with the signal amplitudes and the tilts of the non-oscillatory components, provide constraints to the parameters of the flat-direction model, as well as the strength of the direct coupling between the inflaton and AD field.

\section{Summary and discussions}\label{Sec. conclusion}
We have revisited the mechanism of Affleck-Dine (AD) baryogenesis from the flat-direction potential \eqref{def_UFD} for a charged scalar $\sigma$ (dubbed the AD field) motivated by a supersymmetric extension of the Standard Model with initial conditions developed during the epoch of inflation. Based on the instantaneous reheating assumption, we have studied the relation between the final baryon density isocurvature (BDI) perturbation and the initial quantum fluctuations in the AD field during inflation using both the separated universe approach (Section~\ref{Sec_Separated_Universe}) and the linear perturbation theory (Section~\ref{Sec_Linear_Perturbation}). The numerical relation allows us to update the viable parameter space for the flat-direction model by the recent constraints on the primordial power spectra.

The main goal of this paper is to explore how the physics of this baryogenesis mechanism, which operates at the inflationary energy scale, may be directly probed by cosmological observables. We have shown that the properties of AD scalar field may be probed by a variety of observables imprinted during inflation, in particular, those involving primordial features.

If the curvature perturbation is approximately scale-invariant (no primordial feature), the angular mass $m_\theta$ of the AD scalar condensate can be probed by the tilt of the amplitude of the BDI power spectrum. 
The angular mass $m_\theta$ of the AD scalar condensate given by \eqref{def_m_theta} during inflation can be probed by the overall blue tilt of the BDI power spectrum index in $P_{\rm BDI} \sim k^{3-2\nu_\theta}$ with $\nu_\theta = (9/4 - m_\theta^2/H_I^2)^{1/2}$ (see the upper panel of Figure~\ref{fig.spectral_correction_coupling}). The angular mass is mainly governed by the $c_A$ parameter and the mass range $0 \leq m_\theta/H_I < 3/2$ must be sufficiently light or else the isocurvature perturbations decay away by the end of inflation.

More importantly, we have shown that the presence of primordial feature signals in the density perturbations may allow additional properties of this baryogenesis mechanism to be probed. Such feature signals originate from sudden changes in the inflaton sector. Through gravitational or direct couplings between the inflaton and AD sectors, these changes can excite classical oscillations of the massive component of the AD field. The resulting oscillations generate a characteristic clock signal, providing an observable signature of this massive field, whose mass is much larger than the Hubble scale and would otherwise be difficult to probe.

First, there exists a minimal signal in the BDI power spectrum induced by sharp feature effects in the background Hubble parameter. However, the amplitude of this signal is typically very small, amounting only to a tiny correction to the non-oscillatory component of the BDI power spectrum.

Next, we have considered an example of direct couplings between the inflaton and the AD field within an effective field theory framework. We have shown that such couplings can induce significantly larger correlated effects in both curvature and isocurvature perturbations. The resulting observables include both sharp feature signals and clock signals. The latter provide a direct signature of the massive radial component of the AD field and indicate the presence of a local potential minimum away from the field origin during inflation. This serves as an observational signature of the potential shape following the lifting of the flat direction by supersymmetry-breaking terms during inflation.
The frequency of oscillations in the wave package of the clock signal components (see Figure~\ref{fig.spectral_correction_coupling}, \ref{fig.spectral_correction_coupling_MR}, \ref{fig:SignalK_zeta_smoothed}) can reveal the radial mass $m_R/H_I > 1$ of the flat-direction potential given by \eqref{def_m_R}. The radial mass must be larger than the Hubble scale to prevent $U(1)$ from being restored by high-energy quantum fluctuations during inflation, and it is mainly controlled by the $n$, $\xi$, $c_A$ parameters of the flat direction. 

There are several directions in which this work can be extended or modified.

We have used a simple example to illustrate how primordial features can be used to probe mechanisms of baryogenesis. Certain details may change if we study a different parameter space.
For example, in this model, we assumed (1) sub-dominance of kinetic couplings with inflaton in the flat-direction potential and (2) instantaneous reheating. Both conditions ensure the conventional initial conditions at the end of inflation and the familiar post-inflationary relaxation of the AD field for baryogenesis. These conditions may be violated without affecting the viability of the generation of baryon asymmetry. It is just that parameters of the non-canonical kinetic couplings (such as $c_6$) will become important in the determination of the final baryon asymmetry, which demands extended analysis of the model in a generalized framework. In fact, a finite duration of the preheating phase led by inflaton oscillations is considered in \cite{Dine:1995kz}, where the fractional energy density of the AD field over the inflaton, $\rho_{\rm AD}/\rho_I$, can maintain a constant value in such an effective matter domination era. If the non-canonical coupling \eqref{def_phi_sigma_couplings} exists, it is likely that such a coupling will suddenly dominate the flat-direction potential around the end of inflation since the background energy is largely transited to the kinetic term of inflaton during the coherent oscillation phase. In this case, the post-inflationary evolution of the AD field can be drastically changed due to the dynamical transition of inflaton \cite{Wu:2019ohx}. The sign of the effective mass $m_{c_6}^2$ in the potential in the flat-direction also plays an important role in the post-inflationary evolution. We leave the investigation of these interesting scenarios to future efforts.

We only studied the imprints of primordial features and the baryogenesis model in the power spectra of the density perturbations. These signals should also have correlated counterparts in primordial non-Gaussianities, as generally demonstrated in the inflationary context~\cite{Chen:2006xjb,Kawasaki:2008jy,Chen:2008wn,Adshead:2011jq,Hazra:2012yn, Bartolo:2013exa,Fergusson:2014hya,Fergusson:2014tza}. In addition, non-Gaussian correlation functions are also known to be able to encode a rich set of information about particle spectra and interactions during inflation, as inflation effectively acts as a “cosmological collider” operating at the inflationary Hubble scale~\cite{Chen:2009we,Chen:2009zp,Baumann:2011nk,Noumi:2012vr,Arkani-Hamed:2015bza}. (See e.g.~\cite{Chen:2016uwp, Kumar:2018jxz, Hook:2019vcn, Wang:2019gbi, Liu:2019fag, Li:2020xwr, Cui:2021iie, Bodas:2024hih, Jiang:2025mlm,Fujikura:2025xgl} for applications probing particle physics, and~\cite{Cabass:2024wob, Sohn:2024xzd, Anbajagane:2025uro, Suman:2025vuf, Suman:2025tpv} for recent observational constraints.)
If primordial features, such as the one studied in this paper, are present during inflation, they would inject extra energy to the cosmological collider. The amount of this extra energy is determined by the sharpness of the primordial features and could be orders of magnitude larger than $H$. Consequently, primordial features can not only excite classical oscillations of massive fields and leave their direct signatures and their mass parameter information in terms of clock signals \cite{Chen:2011zf,Chen:2011tu,Chen:2012ja, Saito:2012pd,Battefeld:2013xka,Saito:2013aqa, Gao:2013ota,Noumi:2013cfa, Chen:2014joa,Chen:2014cwa,Huang:2016quc,Domenech:2018bnf, Braglia:2021ckn,Braglia:2021sun,Braglia:2021rej, Bodas:2022zca, Chen:2023txq, Co:2024cji, Quintin:2024boj} (as we have utilized in this work), but also quantum-mechanically excite other particles, including UV degrees of freedom with mass much larger than $H$ (which are not studied in this work), and leave their imprints in shapes of primordial non-Gaussianities as in “classical cosmological collider physics” \cite{Chen:2022vzh,Pajer:2024ckd,Wang:2025qww,Jazayeri:2025vlv}. These can be interesting directions to explore in the context of the baryogenesis models.

Observationally, it is interesting to constrain isocurvature modes that contain oscillatory features. We note that observable signatures of sharp features or oscillating massive fields may also arise in CDI within some dark-matter models \cite{Chen:2023txq, Co:2024cji}. 
Because of the degeneracy between BDI and CDI, the CMB is sensitive only to their combined contribution, namely the total matter density isocurvature (MDI).
Although there can be differences in their signals in MDI between a specific dark-matter model and a specific baryogenesis model, these differences may be model-dependent. In general, distinguishing similar types of BDI and CDI signals in MDI remains an important observational challenge.

\acknowledgments
The authors thank Priyesh Chakraborty, Gongjun Choi, Kin-Wang Ng and Fuminobu Takahashi for many helpful discussions and comments. Y.-P. Wu acknowledges the hospitality of the Harvard-Smithsonian Center for Astrophysics during the development of this project. The project has received funding from the National Science and Technology Council thematic research program under grant agreement No. NSTC 114-2811-M-001-049.

\appendix
\section{The Hubble-induced coefficients}\label{Appendix_Hubble_coefficients}
The Hubble-induced terms in the flat-direction effective potential \eqref{def_UFD} can arise from non-canonical couplings between the inflaton sector $I$ and the AD field $\sigma$ in the K\"{a}hler potential or higher-order interactions in the superpotential. To make our discussion concrete, let us specify a common origin of these Hubble-induced terms led by the coefficients $c_H$ and $c_A$ based on the framework in supergravity considered in \cite{Dine:1995kz,Dine:1995uk}.

To see the origin of $c_H$, we first consider the presence of a quartic coupling between $I$ and $\sigma$ in the K\"{a}hler potential given by:
\begin{align}
	K = I\bar{I} + \sigma \bar{\sigma} + \frac{c_I}{M_P^2}  I\bar{I}  \sigma \bar{\sigma}, 
\end{align}
where the $c_I$ term also gives rise to non-canonical kinetic terms as investigated in Section~\ref{Sec_SUSY_breaking}. If the inflaton potential arising from $F$ terms dominates the background energy density during inflation, the superpotential can be expressed as
\begin{align}
	W = W_0(I) + W_{\rm AD}(I, \sigma),
\end{align}
with $W_{\rm AD} \ll W_0$. The $F$ term structure implies $W_0 = c_0 H M_P^2$ and $\partial_IW_0 = c_1 H M_P$, where $c_0$ and $c_1$ are some constants. However, $c_0$ and $c_1$ are not independent, as they are related by the background energy density through $F$ terms of the supergravity scalar potential \cite{Dine:1995kz,Dine:1995uk}:
\begin{align}
	V_F = e^{K/M_P^2} \left(g^{i\bar{j}}D_iWD_{\bar{j}}W^\ast - 3\frac{\vert W\vert^2}{M_P^2}\right),
\end{align}
where $D_i W = \partial_i W + K_i W/M_P^2$ is the K\"{a}hler derivative and $K_i \equiv \partial_i K$. The supergravity corrections are significantly simplified in the limit with $\vert I\vert/M_P \ll 1$, where $e^{K/M_P^2} \approx 1+ \vert\sigma\vert^2/M_P^2$. In this limit we find $V_{F0} = \vert \partial_I W_0\vert^2 -3\vert W_0\vert^2/M_P^2 \approx 3M_P^2H^2$, which indicates $\vert c_1\vert^2 - 3\vert c_0\vert^2 = 3$ with $\vert c_1\vert \sim \mathcal{O}(1)$ and $\vert c_0 \vert \ll 1$.

Working in the small-filed limit with $\vert I\vert/M_P \ll 1$ for simplicity, we can observe leading contribution to the AD field mass in the scalar potential comes from 
\begin{align}\label{Hubble_mass_cI}
	e^{K/M_P^2}V_{F0}, \quad g^{I\bar{I}} \vert D_IW\vert^2 \supset -c_I \frac{\vert\sigma\vert^2}{M_P^2} \vert\partial_IW_0\vert^2, \quad
	g^{\sigma\bar{\sigma}}\vert D_\sigma W\vert^2 \supset \frac{\vert W_0\vert^2}{M_P^4} \vert\sigma\vert^2,
\end{align}
and the collection of these terms gives
\begin{align}\label{def_cH}
	V_{F} \supset \left((1-c_I)\vert c_1\vert^2 - 2\vert c_0\vert^2\right) H^2\vert\sigma\vert^2 \equiv c_H H^2\vert\sigma\vert^2. 
\end{align}
In the limit of $c_0 =0$, we find $c_H = 3(1-c_I)$, and $c_I \gg 1$ can achieve $\xi = - c_H \approx 3c_I \gg 0$, which is desired for AD baryogenesis. 

In general, the Hubble-induced $c_H$ term is the summation of all contributions from supersymmetry-breaking sectors that couple with the AD field. We provide an example of the generalized scenario in Section~\ref{Sec_SUSY_breaking}. For the generation of resonant clock signals with primordial features, we consider the general origin of $c_H$.

On the other hand, in order to see the origin of the $c_A$ coefficient, we can specify the subdominant part of the superpotential as
\begin{align}
	W_{\rm AD}(I,\sigma) = \frac{\lambda_0}{n}\frac{\sigma^n}{\Lambda^{n-3}}  + \frac{\lambda_1}{n} \frac{I}{M_P} \frac{\sigma^n}{\Lambda^{n-3}},
\end{align}
with free parameters $\lambda_0$ and $\lambda_1$. In the small-filed limit with $\vert I\vert/M_P \ll 1$, we drop all terms that are suppress by the ratio of $I/M_P$. Collecting all terms at the order of $\sigma^n/\Lambda^{n-3}$, we find
\begin{align}
	\nonumber
	 g^{I\bar{I}} \vert D_IW\vert^2 &\supset \frac{\lambda_1}{n M_P}\frac{\sigma^n}{\Lambda^{n-3}} \partial_{\bar{I}}W_0^\ast +h.c., \\\nonumber
	  g^{\sigma\bar{\sigma}} \vert D_\sigma W\vert^2 &\supset \lambda_0\frac{\sigma^n}{\Lambda^{n-3}} \frac{W_0^\ast}{M_P^2} + h.c., \\
	  -3\frac{\vert W\vert^2}{M_P^2} &\supset -3\frac{\lambda_0}{n}\frac{\sigma^n}{\Lambda^{n-3}} \frac{W_0^\ast}{M_P^2} + h.c.,
\end{align}
and therefore
\begin{align}
	V_F \supset \left[c_0^\ast\lambda_0 \left(1-\frac{3}{n}\right) +c_1^\ast \frac{\lambda_1}{n}\right] \frac{H}{\Lambda^{n-3}} \sigma^n 
	\equiv c_A \frac{\lambda}{n}  \frac{H}{\Lambda^{n-3}} \sigma^n.
\end{align}
Note that a cubic coupling in the K\"{a}hler potential $K \sim I \vert\sigma\vert^2/M_P$ can also lead to non-trivial contribution to the Hubble-induced $c_A$ term.

Since the Hubble-induced $c_A$ and $c_H$ terms originate from coupling to the inflaton superpotential $W_0(I)$, after inflaton decays, we assume they immediately go to zero  $c_H\rightarrow 0$, $c_A \rightarrow 0$ in the post-inflationary epochs. We have also checked another limit in which $c_A$ and $c_H$ remain constants after inflation, and we get qualitatively similar results in baryon density isocurvature perturbations.

\section{Unwanted minima in radiation domination}\label{Appendix_unwanted_minima}
\begin{figure}[]
	\begin{center}
		\includegraphics[width=7.5 cm]{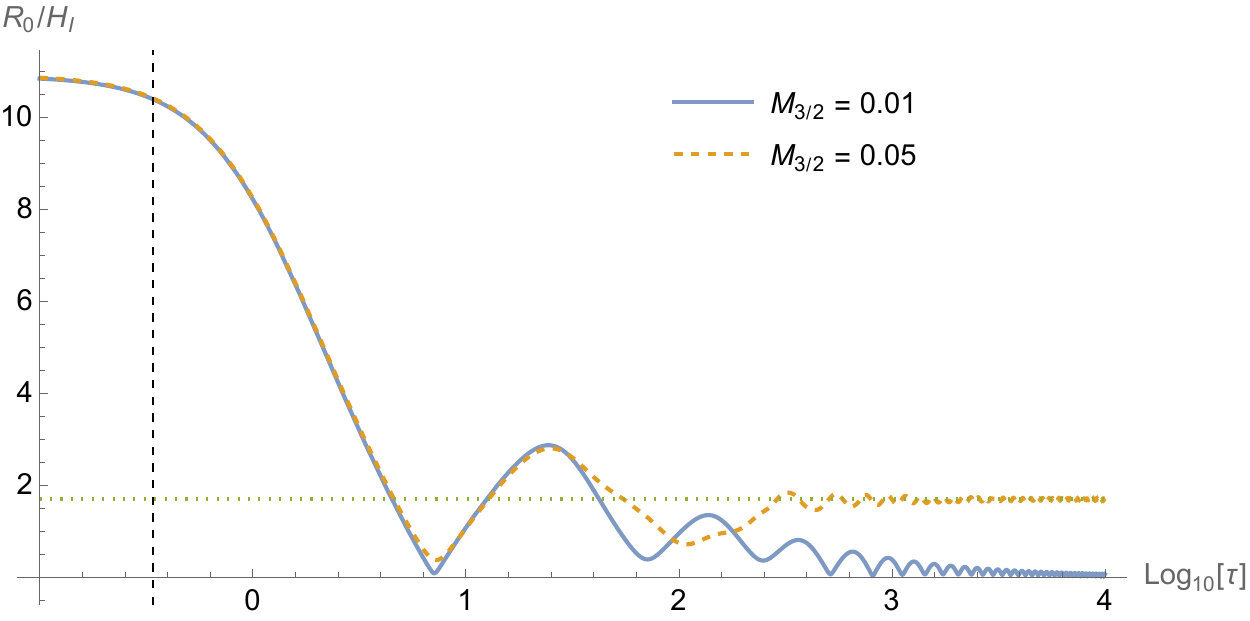}
		\hfill
		\includegraphics[width=7.5 cm]{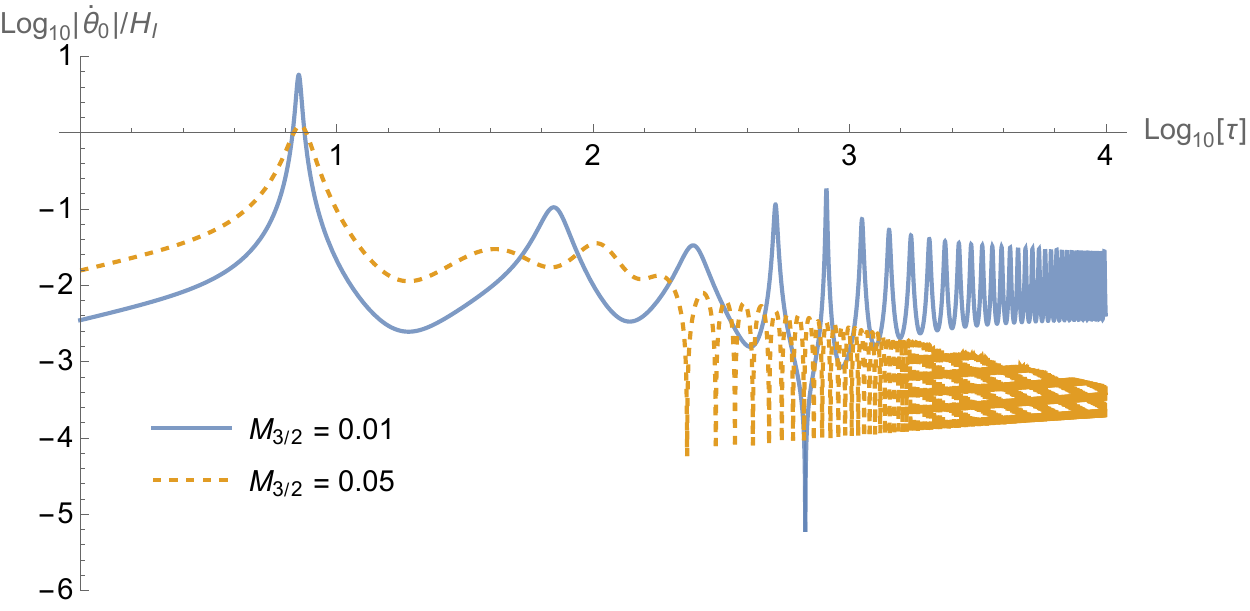}
	\end{center}
	\caption{\label{fig.unwanted_minima} The evolution of $R_0$ (left panel) and $\dot{\theta}_0$ (right panel) in instantaneous reheating with $c_H = c_A = 0$. Model parameters with $n = 4$, $\Lambda/H_I = 100$, $m_\sigma/H_I = 0.01$, $\xi = \lambda = A = 1$, $c_A = 0.1$, $\delta = \pi/2$ are used and $H_I = 10^{11}$ GeV. The radial mode relaxes to the conventional (unwanted) minimum at $R_0 = 0$ ($R_0 > 0$) in the case with $M_{3/2} \equiv m_{3/2}/H_I = 0.01$ ($M_{3/2} = 0.05$), respectively. The case with $M_{3/2} = 0.05$ violates the condition \eqref{condition_unwanted_VEV}. In the left panel, the dotted line is given by \eqref{Xmin_unwanted_VEV} and the vertical dashed line is $\tau = \tau_{\rm rex}$ given by \eqref{def_tau_rex}.}
\end{figure}
\begin{figure}[]
	\begin{center}
		\includegraphics[width=10 cm]{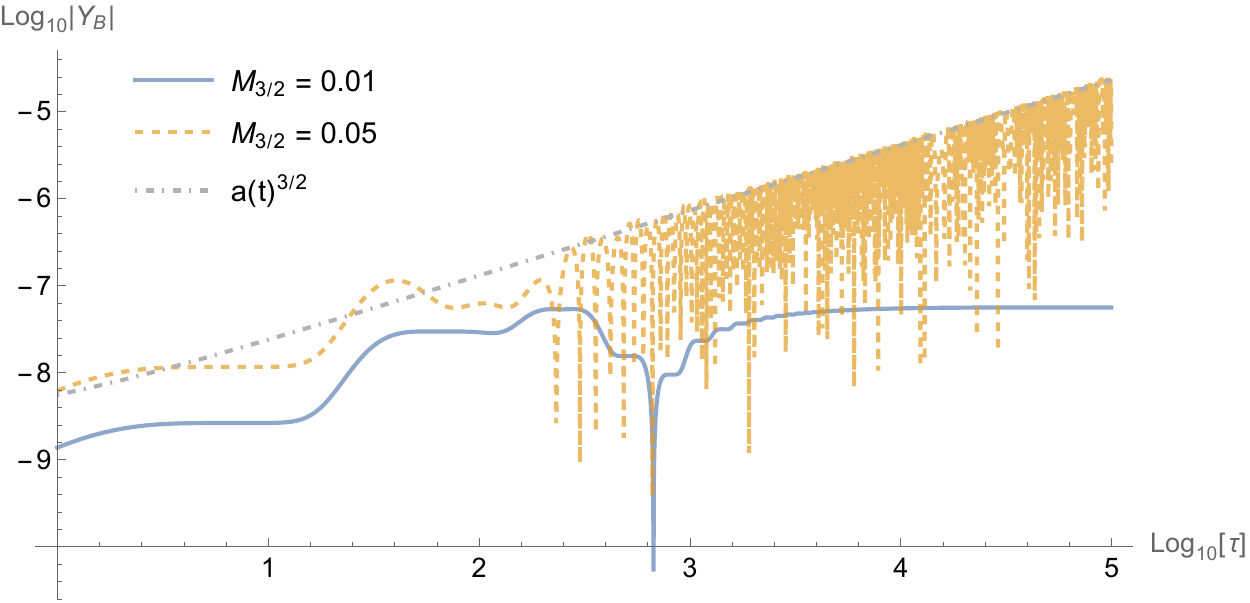}
	\end{center}
	\caption{\label{fig.unwanted_minima_YB} The corresponding baryon asymmetry $Y_B$ of Figure~\ref{fig.unwanted_minima}. The case with $M_{3/2} = 0.05$ violates the condition \eqref{condition_unwanted_VEV}.}
\end{figure}
	
If low-energy soft terms arise from supersymmetry breaking of some hidden sectors, both $m_\sigma$ and $m_{3/2}$ in \eqref{def_UFD} should be of the order of the electroweak scale. This is the reason why $m_\sigma \sim m_{3/2}$ is considered in \cite{Dine:1995kz} for the late-stage evolution of the AD field when $t > t_{3/2}$. However, it is notable that if $m_{3/2}$ is comparable to $m_\sigma$, the $B$ violating $A$ term may develop new potential minima away from the radial origin ($R = 0$) at later times after the decay of the $\xi$ term. Once the AD field is trapped inside one of these new minima at $R > 0$, the final baryon asymmetry is no longer conserved and will grow unstably.

To avoid the presence of these unfamiliar minima led by the $A$ term, we require
\begin{align}\label{condition_unwanted_VEV}
	m_\sigma > \frac{A}{2\sqrt{n-1}} m_{3/2}.
\end{align}
This condition ensures that the late-time solution of \eqref{def_Xmin} is imaginary so that non-trivial solutions of $R_{\rm min}$ given by \eqref{def_Rmin} do not exist.

To see this instability, it is more convenient to use the zero-temperature potential with $\Delta U =0$ for demanstration. Such an instablity developed by the low-energy $A$ term generally exist even with finite-temperature effects since $\Delta U$ eventually decays after sufficiently long time.
As shown in Figure~\ref{fig.unwanted_minima}, if the condition \eqref{condition_unwanted_VEV} is violated, the AD field may relax to one of these additional minima at $\cos(n\theta + \delta) = -1$ such that $\alpha \approx A m_{3/2}/H$ in the late-time limit. With the decay of the $\xi$ and $c_A$ terms in radiation domination, the radial VEV  given by \eqref{def_Rmin} has a time dependence 
\begin{align}\label{Xmin_unwanted_VEV}
		X_{\rm min}H(t) \approx \left[\frac{A^2 m_{3/2}^2}{4(n-1)^2} -\frac{m_\sigma^2}{n -1} \right]^{1/2} + \frac{A m_{3/2}}{2(n-1)},
\end{align}
which approaches a constant value. This is a non-trivial VEV at $R_{\rm min} > 0$. 

If the AD field gets stuck in one of the minima created by the $A$ term, the radial mode becomes a constant with $R_0 \sim a^0$ while the angular mode oscillates as a massive scalar with $\dot{\theta}_0 \sim a^{-3/2}$ (see the right panel of Figure~\ref{fig.unwanted_minima}). The angular momentum conservation $\partial_t (a^3R_0^2\dot{\theta}_0) \approx 0$ then gives $Y_B \sim a^3 R_0^2 \dot{\theta}_0 \sim a^{3/2} \sim t^{3/4}$, as shown by the dot-dashed line in Figure~\ref{fig.unwanted_minima_YB}.

\section{Supplementary materials in smoothed step case}\label{Append_Gaussian_step}
In this appendix, we present the equations of motions with the smoothed version of the step potential, \eqref{Eq:Smoothed_potential} and \eqref{Gaussian_smeared_step_function}. These are used in numerical calculations.

Neglecting the gravitational modulation of the background Hubble parameter, the inflaton motion induced by the smoothed step feature satisfies
\begin{align}
    \phi_1^{\prime\prime} +3\phi_1^{\prime}+ \frac{B}{b\sqrt{\pi}} \frac{\tilde{V}_0}{H_I\dot{\phi}_0}e^{-(N-N_\ast)^2/b^2} =0, \quad \tilde{V}_0 \equiv V_0\left(1+c_6\frac{R_{\rm end}^2}{\Lambda^2}\right)^{-1},
\end{align}
where $\phi^\prime \equiv \partial_N\phi$ is the derivative with respect to the $e$-folding number $N$. The solution that matches the condition $\phi_1(-\infty) = \phi_1^\prime(-\infty) = 0$ reads
\begin{align}
    \phi_1(N) &= \frac{B\tilde{V}_0}{6H_I\dot{\phi}_0} \left\{\text{erfc}\left(\frac{N-N_\ast}{b}\right) -2 +e^{\frac{9b^2}{4}-3(N-N_\ast)} \text{erfc}\left(\frac{3b}{2}-\frac{N-N_\ast}{b}\right)\right\},\\
    \phi_1^\prime(N) &= \frac{B\tilde{V}_0}{6H_I\dot{\phi}_0} \left\{2e^{\frac{9b^2}{4}-3(N-N_\ast)}\delta_b\left(3b^2/2-N+N_\ast\right)  \right.\\\nonumber
    &\left. \qquad\qquad\qquad-2\delta_b(N-N_\ast)-3e^{\frac{9b^2}{4}-3(N-N_\ast)}\text{erfc}\left(\frac{3b}{2}-\frac{N-N_\ast}{b}\right)\right\},
\end{align}
where $\text{erfc}(x) = 1-\text{erf}(x)$ is the complementary error function.

\begin{figure}
    \centering
    \includegraphics[width=0.49\linewidth]{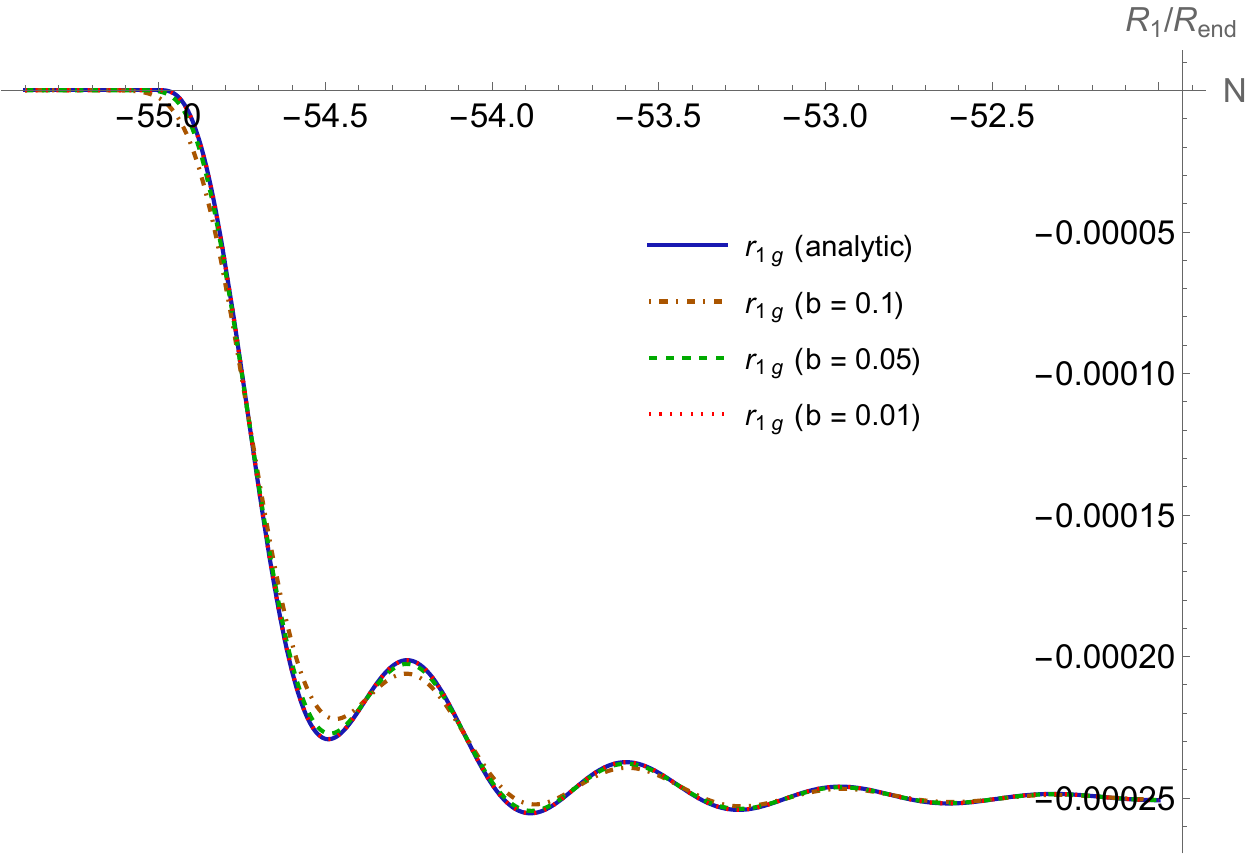}
    \includegraphics[width=0.49\linewidth]{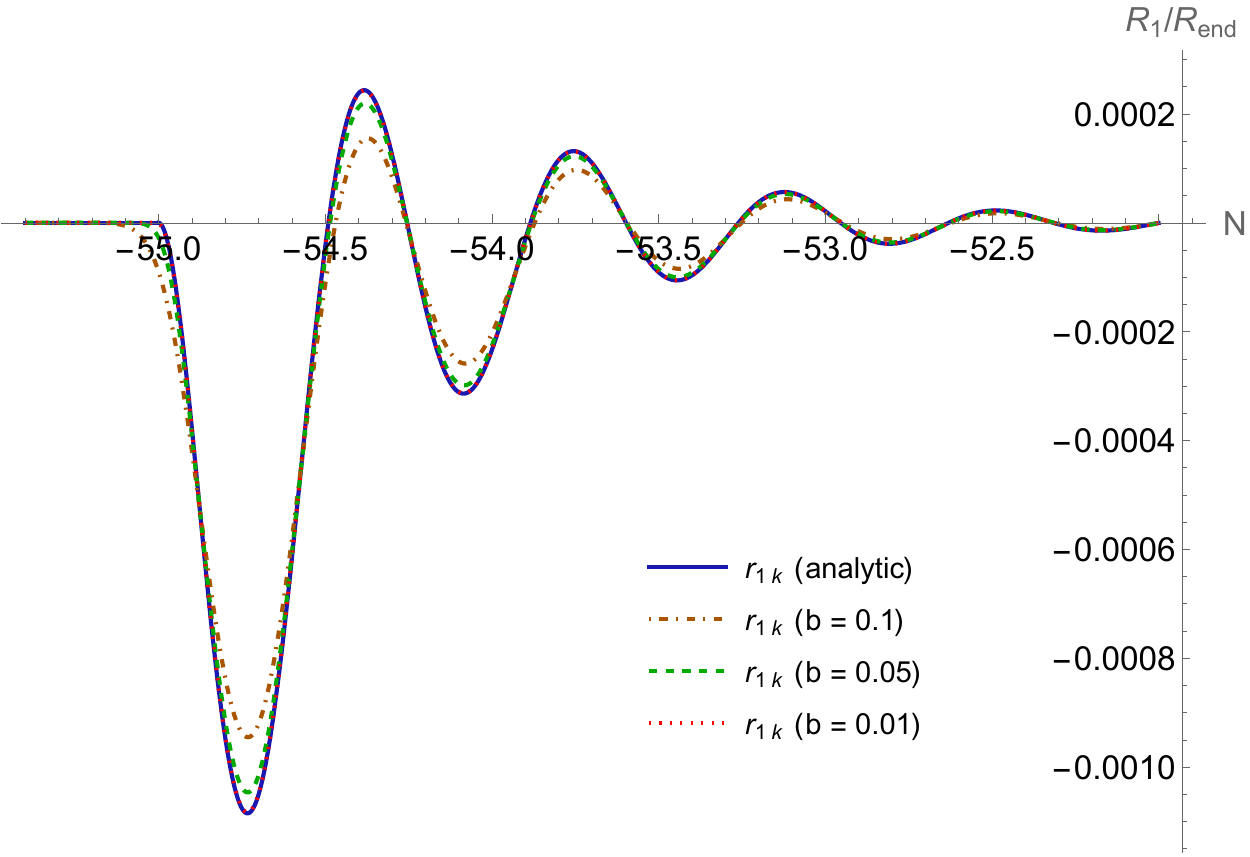}
    \caption{Radial oscillations induced by the smoothed step feature through the gravitational coupling (right panel) and the kinetic coupling with inflaton (left panel), where $r_{\rm 1g} \equiv R_{\rm 1g}/R_{\rm end}$ and $r_{\rm 1k} \equiv R_{\rm 1k}/R_{\rm end}$, respectively. Numerical results with various choices of the sharpness parameter $b$ are compared with the analytic solutions \eqref{R1_gravitational} and \eqref{sol_R1_step} based on the Heaviside step feature. We use $n = 4$, $M_R = 10$, and $B = 10^{-3}$ in the left panel, and $n = 4$, $M_R = 10$, $B \times M_P^2/\Lambda^2 = -0.1$ with $c_6 = 1/6$ for the kinetic coupling effect in the right panel. }
    \label{fig:R1k_smoothed}
\end{figure}

In this case, the background Hubble modulation \eqref{def_Delta_H} becomes
\begin{align}\label{Delta_H_smoothed}
    \frac{\Delta H}{H_I} =& \frac{B}{12}\left\{2e^{\frac{9b^2}{4}-3(N-N_\ast)}\delta_b\left(3b^2/2-N+N_\ast\right)-2\delta_b(N-N_\ast)  \right.\\\nonumber
    &\left. \qquad-3e^{\frac{9b^2}{4}-3(N-N_\ast)}\text{erfc}\left(\frac{3b}{2}-\frac{N-N_\ast}{b}\right)\right\} +\frac{B}{2}\Theta_b(N-N_\ast).
\end{align}
This leads to a smoothed radial VEV modulation, $\Delta R_{\rm end}/R_{\rm end} = \frac{\Delta H}{(n-2)H_I}$, during inflation, which is the source term of the radial mode oscillation induced by gravitational couplings. One can solve \eqref{eom_R1g} to obtain the radial motion $R_{\rm 1g}$ driven by the smoothed step feature. Results of the smoothed radial motion from gravitational effect can be found in the left panel of Figure~\ref{fig:R1k_smoothed}.

We now consider the feature induced signals from the kinetic coupling. The radial oscillation driven by the kinetic-coupling induced mass term $\Delta M_{c_6} = -2c_6 \dot{\phi}_0 \phi_1^\prime/\Lambda^2H_I$ is governed by the equation
\begin{align}\label{eom_r1k_smoothed}
    r_{1k}^{\prime\prime} + 3r_{1k}^\prime +M_R^2 r_{1k} =& c_6B\frac{ M_P^2}{\Lambda^2}\left\{-2e^{\frac{9b^2}{4}-3(N-N_\ast)}\delta_b\left(3b^2/2-N+N_\ast\right)\right. \\\nonumber
    &\left.  +2\delta_b(N-N_\ast)+3e^{\frac{9b^2}{4}-3(N-N_\ast)}\text{erfc}\left(\frac{3b}{2}-\frac{N-N_\ast}{b}\right) \right\},
\end{align}
where $r_{1k} \equiv R_{1k}/R_{\rm end}$
We compare the numerical solutions of $R_{1k}$ from the smoothed step feature with various choices of the sharpness value $b$ with the analytic solution \eqref{sol_R1_step}, which corresponds to the $b\rightarrow 0$ limit, in Figure~\ref{fig:R1k_smoothed}. We find that the results of the Heaviside step feature can be well reproduced with the choice of $b = 0.01$ or smaller.

In terms of the numerical solutions of $R_{1k}$ from the smoothed step feature, we can solve the inflaton perturbation $\delta\phi$ via \eqref{eom_dphi_N} with the Gaussian smoothed effective mass term $\Delta M_\phi$ to obtain the feature induced signals in the power spectrum of the curvature perturbation. We can neglect the contributions from gravitational couplings with a sufficiently large direct coupling. Results in Figure~\ref{fig:SignalK_zeta_smoothed} indicate that, comparing to the infinitely sharp step, a smoothed step feature reduces the amplitudes of both the sharp-feature signal from $\Delta M_\phi$ and the resonant clock signal driven by the kinetic coupling. The full signal can be a superposition of the two with a sufficiently large coupling strength $6c_6M_P^2/\Lambda^2 \gg 1$.

\bibliographystyle{JHEP}
\bibliography{draft_2025Sep}

@article{Affleck:1984fy,
    author = "Affleck, Ian and Dine, Michael",
    title = "{A New Mechanism for Baryogenesis}",
    reportNumber = "Print-84-0574 (PRINCETON)",
    doi = "10.1016/0550-3213(85)90021-5",
    journal = "Nucl. Phys. B",
    volume = "249",
    pages = "361--380",
    year = "1985"
}

@article{Dine:1995kz,
    author = "Dine, Michael and Randall, Lisa and Thomas, Scott D.",
    title = "{Baryogenesis from flat directions of the supersymmetric standard model}",
    eprint = "hep-ph/9507453",
    archivePrefix = "arXiv",
    reportNumber = "SLAC-PUB-6846, SLAC-PUB-95-6846, MIT-CTP-2441, SCIPP-95-20",
    doi = "10.1016/0550-3213(95)00538-2",
    journal = "Nucl. Phys. B",
    volume = "458",
    pages = "291--326",
    year = "1996"
}

@article{Dine:2003ax,
    author = "Dine, Michael and Kusenko, Alexander",
    title = "{The Origin of the matter - antimatter asymmetry}",
    eprint = "hep-ph/0303065",
    archivePrefix = "arXiv",
    reportNumber = "SCIPP-2003-08, UCLA-03-TEP-08",
    doi = "10.1103/RevModPhys.76.1",
    journal = "Rev. Mod. Phys.",
    volume = "76",
    pages = "1",
    year = "2003"
}

@article{Bell:2011tn,
    author = "Bell, Nicole F. and Petraki, Kalliopi and Shoemaker, Ian M. and Volkas, Raymond R.",
    title = "{Pangenesis in a Baryon-Symmetric Universe: Dark and Visible Matter via the Affleck-Dine Mechanism}",
    eprint = "1105.3730",
    archivePrefix = "arXiv",
    primaryClass = "hep-ph",
    doi = "10.1103/PhysRevD.84.123505",
    journal = "Phys. Rev. D",
    volume = "84",
    pages = "123505",
    year = "2011"
}

@article{vonHarling:2012yn,
    author = "von Harling, Benedict and Petraki, Kalliopi and Volkas, Raymond R.",
    title = "{Affleck-Dine dynamics and the dark sector of pangenesis}",
    eprint = "1201.2200",
    archivePrefix = "arXiv",
    primaryClass = "hep-ph",
    doi = "10.1088/1475-7516/2012/05/021",
    journal = "JCAP",
    volume = "05",
    pages = "021",
    year = "2012"
}

@article{Petraki:2013wwa,
    author = "Petraki, Kalliopi and Volkas, Raymond R.",
    title = "{Review of asymmetric dark matter}",
    eprint = "1305.4939",
    archivePrefix = "arXiv",
    primaryClass = "hep-ph",
    reportNumber = "NIKHEF-2013-016",
    doi = "10.1142/S0217751X13300287",
    journal = "Int. J. Mod. Phys. A",
    volume = "28",
    pages = "1330028",
    year = "2013"
}

@article{Wu:2021gtd,
    author = "Wu, Yi-Peng and Pinetti, Elena and Silk, Joseph",
    title = "{Cosmic Coincidences of Primordial-Black-Hole Dark Matter}",
    eprint = "2109.09875",
    archivePrefix = "arXiv",
    primaryClass = "astro-ph.CO",
    reportNumber = "FERMILAB-PUB-21-784-T",
    doi = "10.1103/PhysRevLett.128.031102",
    journal = "Phys. Rev. Lett.",
    volume = "128",
    number = "3",
    pages = "031102",
    year = "2022"
}

@article{Wu:2021mwy,
    author = "Wu, Yi-Peng and Pinetti, Elena and Petraki, Kalliopi and Silk, Joseph",
    title = "{Baryogenesis from ultra-slow-roll inflation}",
    eprint = "2109.00118",
    archivePrefix = "arXiv",
    primaryClass = "hep-ph",
    reportNumber = "FERMILAB-PUB-21-783-T",
    doi = "10.1007/JHEP01(2022)015",
    journal = "JHEP",
    volume = "01",
    pages = "015",
    year = "2022"
}

@article{Balaji:2022rsy,
    author = "Balaji, Shyam and Silk, Joseph and Wu, Yi-Peng",
    title = "{Induced gravitational waves from the cosmic coincidence}",
    eprint = "2202.00700",
    archivePrefix = "arXiv",
    primaryClass = "astro-ph.CO",
    doi = "10.1088/1475-7516/2022/06/008",
    journal = "JCAP",
    volume = "06",
    number = "06",
    pages = "008",
    year = "2022"
}

@article{Co:2019jts,
    author = "Co, Raymond T. and Hall, Lawrence J. and Harigaya, Keisuke",
    title = "{Axion Kinetic Misalignment Mechanism}",
    eprint = "1910.14152",
    archivePrefix = "arXiv",
    primaryClass = "hep-ph",
    reportNumber = "LCTP-19-28",
    doi = "10.1103/PhysRevLett.124.251802",
    journal = "Phys. Rev. Lett.",
    volume = "124",
    number = "25",
    pages = "251802",
    year = "2020"
}

@article{Co:2019wyp,
    author = "Co, Raymond T. and Harigaya, Keisuke",
    title = "{Axiogenesis}",
    eprint = "1910.02080",
    archivePrefix = "arXiv",
    primaryClass = "hep-ph",
    reportNumber = "LCTP-19-27",
    doi = "10.1103/PhysRevLett.124.111602",
    journal = "Phys. Rev. Lett.",
    volume = "124",
    number = "11",
    pages = "111602",
    year = "2020"
}

@article{Co:2020jtv,
    author = "Co, Raymond T. and Fernandez, Nicolas and Ghalsasi, Akshay and Hall, Lawrence J. and Harigaya, Keisuke",
    title = "{Lepto-Axiogenesis}",
    eprint = "2006.05687",
    archivePrefix = "arXiv",
    primaryClass = "hep-ph",
    doi = "10.1007/JHEP03(2021)017",
    journal = "JHEP",
    volume = "03",
    pages = "017",
    year = "2021"
}

@article{ATLAS:2024fub,
    author = "Aad, Georges and others",
    collaboration = "ATLAS",
    title = "{Search for electroweak production of supersymmetric particles in final states with two {\ensuremath{\tau}}-leptons in $ \sqrt{s} $ = 13 TeV pp collisions with the ATLAS detector}",
    eprint = "2402.00603",
    archivePrefix = "arXiv",
    primaryClass = "hep-ex",
    reportNumber = "CERN-EP-2023-295",
    doi = "10.1007/JHEP05(2024)150",
    journal = "JHEP",
    volume = "05",
    pages = "150",
    year = "2024"
}

@article{Ghosh:2024joj,
    author = "Ghosh, Kirtiman and Huitu, Katri and Sahu, Rameswar",
    title = "{Revisiting the LHC constraints on gauge-mediated supersymmetry breaking scenarios}",
    eprint = "2411.09650",
    archivePrefix = "arXiv",
    primaryClass = "hep-ph",
    doi = "10.1103/PhysRevD.111.075011",
    journal = "Phys. Rev. D",
    volume = "111",
    number = "7",
    pages = "075011",
    year = "2025"
}

@article{Constantin:2025mex,
    author = "Constantin, L. and Kraml, S. and Mahmoudi, F.",
    title = "{The LHC has ruled out supersymmetry {\textendash} really?}",
    eprint = "2505.11251",
    archivePrefix = "arXiv",
    primaryClass = "hep-ph",
    reportNumber = "CERN-TH-2025-100",
    doi = "10.1016/j.nuclphysb.2025.117012",
    journal = "Nucl. Phys. B",
    volume = "1018",
    pages = "117012",
    year = "2025"
}

@article{Planck:2018jri,
    author = "Akrami, Y. and others",
    collaboration = "Planck",
    title = "{Planck 2018 results. X. Constraints on inflation}",
    eprint = "1807.06211",
    archivePrefix = "arXiv",
    primaryClass = "astro-ph.CO",
    doi = "10.1051/0004-6361/201833887",
    journal = "Astron. Astrophys.",
    volume = "641",
    pages = "A10",
    year = "2020"
}

@article{Chen:2010xka,
    author = "Chen, Xingang",
    title = "{Primordial Non-Gaussianities from Inflation Models}",
    eprint = "1002.1416",
    archivePrefix = "arXiv",
    primaryClass = "astro-ph.CO",
    doi = "10.1155/2010/638979",
    journal = "Adv. Astron.",
    volume = "2010",
    pages = "638979",
    year = "2010"
}

@article{Chluba:2015bqa,
    author = "Chluba, Jens and Hamann, Jan and Patil, Subodh P.",
    title = "{Features and New Physical Scales in Primordial Observables: Theory and Observation}",
    eprint = "1505.01834",
    archivePrefix = "arXiv",
    primaryClass = "astro-ph.CO",
    reportNumber = "CERN-PH-TH-2015-096",
    doi = "10.1142/S0218271815300232",
    journal = "Int. J. Mod. Phys. D",
    volume = "24",
    number = "10",
    pages = "1530023",
    year = "2015"
}

@article{Slosar:2019gvt,
    author = "Slosar, Anze and others",
    title = "{Scratches from the Past: Inflationary Archaeology through Features in the Power Spectrum of Primordial Fluctuations}",
    eprint = "1903.09883",
    archivePrefix = "arXiv",
    primaryClass = "astro-ph.CO",
    journal = "Bull. Am. Astron. Soc.",
    volume = "51",
    number = "3",
    pages = "98",
    year = "2019"
}

@article{Achucarro:2022qrl,
    author = "Ach{\'u}carro, Ana and others",
    title = "{Inflation: Theory and Observations}",
    eprint = "2203.08128",
    archivePrefix = "arXiv",
    primaryClass = "astro-ph.CO",
    month = "3",
    year = "2022"
}

@article{Peiris:2003ff,
    author = "Peiris, H. V. and others",
    collaboration = "WMAP",
    title = "{First year Wilkinson Microwave Anisotropy Probe (WMAP) observations: Implications for inflation}",
    eprint = "astro-ph/0302225",
    archivePrefix = "arXiv",
    doi = "10.1086/377228",
    journal = "Astrophys. J. Suppl.",
    volume = "148",
    pages = "213--231",
    year = "2003"
}

@article{Akrami:2018odb,
    author = "Akrami, Y. and others",
    collaboration = "Planck",
    title = "{Planck 2018 results. X. Constraints on inflation}",
    eprint = "1807.06211",
    archivePrefix = "arXiv",
    primaryClass = "astro-ph.CO",
    doi = "10.1051/0004-6361/201833887",
    journal = "Astron. Astrophys.",
    volume = "641",
    pages = "A10",
    year = "2020"
}

@article{Adams:2001vc,
    author = "Adams, Jennifer A. and Cresswell, Bevan and Easther, Richard",
    title = "{Inflationary perturbations from a potential with a step}",
    eprint = "astro-ph/0102236",
    archivePrefix = "arXiv",
    reportNumber = "CU-TP-1005",
    doi = "10.1103/PhysRevD.64.123514",
    journal = "Phys. Rev. D",
    volume = "64",
    pages = "123514",
    year = "2001"
}

@article{Bean:2008na,
    author = "Bean, Rachel and Chen, Xingang and Hailu, Girma and Tye, S. -H. Henry and Xu, Jiajun",
    title = "{Duality Cascade in Brane Inflation}",
    eprint = "0802.0491",
    archivePrefix = "arXiv",
    primaryClass = "hep-th",
    reportNumber = "MIT-CTP-3921, CAS-KITPC-ITP-046",
    doi = "10.1088/1475-7516/2008/03/026",
    journal = "JCAP",
    volume = "03",
    pages = "026",
    year = "2008"
}

@article{Mortonson:2009qv,
    author = "Mortonson, Michael J. and Dvorkin, Cora and Peiris, Hiranya V. and Hu, Wayne",
    title = "{CMB polarization features from inflation versus reionization}",
    eprint = "0903.4920",
    archivePrefix = "arXiv",
    primaryClass = "astro-ph.CO",
    doi = "10.1103/PhysRevD.79.103519",
    journal = "Phys. Rev. D",
    volume = "79",
    pages = "103519",
    year = "2009"
}

@article{Hazra:2010ve,
    author = "Hazra, Dhiraj Kumar and Aich, Moumita and Jain, Rajeev Kumar and Sriramkumar, L. and Souradeep, Tarun",
    title = "{Primordial features due to a step in the inflaton potential}",
    eprint = "1005.2175",
    archivePrefix = "arXiv",
    primaryClass = "astro-ph.CO",
    doi = "10.1088/1475-7516/2010/10/008",
    journal = "JCAP",
    volume = "10",
    pages = "008",
    year = "2010"
}

@article{Hazra:2014goa,
    author = "Hazra, Dhiraj Kumar and Shafieloo, Arman and Smoot, George F. and Starobinsky, Alexei A.",
    title = "{Wiggly Whipped Inflation}",
    eprint = "1405.2012",
    archivePrefix = "arXiv",
    primaryClass = "astro-ph.CO",
    doi = "10.1088/1475-7516/2014/08/048",
    journal = "JCAP",
    volume = "08",
    pages = "048",
    year = "2014"
}

@article{Miranda:2014fwa,
    author = "Miranda, Vin{\'\i}cius and Hu, Wayne and Dvorkin, Cora",
    title = "{Polarization Predictions for Inflationary CMB Power Spectrum Features}",
    eprint = "1411.5956",
    archivePrefix = "arXiv",
    primaryClass = "astro-ph.CO",
    doi = "10.1103/PhysRevD.91.063514",
    journal = "Phys. Rev. D",
    volume = "91",
    number = "6",
    pages = "063514",
    year = "2015"
}

@article{Hazra:2014jwa,
    author = "Hazra, Dhiraj Kumar and Shafieloo, Arman and Souradeep, Tarun",
    title = "{Primordial power spectrum from Planck}",
    eprint = "1406.4827",
    archivePrefix = "arXiv",
    primaryClass = "astro-ph.CO",
    doi = "10.1088/1475-7516/2014/11/011",
    journal = "JCAP",
    volume = "11",
    pages = "011",
    year = "2014"
}

@article{Canas-Herrera:2020mme,
    author = "Ca{\~n}as-Herrera, Guadalupe and Torrado, Jes{\'u}s and Ach{\'u}carro, Ana",
    title = "{Bayesian reconstruction of the inflaton{\textquoteright}s speed of sound using CMB data}",
    eprint = "2012.04640",
    archivePrefix = "arXiv",
    primaryClass = "astro-ph.CO",
    doi = "10.1103/PhysRevD.103.123531",
    journal = "Phys. Rev. D",
    volume = "103",
    pages = "123531",
    year = "2021"
}

@article{Braglia:2021ckn,
    author = "Braglia, Matteo and Chen, Xingang and Hazra, Dhiraj Kumar",
    title = "{Comparing multi-field primordial feature models with the Planck data}",
    eprint = "2103.03025",
    archivePrefix = "arXiv",
    primaryClass = "astro-ph.CO",
    doi = "10.1088/1475-7516/2021/06/005",
    journal = "JCAP",
    volume = "06",
    pages = "005",
    year = "2021"
}

@article{Braglia:2021sun,
    author = "Braglia, Matteo and Chen, Xingang and Hazra, Dhiraj Kumar",
    title = "{Uncovering the history of cosmic inflation from anomalies in cosmic microwave background spectra}",
    eprint = "2106.07546",
    archivePrefix = "arXiv",
    primaryClass = "astro-ph.CO",
    reportNumber = "IFT-UAM/CSIC-21-128",
    doi = "10.1140/epjc/s10052-022-10461-3",
    journal = "Eur. Phys. J. C",
    volume = "82",
    number = "5",
    pages = "498",
    year = "2022"
}

@article{Braglia:2021rej,
    author = "Braglia, Matteo and Chen, Xingang and Hazra, Dhiraj Kumar",
    title = "{Primordial standard clock models and CMB residual anomalies}",
    eprint = "2108.10110",
    archivePrefix = "arXiv",
    primaryClass = "astro-ph.CO",
    reportNumber = "IFT-UAM/CSIC-21-129",
    doi = "10.1103/PhysRevD.105.103523",
    journal = "Phys. Rev. D",
    volume = "105",
    number = "10",
    pages = "103523",
    year = "2022"
}

@article{Braglia:2022ftm,
    author = "Braglia, Matteo and Chen, Xingang and Hazra, Dhiraj Kumar and Pinol, Lucas",
    title = "{Back to the features: assessing the discriminating power of future CMB missions on inflationary models}",
    eprint = "2210.07028",
    archivePrefix = "arXiv",
    primaryClass = "astro-ph.CO",
    reportNumber = "IFT-UAM/CSIC-22-125",
    doi = "10.1088/1475-7516/2023/03/014",
    journal = "JCAP",
    volume = "03",
    pages = "014",
    year = "2023"
}

@article{Petretti:2024mjy,
    author = "Petretti, Catherine and Braglia, Matteo and Chen, Xingang and Kumar Hazra, Dhiraj and Paban, Sonia",
    title = "{Investigating the origin of CMB large-scale features using LiteBIRD and CMB-S4}",
    eprint = "2411.03459",
    archivePrefix = "arXiv",
    primaryClass = "astro-ph.CO",
    doi = "10.1088/1475-7516/2025/06/035",
    journal = "JCAP",
    volume = "06",
    number = "035",
    pages = "035",
    year = "2025"
}

@article{Huang:2012mr,
    author = "Huang, Zhiqi and Verde, Licia and Vernizzi, Filippo",
    title = "{Constraining inflation with future galaxy redshift surveys}",
    eprint = "1201.5955",
    archivePrefix = "arXiv",
    primaryClass = "astro-ph.CO",
    doi = "10.1088/1475-7516/2012/04/005",
    journal = "JCAP",
    volume = "04",
    pages = "005",
    year = "2012"
}

@article{Hazra:2012vs,
    author = "Hazra, Dhiraj Kumar",
    title = "{Changes in the halo formation rates due to features in the primordial spectrum}",
    eprint = "1210.7170",
    archivePrefix = "arXiv",
    primaryClass = "astro-ph.CO",
    doi = "10.1088/1475-7516/2013/03/003",
    journal = "JCAP",
    volume = "03",
    pages = "003",
    year = "2013"
}

@article{Chen:2016vvw,
    author = "Chen, Xingang and Dvorkin, Cora and Huang, Zhiqi and Namjoo, Mohammad Hossein and Verde, Licia",
    title = "{The Future of Primordial Features with Large-Scale Structure Surveys}",
    eprint = "1605.09365",
    archivePrefix = "arXiv",
    primaryClass = "astro-ph.CO",
    doi = "10.1088/1475-7516/2016/11/014",
    journal = "JCAP",
    volume = "11",
    pages = "014",
    year = "2016"
}

@article{Ballardini:2016hpi,
    author = "Ballardini, Mario and Finelli, Fabio and Fedeli, Cosimo and Moscardini, Lauro",
    title = "{Probing primordial features with future galaxy surveys}",
    eprint = "1606.03747",
    archivePrefix = "arXiv",
    primaryClass = "astro-ph.CO",
    doi = "10.1088/1475-7516/2016/10/041",
    journal = "JCAP",
    volume = "10",
    pages = "041",
    year = "2016",
    note = "[Erratum: JCAP 04, E01 (2018)]"
}

@article{Palma:2017wxu,
    author = "Palma, Gonzalo A. and Sapone, Domenico and Sypsas, Spyros",
    title = "{Constraints on inflation with LSS surveys: features in the primordial power spectrum}",
    eprint = "1710.02570",
    archivePrefix = "arXiv",
    primaryClass = "astro-ph.CO",
    doi = "10.1088/1475-7516/2018/06/004",
    journal = "JCAP",
    volume = "06",
    pages = "004",
    year = "2018"
}

@article{LHuillier:2017lgm,
    author = "L'Huillier, Benjamin and Shafieloo, Arman and Hazra, Dhiraj Kumar and Smoot, George F. and Starobinsky, Alexei A.",
    title = "{Probing features in the primordial perturbation spectrum with large-scale structure data}",
    eprint = "1710.10987",
    archivePrefix = "arXiv",
    primaryClass = "astro-ph.CO",
    doi = "10.1093/mnras/sty745",
    journal = "Mon. Not. Roy. Astron. Soc.",
    volume = "477",
    number = "2",
    pages = "2503--2512",
    year = "2018"
}

@article{Ballardini:2017qwq,
    author = "Ballardini, Mario and Finelli, Fabio and Maartens, Roy and Moscardini, Lauro",
    title = "{Probing primordial features with next-generation photometric and radio surveys}",
    eprint = "1712.07425",
    archivePrefix = "arXiv",
    primaryClass = "astro-ph.CO",
    doi = "10.1088/1475-7516/2018/04/044",
    journal = "JCAP",
    volume = "04",
    pages = "044",
    year = "2018"
}

@article{Vasudevan:2019ewf,
    author = "Vasudevan, Anagha and Ivanov, Mikhail M. and Sibiryakov, Sergey and Lesgourgues, Julien",
    title = "{Time-sliced perturbation theory with primordial non-Gaussianity and effects of large bulk flows on inflationary oscillating features}",
    eprint = "1906.08697",
    archivePrefix = "arXiv",
    primaryClass = "astro-ph.CO",
    reportNumber = "CERN-TH-2019-091, INR-TH-2019-012, TTK-19-22",
    doi = "10.1088/1475-7516/2019/09/037",
    journal = "JCAP",
    volume = "09",
    pages = "037",
    year = "2019"
}

@article{Beutler:2019ojk,
    author = "Beutler, Florian and Biagetti, Matteo and Green, Daniel and Slosar, An{\v{z}}e and Wallisch, Benjamin",
    title = "{Primordial Features from Linear to Nonlinear Scales}",
    eprint = "1906.08758",
    archivePrefix = "arXiv",
    primaryClass = "astro-ph.CO",
    doi = "10.1103/PhysRevResearch.1.033209",
    journal = "Phys. Rev. Res.",
    volume = "1",
    number = "3",
    pages = "033209",
    year = "2019"
}

@article{Ballardini:2019tuc,
    author = "Ballardini, Mario and Murgia, Riccardo and Baldi, Marco and Finelli, Fabio and Viel, Matteo",
    title = "{Non-linear damping of superimposed primordial oscillations on the matter power spectrum in galaxy surveys}",
    eprint = "1912.12499",
    archivePrefix = "arXiv",
    primaryClass = "astro-ph.CO",
    doi = "10.1088/1475-7516/2020/04/030",
    journal = "JCAP",
    volume = "04",
    number = "04",
    pages = "030",
    year = "2020"
}

@article{Debono:2020emh,
    author = "Debono, Ivan and Hazra, Dhiraj Kumar and Shafieloo, Arman and Smoot, George F. and Starobinsky, Alexei A.",
    title = "{Constraints on features in the inflationary potential from future Euclid data}",
    eprint = "2003.05262",
    archivePrefix = "arXiv",
    primaryClass = "astro-ph.CO",
    doi = "10.1093/mnras/staa1765",
    journal = "Mon. Not. Roy. Astron. Soc.",
    volume = "496",
    number = "3",
    pages = "3448--3468",
    year = "2020"
}

@article{Chen:2020ckc,
    author = "Chen, Shi-Fan and Vlah, Zvonimir and White, Martin",
    title = "{Modeling features in the redshift-space halo power spectrum with perturbation theory}",
    eprint = "2007.00704",
    archivePrefix = "arXiv",
    primaryClass = "astro-ph.CO",
    doi = "10.1088/1475-7516/2020/11/035",
    journal = "JCAP",
    volume = "11",
    pages = "035",
    year = "2020"
}

@article{Li:2021jvz,
    author = "Li, Yuhao and Zhu, Hong-Ming and Li, Baojiu",
    title = "{Non-linear reconstruction of features in the primordial power spectrum from large-scale structure}",
    eprint = "2102.09007",
    archivePrefix = "arXiv",
    primaryClass = "astro-ph.CO",
    doi = "10.1093/mnras/stac1544",
    journal = "Mon. Not. Roy. Astron. Soc.",
    volume = "514",
    number = "3",
    pages = "4363--4378",
    year = "2022"
}

@article{Chen:2006xjb,
    author = "Chen, Xingang and Easther, Richard and Lim, Eugene A.",
    title = "{Large Non-Gaussianities in Single Field Inflation}",
    eprint = "astro-ph/0611645",
    archivePrefix = "arXiv",
    doi = "10.1088/1475-7516/2007/06/023",
    journal = "JCAP",
    volume = "06",
    pages = "023",
    year = "2007"
}

@article{Chen:2011zf,
    author = "Chen, Xingang",
    title = "{Primordial Features as Evidence for Inflation}",
    eprint = "1104.1323",
    archivePrefix = "arXiv",
    primaryClass = "hep-th",
    doi = "10.1088/1475-7516/2012/01/038",
    journal = "JCAP",
    volume = "01",
    pages = "038",
    year = "2012"
}

@article{Chen:2014cwa,
    author = "Chen, Xingang and Namjoo, Mohammad Hossein and Wang, Yi",
    title = "{Models of the Primordial Standard Clock}",
    eprint = "1411.2349",
    archivePrefix = "arXiv",
    primaryClass = "astro-ph.CO",
    doi = "10.1088/1475-7516/2015/02/027",
    journal = "JCAP",
    volume = "02",
    pages = "027",
    year = "2015"
}

@article{Quintin:2024boj,
    author = "Quintin, Jerome and Chen, Xingang and Ebadi, Reza",
    title = "{Fingerprints of a non-inflationary universe from massive fields}",
    eprint = "2405.11016",
    archivePrefix = "arXiv",
    primaryClass = "astro-ph.CO",
    doi = "10.1088/1475-7516/2024/09/026",
    journal = "JCAP",
    volume = "09",
    pages = "026",
    year = "2024"
}

@article{Wu:2020pej,
    author = "Wu, Yi-Peng and Petraki, Kalliopi",
    title = "{Stochastic Baryogenesis}",
    eprint = "2008.08549",
    archivePrefix = "arXiv",
    primaryClass = "hep-ph",
    reportNumber = "Nikhef-2020-025",
    doi = "10.1088/1475-7516/2021/01/022",
    journal = "JCAP",
    volume = "01",
    pages = "022",
    year = "2021"
}

@article{Planck:2018vyg,
    author = "Aghanim, N. and others",
    collaboration = "Planck",
    title = "{Planck 2018 results. VI. Cosmological parameters}",
    eprint = "1807.06209",
    archivePrefix = "arXiv",
    primaryClass = "astro-ph.CO",
    doi = "10.1051/0004-6361/201833910",
    journal = "Astron. Astrophys.",
    volume = "641",
    pages = "A6",
    year = "2020",
    note = "[Erratum: Astron.Astrophys. 652, C4 (2021)]"
}

@article{Kusenko:1997si,
    author = "Kusenko, Alexander and Shaposhnikov, Mikhail E.",
    title = "{Supersymmetric Q balls as dark matter}",
    eprint = "hep-ph/9709492",
    archivePrefix = "arXiv",
    reportNumber = "CERN-TH-97-259",
    doi = "10.1016/S0370-2693(97)01375-0",
    journal = "Phys. Lett. B",
    volume = "418",
    pages = "46--54",
    year = "1998"
}

@article{BICEP:2021xfz,
    author = "Ade, P. A. R. and others",
    collaboration = "BICEP, Keck",
    title = "{Improved Constraints on Primordial Gravitational Waves using Planck, WMAP, and BICEP/Keck Observations through the 2018 Observing Season}",
    eprint = "2110.00483",
    archivePrefix = "arXiv",
    primaryClass = "astro-ph.CO",
    doi = "10.1103/PhysRevLett.127.151301",
    journal = "Phys. Rev. Lett.",
    volume = "127",
    number = "15",
    pages = "151301",
    year = "2021"
}

@article{Hook:2015foa,
    author = "Hook, Anson",
    title = "{Baryogenesis in a CP invariant theory}",
    eprint = "1508.05094",
    archivePrefix = "arXiv",
    primaryClass = "hep-ph",
    doi = "10.1007/JHEP11(2015)143",
    journal = "JHEP",
    volume = "11",
    pages = "143",
    year = "2015"
}

@article{Pajer:2024ckd,
    author = "Pajer, Enrico and Wang, Dong-Gang and Zhang, Bowei",
    title = "{The UV Sensitivity of Axion Monodromy Inflation}",
    eprint = "2412.05762",
    archivePrefix = "arXiv",
    primaryClass = "hep-th",
    month = "12",
    year = "2024"
}

@article{Chen:2023txq,
    author = "Chen, Xingang and Fan, JiJi and Li, Lingfeng",
    title = "{New inflationary probes of axion dark matter}",
    eprint = "2303.03406",
    archivePrefix = "arXiv",
    primaryClass = "hep-ph",
    doi = "10.1007/JHEP12(2023)197",
    journal = "JHEP",
    volume = "12",
    pages = "197",
    year = "2023"
}

@article{Antoniadis:2011ib,
    author = "Antoniadis, Ignatios and Mazur, Pawel O. and Mottola, Emil",
    title = "{Conformal Invariance, Dark Energy, and CMB Non-Gaussianity}",
    eprint = "1103.4164",
    archivePrefix = "arXiv",
    primaryClass = "gr-qc",
    reportNumber = "LA-UR-11-10115, CERN-PH-TH-2011-057",
    doi = "10.1088/1475-7516/2012/09/024",
    journal = "JCAP",
    volume = "09",
    pages = "024",
    year = "2012"
}

@article{Ng:2021hll,
    author = "Ng, Kin-Wang and Wu, Yi-Peng",
    title = "{Constant-rate inflation: primordial black holes from conformal weight transitions}",
    eprint = "2102.05620",
    archivePrefix = "arXiv",
    primaryClass = "astro-ph.CO",
    doi = "10.1007/JHEP11(2021)076",
    journal = "JHEP",
    volume = "11",
    pages = "076",
    year = "2021"
}

@article{Wu:2019ohx,
    author = "Wu, Yi-Peng and Yang, Louis and Kusenko, Alexander",
    title = "{Leptogenesis from spontaneous symmetry breaking during inflation}",
    eprint = "1905.10537",
    archivePrefix = "arXiv",
    primaryClass = "hep-ph",
    reportNumber = "RESCEU 6/19, IPMU19-0079",
    doi = "10.1007/JHEP12(2019)088",
    journal = "JHEP",
    volume = "12",
    pages = "088",
    year = "2019"
}

@article{Chen:2008wn,
    author = "Chen, Xingang and Easther, Richard and Lim, Eugene A.",
    title = "{Generation and Characterization of Large Non-Gaussianities in Single Field Inflation}",
    eprint = "0801.3295",
    archivePrefix = "arXiv",
    primaryClass = "astro-ph",
    doi = "10.1088/1475-7516/2008/04/010",
    journal = "JCAP",
    volume = "04",
    pages = "010",
    year = "2008"
}

@article{Adshead:2011jq,
    author = "Adshead, Peter and Dvorkin, Cora and Hu, Wayne and Lim, Eugene A.",
    title = "{Non-Gaussianity from Step Features in the Inflationary Potential}",
    eprint = "1110.3050",
    archivePrefix = "arXiv",
    primaryClass = "astro-ph.CO",
    doi = "10.1103/PhysRevD.85.023531",
    journal = "Phys. Rev. D",
    volume = "85",
    pages = "023531",
    year = "2012"
}

@article{Hazra:2012yn,
    author = "Hazra, Dhiraj Kumar and Sriramkumar, L. and Martin, Jerome",
    title = "{BINGO: A code for the efficient computation of the scalar bi-spectrum}",
    eprint = "1201.0926",
    archivePrefix = "arXiv",
    primaryClass = "astro-ph.CO",
    doi = "10.1088/1475-7516/2013/05/026",
    journal = "JCAP",
    volume = "05",
    pages = "026",
    year = "2013"
}

@article{Bartolo:2013exa,
    author = "Bartolo, Nicola and Cannone, Dario and Matarrese, Sabino",
    title = "{The Effective Field Theory of Inflation Models with Sharp Features}",
    eprint = "1307.3483",
    archivePrefix = "arXiv",
    primaryClass = "astro-ph.CO",
    doi = "10.1088/1475-7516/2013/10/038",
    journal = "JCAP",
    volume = "10",
    pages = "038",
    year = "2013"
}

@article{Fergusson:2014hya,
    author = "Fergusson, J. R. and Gruetjen, H. F. and Shellard, E. P. S. and Liguori, M.",
    title = "{Combining power spectrum and bispectrum measurements to detect oscillatory features}",
    eprint = "1410.5114",
    archivePrefix = "arXiv",
    primaryClass = "astro-ph.CO",
    doi = "10.1103/PhysRevD.91.023502",
    journal = "Phys. Rev. D",
    volume = "91",
    number = "2",
    pages = "023502",
    year = "2015"
}

@article{Fergusson:2014tza,
    author = "Fergusson, J. R. and Gruetjen, H. F. and Shellard, E. P. S. and Wallisch, B.",
    title = "{Polyspectra searches for sharp oscillatory features in cosmic microwave sky data}",
    eprint = "1412.6152",
    archivePrefix = "arXiv",
    primaryClass = "astro-ph.CO",
    doi = "10.1103/PhysRevD.91.123506",
    journal = "Phys. Rev. D",
    volume = "91",
    number = "12",
    pages = "123506",
    year = "2015"
}

@article{Chen:2009we,
    author = "Chen, Xingang and Wang, Yi",
    title = "{Large non-Gaussianities with Intermediate Shapes from Quasi-Single Field Inflation}",
    eprint = "0909.0496",
    archivePrefix = "arXiv",
    primaryClass = "astro-ph.CO",
    reportNumber = "MIT-CTP-4071",
    doi = "10.1103/PhysRevD.81.063511",
    journal = "Phys. Rev. D",
    volume = "81",
    pages = "063511",
    year = "2010"
}

@article{Chen:2009zp,
    author = "Chen, Xingang and Wang, Yi",
    title = "{Quasi-Single Field Inflation and Non-Gaussianities}",
    eprint = "0911.3380",
    archivePrefix = "arXiv",
    primaryClass = "hep-th",
    doi = "10.1088/1475-7516/2010/04/027",
    journal = "JCAP",
    volume = "04",
    pages = "027",
    year = "2010"
}

@article{Baumann:2011nk,
    author = "Baumann, Daniel and Green, Daniel",
    title = "{Signatures of Supersymmetry from the Early Universe}",
    eprint = "1109.0292",
    archivePrefix = "arXiv",
    primaryClass = "hep-th",
    doi = "10.1103/PhysRevD.85.103520",
    journal = "Phys. Rev. D",
    volume = "85",
    pages = "103520",
    year = "2012"
}

@article{Noumi:2012vr,
    author = "Noumi, Toshifumi and Yamaguchi, Masahide and Yokoyama, Daisuke",
    title = "{Effective field theory approach to quasi-single field inflation and effects of heavy fields}",
    eprint = "1211.1624",
    archivePrefix = "arXiv",
    primaryClass = "hep-th",
    reportNumber = "UT-KOMABA-12-9, TIT-HEP-625",
    doi = "10.1007/JHEP06(2013)051",
    journal = "JHEP",
    volume = "06",
    pages = "051",
    year = "2013"
}

@article{Arkani-Hamed:2015bza,
    author = "Arkani-Hamed, Nima and Maldacena, Juan",
    title = "{Cosmological Collider Physics}",
    eprint = "1503.08043",
    archivePrefix = "arXiv",
    primaryClass = "hep-th",
    month = "3",
    year = "2015"
}

@article{Chen:2016uwp,
    author = "Chen, Xingang and Wang, Yi and Xianyu, Zhong-Zhi",
    title = "{Standard Model Background of the Cosmological Collider}",
    eprint = "1610.06597",
    archivePrefix = "arXiv",
    primaryClass = "hep-th",
    doi = "10.1103/PhysRevLett.118.261302",
    journal = "Phys. Rev. Lett.",
    volume = "118",
    number = "26",
    pages = "261302",
    year = "2017"
}

@article{Kumar:2018jxz,
    author = "Kumar, Soubhik and Sundrum, Raman",
    title = "{Seeing Higher-Dimensional Grand Unification In Primordial Non-Gaussianities}",
    eprint = "1811.11200",
    archivePrefix = "arXiv",
    primaryClass = "hep-ph",
    reportNumber = "UMD-PP-018-09",
    doi = "10.1007/JHEP04(2019)120",
    journal = "JHEP",
    volume = "04",
    pages = "120",
    year = "2019"
}

@article{Hook:2019vcn,
    author = "Hook, Anson and Huang, Junwu and Racco, Davide",
    title = "{Minimal signatures of the Standard Model in non-Gaussianities}",
    eprint = "1908.00019",
    archivePrefix = "arXiv",
    primaryClass = "hep-ph",
    doi = "10.1103/PhysRevD.101.023519",
    journal = "Phys. Rev. D",
    volume = "101",
    number = "2",
    pages = "023519",
    year = "2020"
}

@article{Wang:2019gbi,
    author = "Wang, Lian-Tao and Xianyu, Zhong-Zhi",
    title = "{In Search of Large Signals at the Cosmological Collider}",
    eprint = "1910.12876",
    archivePrefix = "arXiv",
    primaryClass = "hep-ph",
    doi = "10.1007/JHEP02(2020)044",
    journal = "JHEP",
    volume = "02",
    pages = "044",
    year = "2020"
}

@article{Liu:2019fag,
    author = "Liu, Tao and Tong, Xi and Wang, Yi and Xianyu, Zhong-Zhi",
    title = "{Probing P and CP Violations on the Cosmological Collider}",
    eprint = "1909.01819",
    archivePrefix = "arXiv",
    primaryClass = "hep-ph",
    doi = "10.1007/JHEP04(2020)189",
    journal = "JHEP",
    volume = "04",
    pages = "189",
    year = "2020"
}

@article{Li:2020xwr,
    author = "Li, Lingfeng and Lu, Shiyun and Wang, Yi and Zhou, Siyi",
    title = "{Cosmological Signatures of Superheavy Dark Matter}",
    eprint = "2002.01131",
    archivePrefix = "arXiv",
    primaryClass = "hep-ph",
    doi = "10.1007/JHEP07(2020)231",
    journal = "JHEP",
    volume = "07",
    pages = "231",
    year = "2020"
}

@article{Cui:2021iie,
    author = "Cui, Yanou and Xianyu, Zhong-Zhi",
    title = "{Probing Leptogenesis with the Cosmological Collider}",
    eprint = "2112.10793",
    archivePrefix = "arXiv",
    primaryClass = "hep-ph",
    doi = "10.1103/PhysRevLett.129.111301",
    journal = "Phys. Rev. Lett.",
    volume = "129",
    number = "11",
    pages = "111301",
    year = "2022"
}

@article{Chen:2022vzh,
    author = "Chen, Xingang and Ebadi, Reza and Kumar, Soubhik",
    title = "{Classical cosmological collider physics and primordial features}",
    eprint = "2205.01107",
    archivePrefix = "arXiv",
    primaryClass = "hep-ph",
    doi = "10.1088/1475-7516/2022/08/083",
    journal = "JCAP",
    volume = "08",
    pages = "083",
    year = "2022"
}

@article{Wu:2024wti,
    author = "Wu, Yi-Peng",
    title = "{The cosmological collider in R $^{2}$ inflation}",
    eprint = "2404.05031",
    archivePrefix = "arXiv",
    primaryClass = "astro-ph.CO",
    doi = "10.1088/1475-7516/2024/07/010",
    journal = "JCAP",
    volume = "07",
    pages = "010",
    year = "2024"
}

@article{Sohn:2024xzd,
    author = "Sohn, Wuhyun and Wang, Dong-Gang and Fergusson, James R. and Shellard, E. P. S.",
    title = "{Searching for cosmological collider in the Planck CMB data}",
    eprint = "2404.07203",
    archivePrefix = "arXiv",
    primaryClass = "astro-ph.CO",
    doi = "10.1088/1475-7516/2024/09/016",
    journal = "JCAP",
    volume = "09",
    pages = "016",
    year = "2024"
}

@article{Chen:2011tu,
    author = "Chen, Xingang",
    title = "{Fingerprints of Primordial Universe Paradigms as Features in Density Perturbations}",
    eprint = "1106.1635",
    archivePrefix = "arXiv",
    primaryClass = "astro-ph.CO",
    doi = "10.1016/j.physletb.2011.11.009",
    journal = "Phys. Lett. B",
    volume = "706",
    pages = "111--115",
    year = "2011"
}

@article{Chen:2012ja,
    author = "Chen, Xingang and Ringeval, Christophe",
    title = "{Searching for Standard Clocks in the Primordial Universe}",
    eprint = "1205.6085",
    archivePrefix = "arXiv",
    primaryClass = "astro-ph.CO",
    doi = "10.1088/1475-7516/2012/08/014",
    journal = "JCAP",
    volume = "08",
    pages = "014",
    year = "2012"
}

@article{Saito:2012pd,
    author = "Saito, Ryo and Nakashima, Masahiro and Takamizu, Yu-ichi and Yokoyama, Jun'ichi",
    title = "{Resonant Signatures of Heavy Scalar Fields in the Cosmic Microwave Background}",
    eprint = "1206.2164",
    archivePrefix = "arXiv",
    primaryClass = "astro-ph.CO",
    reportNumber = "YITP-12-50",
    doi = "10.1088/1475-7516/2012/11/036",
    journal = "JCAP",
    volume = "11",
    pages = "036",
    year = "2012"
}

@article{Battefeld:2013xka,
    author = "Battefeld, Thorsten and Niemeyer, Jens C. and Vlaykov, Dimitar",
    title = "{Probing Two-Field Open Inflation by Resonant Signals in Correlation Functions}",
    eprint = "1302.3877",
    archivePrefix = "arXiv",
    primaryClass = "astro-ph.CO",
    doi = "10.1088/1475-7516/2013/05/006",
    journal = "JCAP",
    volume = "05",
    pages = "006",
    year = "2013"
}

@article{Saito:2013aqa,
    author = "Saito, Ryo and Takamizu, Yu-ichi",
    title = "{Localized Features in Non-Gaussianity from Heavy Physics}",
    eprint = "1303.3839",
    archivePrefix = "arXiv",
    primaryClass = "astro-ph.CO",
    doi = "10.1088/1475-7516/2013/06/031",
    journal = "JCAP",
    volume = "06",
    pages = "031",
    year = "2013"
}

@article{Gao:2013ota,
    author = "Gao, Xian and Langlois, David and Mizuno, Shuntaro",
    title = "{Oscillatory features in the curvature power spectrum after a sudden turn of the inflationary trajectory}",
    eprint = "1306.5680",
    archivePrefix = "arXiv",
    primaryClass = "hep-th",
    doi = "10.1088/1475-7516/2013/10/023",
    journal = "JCAP",
    volume = "10",
    pages = "023",
    year = "2013"
}

@article{Noumi:2013cfa,
    author = "Noumi, Toshifumi and Yamaguchi, Masahide",
    title = "{Primordial spectra from sudden turning trajectory}",
    eprint = "1307.7110",
    archivePrefix = "arXiv",
    primaryClass = "hep-th",
    reportNumber = "RIKEN-MP-73",
    doi = "10.1088/1475-7516/2013/12/038",
    journal = "JCAP",
    volume = "12",
    pages = "038",
    year = "2013"
}

@article{Chen:2014joa,
    author = "Chen, Xingang and Namjoo, Mohammad Hossein",
    title = "{Standard Clock in Primordial Density Perturbations and Cosmic Microwave Background}",
    eprint = "1404.1536",
    archivePrefix = "arXiv",
    primaryClass = "astro-ph.CO",
    doi = "10.1016/j.physletb.2014.11.002",
    journal = "Phys. Lett. B",
    volume = "739",
    pages = "285--292",
    year = "2014"
}

@article{Huang:2016quc,
    author = "Huang, Qing-Guo and Pi, Shi",
    title = "{Power-law modulation of the scalar power spectrum from a heavy field with a monomial potential}",
    eprint = "1610.00115",
    archivePrefix = "arXiv",
    primaryClass = "hep-th",
    doi = "10.1088/1475-7516/2018/04/001",
    journal = "JCAP",
    volume = "04",
    pages = "001",
    year = "2018"
}

@article{Domenech:2018bnf,
    author = "Dom{\`e}nech, Guillem and Rubio, Javier and Wons, Julius",
    title = "{Mimicking features in alternatives to inflation with interacting spectator fields}",
    eprint = "1811.08224",
    archivePrefix = "arXiv",
    primaryClass = "astro-ph.CO",
    reportNumber = "HIP-2018-28/TH",
    doi = "10.1016/j.physletb.2019.01.039",
    journal = "Phys. Lett. B",
    volume = "790",
    pages = "263--269",
    year = "2019"
}

@article{Wang:2025qww,
    author = "Wang, Dong-Gang and Zhang, Bowei",
    title = "{Bootstrapping the cosmological collider with resonant features}",
    eprint = "2505.19066",
    archivePrefix = "arXiv",
    primaryClass = "hep-th",
    doi = "10.1007/JHEP09(2025)122",
    journal = "JHEP",
    volume = "09",
    pages = "122",
    year = "2025"
}

@article{Jazayeri:2025vlv,
    author = "Jazayeri, Sadra and Tong, Xi and Zhu, Yuhang",
    title = "{Every Wrinkle Carries A Memory: An Integro-differential Bootstrap for Features in Cosmological Correlators}",
    eprint = "2511.00152",
    archivePrefix = "arXiv",
    primaryClass = "hep-th",
    month = "10",
    year = "2025"
}

@article{Starobinsky:1985ibc,
    author = "Starobinsky, Alexei A.",
    title = "{Multicomponent de Sitter (Inflationary) Stages and the Generation of Perturbations}",
    journal = "JETP Lett.",
    volume = "42",
    pages = "152--155",
    year = "1985"
}

@article{Sasaki:1995aw,
    author = "Sasaki, Misao and Stewart, Ewan D.",
    title = "{A General analytic formula for the spectral index of the density perturbations produced during inflation}",
    eprint = "astro-ph/9507001",
    archivePrefix = "arXiv",
    reportNumber = "LANCS-TH-9504, OU-TAP-22",
    doi = "10.1143/PTP.95.71",
    journal = "Prog. Theor. Phys.",
    volume = "95",
    pages = "71--78",
    year = "1996"
}

@article{Sasaki:1998ug,
    author = "Sasaki, Misao and Tanaka, Takahiro",
    title = "{Superhorizon scale dynamics of multiscalar inflation}",
    eprint = "gr-qc/9801017",
    archivePrefix = "arXiv",
    reportNumber = "OU-TAP-72",
    doi = "10.1143/PTP.99.763",
    journal = "Prog. Theor. Phys.",
    volume = "99",
    pages = "763--782",
    year = "1998"
}

@article{Lyth:2004gb,
    author = "Lyth, David H. and Malik, Karim A. and Sasaki, Misao",
    title = "{A General proof of the conservation of the curvature perturbation}",
    eprint = "astro-ph/0411220",
    archivePrefix = "arXiv",
    reportNumber = "YITP-04-67",
    doi = "10.1088/1475-7516/2005/05/004",
    journal = "JCAP",
    volume = "05",
    pages = "004",
    year = "2005"
}

@article{Suman:2025vuf,
    author = "Suman, Petar and Wang, Dong-Gang and Sohn, Wuhyun and Fergusson, James R. and Shellard, E. P. S.",
    title = "{How Significant are Cosmological Collider Signals in the Planck Data?}",
    eprint = "2511.17500",
    archivePrefix = "arXiv",
    primaryClass = "astro-ph.CO",
    month = "11",
    year = "2025"
}

@article{Bodas:2024hih,
    author = "Bodas, Arushi and Broadberry, Edward and Sundrum, Raman",
    title = "{Grand unification at the cosmological collider with chemical potential}",
    eprint = "2409.07524",
    archivePrefix = "arXiv",
    primaryClass = "hep-ph",
    reportNumber = "FERMILAB-PUB-24-0568-V",
    doi = "10.1007/JHEP01(2025)115",
    journal = "JHEP",
    volume = "01",
    pages = "115",
    year = "2025"
}

@article{Cabass:2024wob,
    author = "Cabass, Giovanni and Philcox, Oliver H. E. and Ivanov, Mikhail M. and Akitsu, Kazuyuki and Chen, Shi-Fan and Simonovi{\'c}, Marko and Zaldarriaga, Matias",
    title = "{BOSS constraints on massive particles during inflation: The cosmological collider in action}",
    eprint = "2404.01894",
    archivePrefix = "arXiv",
    primaryClass = "astro-ph.CO",
    reportNumber = "RBI-ThPhys-2024-21, MIT-CTP/5698",
    doi = "10.1103/PhysRevD.111.063510",
    journal = "Phys. Rev. D",
    volume = "111",
    number = "6",
    pages = "063510",
    year = "2025"
}

@article{Anbajagane:2025uro,
    author = "Anbajagane, Dhayaa and Lee, Hayden",
    title = "{Primordial Physics in the Nonlinear Universe: mapping cosmological collider models to weak-lensing observables}",
    eprint = "2509.02693",
    archivePrefix = "arXiv",
    primaryClass = "astro-ph.CO",
    month = "9",
    year = "2025"
}

@article{Suman:2025tpv,
    author = "Suman, Petar and Wang, Dong-Gang and Sohn, Wuhyun and Fergusson, James R. and Shellard, E. P. S.",
    title = "{Searching for Cosmological Collider in the Planck CMB Data II: collider templates and Modal analysis}",
    eprint = "2512.22085",
    archivePrefix = "arXiv",
    primaryClass = "astro-ph.CO",
    month = "12",
    year = "2025"
}

@article{Jiang:2025mlm,
    author = "Jiang, Yikun and Pimentel, Guilherme L. and Yang, Chen",
    title = "{Strongly Coupled Sectors in Inflation: Gapped Theories of Unparticles}",
    eprint = "2512.23796",
    archivePrefix = "arXiv",
    primaryClass = "hep-th",
    month = "12",
    year = "2025"
}

@article{Fujikura:2025xgl,
    author = "Fujikura, Kohei and Noumi, Toshifumi",
    title = "{Inflationary QCD phase diagram}",
    eprint = "2512.24024",
    archivePrefix = "arXiv",
    primaryClass = "hep-ph",
    month = "12",
    year = "2025"
}

@article{Enqvist:2003gh,
    author = "Enqvist, Kari and Mazumdar, Anupam",
    title = "{Cosmological consequences of MSSM flat directions}",
    eprint = "hep-ph/0209244",
    archivePrefix = "arXiv",
    reportNumber = "HIP-2002-46-TH",
    doi = "10.1016/S0370-1573(03)00119-4",
    journal = "Phys. Rept.",
    volume = "380",
    pages = "99--234",
    year = "2003"
}

@article{Luty:1992un,
    author = "Luty, M. A.",
    title = "{Baryogenesis via leptogenesis}",
    doi = "10.1103/PhysRevD.45.455",
    journal = "Phys. Rev. D",
    volume = "45",
    pages = "455--465",
    year = "1992"
}

@inproceedings{Cline:2006ts,
    author = "Cline, James M.",
    title = "{Baryogenesis}",
    booktitle = "{Les Houches Summer School - Session 86: Particle Physics and Cosmology: The Fabric of Spacetime}",
    eprint = "hep-ph/0609145",
    archivePrefix = "arXiv",
    month = "9",
    year = "2006"
}

@article{Bodas:2022zca,
    author = "Bodas, Arushi and Sundrum, Raman",
    title = "{Primordial clocks within stochastic gravitational wave anisotropies}",
    eprint = "2205.04482",
    archivePrefix = "arXiv",
    primaryClass = "astro-ph.CO",
    doi = "10.1088/1475-7516/2022/10/012",
    journal = "JCAP",
    volume = "10",
    pages = "012",
    year = "2022"
}

@article{Co:2024cji,
    author = "Co, Raymond T. and Lee, Taegyu and Tadepalli, Sai Chaitanya",
    title = "{Non-Gaussianity from explicit U(1)-breaking interactions}",
    eprint = "2412.13246",
    archivePrefix = "arXiv",
    primaryClass = "astro-ph.CO",
    doi = "10.1088/1475-7516/2025/07/070",
    journal = "JCAP",
    volume = "07",
    pages = "070",
    year = "2025"
}

@article{Charng:2008ke,
    author = "Charng, Yeo-Yie and Lee, Da-Shin and Leung, Chung Ngoc and Ng, Kin-Wang",
    title = "{Affleck-Dine Baryogenesis, Split Supersymmetry, and Inflation}",
    eprint = "0802.1328",
    archivePrefix = "arXiv",
    primaryClass = "hep-ph",
    doi = "10.1103/PhysRevD.80.063519",
    journal = "Phys. Rev. D",
    volume = "80",
    pages = "063519",
    year = "2009"
}

@article{Dine:1995uk,
    author = "Dine, Michael and Randall, Lisa and Thomas, Scott D.",
    title = "{Supersymmetry breaking in the early universe}",
    eprint = "hep-ph/9503303",
    archivePrefix = "arXiv",
    reportNumber = "SLAC-PUB-6776, SLAC-PUB-95-6776, SCIPP-95-13, MIT-CTP-2421",
    doi = "10.1103/PhysRevLett.75.398",
    journal = "Phys. Rev. Lett.",
    volume = "75",
    pages = "398--401",
    year = "1995"
}

@article{Ellis:1987rw,
    author = "Ellis, John R. and Enqvist, K. and Nanopoulos, Dimitri V. and Olive, Keith A.",
    title = "{Inflationary Fluctuations, Entropy Generation and Baryogenesis}",
    reportNumber = "UMN-TH-597/87, MAD/TH/87-06, CERN-TH-4667/87",
    doi = "10.1016/0370-2693(87)90620-4",
    journal = "Phys. Lett. B",
    volume = "191",
    pages = "343--348",
    year = "1987"
}

@article{Ng:1988yn,
    author = "Ng, Kin-Wang",
    title = "{Residual F Terms and the Affleck-dine Mechanism for Baryogenesis}",
    reportNumber = "UMN-TH-704/88",
    doi = "10.1016/0550-3213(89)90355-6",
    journal = "Nucl. Phys. B",
    volume = "321",
    pages = "528--540",
    year = "1989"
}

@article{Enqvist:1987wi,
    author = "Enqvist, K. and Ng, K. W. and Olive, Keith A.",
    title = "{Scalar Field Fluctuations and the Baryon Asymmetry of the Universe}",
    reportNumber = "UMN-638-87, HU-TFT-87-48",
    doi = "10.1103/PhysRevD.37.2111",
    journal = "Phys. Rev. D",
    volume = "37",
    pages = "2111",
    year = "1988"
}

@article{Brax:2005jv,
	author = "Brax, Philippe and Martin, Jerome",
	title = "{Shift symmetry and inflation in supergravity}",
	eprint = "hep-th/0504168",
	archivePrefix = "arXiv",
	doi = "10.1103/PhysRevD.72.023518",
	journal = "Phys. Rev. D",
	volume = "72",
	pages = "023518",
	year = "2005"
}

@article{Antusch:2009ef,
	author = "Antusch, Stefan and Dutta, Koushik and Kostka, Philipp M.",
	title = "{SUGRA Hybrid Inflation with Shift Symmetry}",
	eprint = "0902.2934",
	archivePrefix = "arXiv",
	primaryClass = "hep-ph",
	reportNumber = "MPP-2009-18",
	doi = "10.1016/j.physletb.2009.05.043",
	journal = "Phys. Lett. B",
	volume = "677",
	pages = "221--225",
	year = "2009"
}

@article{Arkani-Hamed:2004ymt,
	author = "Arkani-Hamed, Nima and Dimopoulos, Savas",
	title = "{Supersymmetric unification without low energy supersymmetry and signatures for fine-tuning at the LHC}",
	eprint = "hep-th/0405159",
	archivePrefix = "arXiv",
	doi = "10.1088/1126-6708/2005/06/073",
	journal = "JHEP",
	volume = "06",
	pages = "073",
	year = "2005"
}

@article{Giudice:2004tc,
	author = "Giudice, G. F. and Romanino, A.",
	title = "{Split supersymmetry}",
	eprint = "hep-ph/0406088",
	archivePrefix = "arXiv",
	reportNumber = "CERN-PH-TH-2004-100",
	doi = "10.1016/j.nuclphysb.2004.08.001",
	journal = "Nucl. Phys. B",
	volume = "699",
	pages = "65--89",
	year = "2004",
	note = "[Erratum: Nucl.Phys.B 706, 487--487 (2005)]"
}

@article{Arkani-Hamed:2004zhs,
	author = "Arkani-Hamed, N. and Dimopoulos, S. and Giudice, G. F. and Romanino, A.",
	title = "{Aspects of split supersymmetry}",
	eprint = "hep-ph/0409232",
	archivePrefix = "arXiv",
	reportNumber = "CERN-PH-TH-2004-183",
	doi = "10.1016/j.nuclphysb.2004.12.026",
	journal = "Nucl. Phys. B",
	volume = "709",
	pages = "3--46",
	year = "2005"
}

@article{Kasuya:2005sb,
	author = "Kasuya, Shinta and Takahashi, Fuminobu",
	title = "{Smallness of baryon asymmetry from split supersymmetry}",
	eprint = "hep-ph/0501240",
	archivePrefix = "arXiv",
	doi = "10.1103/PhysRevD.71.121303",
	journal = "Phys. Rev. D",
	volume = "71",
	pages = "121303",
	year = "2005"
}

@article{Kasuya:2008xp,
	author = "Kasuya, Shinta and Kawasaki, Masahiro and Takahashi, Fuminobu",
	title = "{Isocurvature fluctuations in Affleck-Dine mechanism and constraints on inflation models}",
	eprint = "0805.4245",
	archivePrefix = "arXiv",
	primaryClass = "hep-ph",
	reportNumber = "IPMU-08-0032",
	doi = "10.1088/1475-7516/2008/10/017",
	journal = "JCAP",
	volume = "10",
	pages = "017",
	year = "2008"
}

@article{Kawasaki:2008jy,
	author = "Kawasaki, Masahiro and Nakayama, Kazunori and Takahashi, Fuminobu",
	title = "{Non-Gaussianity from Baryon Asymmetry}",
	eprint = "0809.2242",
	archivePrefix = "arXiv",
	primaryClass = "hep-ph",
	reportNumber = "IPMU-08-0061, ICRR-REPORT-528",
	doi = "10.1088/1475-7516/2009/01/002",
	journal = "JCAP",
	volume = "01",
	pages = "002",
	year = "2009"
}

@article{Allahverdi:2000zd,
	author = "Allahverdi, Rouzbeh and Campbell, Bruce A. and Ellis, John R.",
	title = "{Reheating and supersymmetric flat direction baryogenesis}",
	eprint = "hep-ph/0001122",
	archivePrefix = "arXiv",
	reportNumber = "ALBERTA-THY-19-99, CERN-TH-99-392",
	doi = "10.1016/S0550-3213(00)00124-3",
	journal = "Nucl. Phys. B",
	volume = "579",
	pages = "355--375",
	year = "2000"
}

@article{Anisimov:2000wx,
	author = "Anisimov, Alexey and Dine, Michael",
	title = "{Some issues in flat direction baryogenesis}",
	eprint = "hep-ph/0008058",
	archivePrefix = "arXiv",
	reportNumber = "SCIPP-00-26",
	doi = "10.1016/S0550-3213(01)00550-8",
	journal = "Nucl. Phys. B",
	volume = "619",
	pages = "729--740",
	year = "2001"
}

@article{Craig:2020jfv,
	author = "Craig, Nathaniel and Levi, Noam and Mariotti, Alberto and Redigolo, Diego",
	title = "{Ripples in Spacetime from Broken Supersymmetry}",
	eprint = "2011.13949",
	archivePrefix = "arXiv",
	primaryClass = "hep-ph",
	doi = "10.1007/JHEP02(2021)184",
	journal = "JHEP",
	volume = "21",
	pages = "184",
	year = "2020"
}
    
\end{document}